\documentclass[pre,twocolumn,secnumarabic,nobalancelastpage,amsmath,amssymb,showpacs,
nofootinbib]{revtex4-1}

\usepackage{graphics}      % standard graphics specifications
\usepackage{graphicx}      % alternative graphics specifications
\usepackage{longtable}     % helps with long table options
\usepackage{url}           % for on-line citations
\usepackage{bm}            % special 'bold-math' package
\usepackage{amsmath}

\begin{document}
\title{Maximum Entropy and the Stress Distribution in Soft Disk Packings Above Jamming}
\author         {Yegang Wu and S. Teitel}
\affiliation    {Department of Physics and Astronomy, University of Rochester, Rochester, New York 14627, USA}
\date{\today}

\begin{abstract}
We show that the maximum entropy hypothesis can successfully explain the distribution of stresses on compact clusters of particles within disordered mechanically stable packings of soft, isotropically stressed, frictionless disks above the jamming transition.  We show that, in our two dimensional case, it becomes necessary to consider not only the stress but also the Maxwell-Cremona force-tile area, as a constraining variable that determines the stress distribution.  The importance of the force-tile area had been suggested by earlier computations on an idealized force-network ensemble.
\end{abstract}
\maketitle

%%%%%%%%%%%%%%%%%%%%%%%%%%%%%%%%%%%%%%%%%%%%%%%%%%%%%%%%%%%%%%%%%%%%%%%%%%%%%
\section{Introduction}
\label{secIntro}

As the density of granular particles increases to a critical packing fraction, $\phi_J$, the system undergoes a {\em jamming transition} from a liquid-like to a solid-like state \cite{Liu+Nagel,OHern}.  
For large particles thermal fluctuations are irrelevant, and in the absence of mechanical agitation, the dense system relaxes into a mechanically stable {\em rigid} but {\em disordered} configuration.
Given a set of macroscopic constrains there are in general a large number of such configurations that are accessible to the system.  A long standing question is whether there is a convenient statistical description for the properties of such quenched configurations.

For hard-core, rough (i.e. frictional), particles, the jamming $\phi_J$ (and hence the system volume at jamming) may span a range of values from random loose packed to random close packed. Edwards and co-workers \cite{Edwards} proposed a statistical description for the distribution of the Voronoi volume of such particles in terms of a maximum entropy hypothesis, assuming that all accessible states are equally likely.
Henkes and co-workers \cite{Henkes1,Henkes2} extended these ideas to consider the distribution of stress on clusters of particles within packings of frictionless soft particles, compressed {\em above} the jamming $\phi_J$.   They denoted their formalism as the {\em stress ensemble}.  Similar ideas were then proposed by Blumenfeld and Edwards \cite{Blumenfeld}.
Subsequently, Tighe and co-workers \cite{Tighe1,Tighe3,Tighe4}, using an idealized model called the force-network ensemble (FNE), argued that in two dimensions the Maxwell-Cremona force-tile area acts as an additional constraining variable, that must be taken into account in order to arrive at a correct maximum entropy description of the stress distribution.  Recent experiments \cite{Lechenault, McNamara,Puckett,Schroter} have sought to test such statistical models.

The main goal of this work is to numerically investigate these statistical ensemble ideas as applied to the distribution of stress, and in particular to test if the analysis of Tighe and co-workers for the idealized FNE, continues to hold in a more realistic model of jammed soft-core particles.
To this end we carry out detailed numerical simulations of a simple model of two dimensional, soft-core, bidisperse frictionless disks, to determine the distribution of stress and force-tile area on compact clusters of particles embedded in a larger, mechanically stable, packing at finite isotropic stress above the jamming transition.  Measuring behavior as a function of both the cluster size and the total system stress, we find that the stress distribution is consistent with the maximum entropy hypothesis, provided one takes both the cluster stress and the force-tile area as constraining variables that characterize the distribution. We find that it remains necessary to consider both variables even as the cluster size gets large, contrary to results reported for the FNE \cite{Tighe3}.

The remainder of this paper is organized as follows.  In Sec.~\ref{secModel} we provide details of our numerical model and simulations, discussing our method to produce jammed packings with a specified isotropic total stress tensor, and defining the construction of our clusters and the quantities measured. In Sec.~\ref{sStress} we analyze our results in the context of the stress ensemble of Henkes et al. \cite{Henkes1,Henkes2}.  We use a ratio of cluster stress distributions at different values of the total system stress to investigate the Boltzmann factor predicted for the distribution, and find that this Boltzmann factor includes a term quadratic in the cluster stress, rather than being linear in the stress as predicted by the stress ensemble.  We compare our results against a simpler Gaussian approximation, and find that the quadratic Boltzmann factor gives a better description.  We discuss the previous results by Henkes et al. \cite{Henkes1,Henkes2} and indicate why they may not have detected the quadratic term which we find here.

In Sec.~\ref{sResultsForceTile} we define the Maxwell-Cremona force tile area, and consider the joint distribution of cluster stress and force-tile area.  Using a ratio of this joint distribution at different values of the total system stress, we find results consistent with a Boltzmann factor that is linear in both stress and force-tile area, thus supporting the maximum entropy hypothesis.  We make comparison between the temperature-like parameters resulting from this ratio analysis and those predicted from fluctuations via the covariance matrix of the constraining variables, and find reasonable, though not perfect, agreement.  We then discuss the relation of our results to previous results of Tighe et al. \cite{Tighe1,Tighe3,Tighe4} for the FNE, and discuss the relation between the Boltzmann factor of the joint distribution, and the quadratic Boltzmann factor of the stress distribution analyzed in the previous section.  Finally, in Sec.~\ref{sSummary} we summarize and discuss our conclusions.

%%%%%%%%%%%%%%%%%%%%%%%%%%%%%%%%%%%%%%%%%%%%%%%%%%%%%%%%%%%%%%%%%%%%%%%%%%%%%%%

\section{Model}
\label{secModel}

\subsection{Global ensemble}
\label{secGlobEnsem}

Our system is a two-dimensional bidisperse mixture of equal numbers of big and small circular, frictionless, disks with diameters $d_b$ and $d_s$ in the ratio $d_b/d_s=1.4$ \cite{OHern}.  Disks $i$ and $j$ interact only when they overlap, in which case they repel with a soft-core harmonic interaction potential,
\begin{equation}
{\cal V}_{ij}(r_{ij})=\left\{
\begin{array}{ll}
\frac{1}{2}k_e(1-r_{ij}/d_{ij})^2, & r_{ij}<d_{ij}\\[8pt]
0, & r_{ij}\ge d_{ij}.
\end{array}
\right.
\label{einteraction}
\end{equation}
Here $r_{ij}=|\mathbf{r}_i-\mathbf{r}_j|$ is the center-to-center distance between the particles, and $d_{ij}=(d_i+d_j)/2$ is the sum of their radii.  We will measure energy in units such that $k_e=1$, and length in units so that the small disk diameter $d_s=1$.

\begin{figure}[h]
\begin{center}
\includegraphics[width=1.8in]{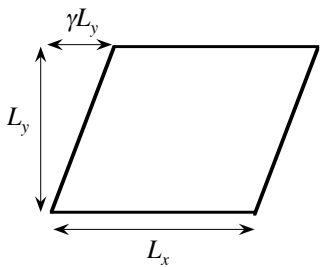}
\caption{Geometry of our system box.  $L_x$ and $L_y$ are the lengths in the $\mathbf{\hat x}$ and $\mathbf{\hat y}$ directions, and $\gamma$ is the skew ratio.  Lees-Edwards boundary conditions are used.
}
\label{box}
\end{center}
\end{figure}

The geometry of our system box is characterized by three parameters, $L_x,L_y,\gamma$, as illustrated in Fig.~\ref{box}. $L_x $ and $L_y$ are the lengths of the box in the $\mathbf{\hat x}$ and $\mathbf{\hat y}$ directions, while $\gamma$ is the skew ratio of the box.  We use Lees-Edwards boundary conditions \cite{LeesEdwards} to periodically repeat this box throughout all space.

We consider here systems with a fixed total number $N$ of disks, and study mechanically stable particle packings above the jamming transition, that have a specified {\em isotropic} total stress tensor $\Sigma_{\alpha\beta}^{(N)}$,
\begin{equation}
\Sigma_{\alpha\beta}^{(N)}=\Gamma_N\delta_{\alpha\beta}, \quad \mathrm{where}\quad\Gamma_N=pV,
\label{eisostress}
\end{equation}
$p$ is the system pressure, and $V=L_xL_y$ is the total system volume (in two dimensions we will use ``volume" as a synonym for area).   Here $\alpha,\beta$ denote the spatial coordinate directions $x,y$.  

To create our isotropic packings, in which the shear stress vanishes, we use a scheme in which we vary the box parameters $L_x,L_y$ and $\gamma$ as we search for mechanically stable states \cite{Dagois}.
We introduce \cite{WuTeitel} a modified energy function $\tilde U$ that depends on the particle positions $\{\mathbf{r}_i\}$, as well as the box parameters $L_x,L_y,\gamma$,
\begin{equation}
\tilde U\equiv U+\Gamma_N(\ln L_x +\ln L_y),\quad U\equiv\sum_{i<j}{\cal V}_{ij}(r_{ij}).
\end{equation}

Noting that the interaction energy $U$ depends implicitly on the box parameters $L_x,L_y,\gamma$ via the boundary conditions, we get the relations,
\begin{equation}
\begin{aligned}
L_x\frac{\partial U}{\partial L_x}=-\Sigma_{xx}^{(N)}+\gamma\Sigma_{xy}^{(N)},&\quad\frac{\partial U}{\partial \gamma}=-\Sigma_{xy}^{(N)},\\ 
L_y\frac{\partial U}{\partial L_y}=-\Sigma_{yy}^{(N)}-\gamma\Sigma_{xy}^{(N)}.&
\end{aligned}
\end{equation}

We then start from an initial configuration of randomly positioned particles in a square box ($L_x=L_y,\gamma=0$) at packing fraction $\phi_\mathrm{init}=0.84$ (just slightly below the jamming transition $\phi_J\approx 0.842$ \cite{VagbergFSS}), and fixing a target value of $\Gamma_N$, we minimize $\tilde U$
with respect to both particle positions and box parameters.  
The resulting local minimum of $\tilde U$
gives a mechanically stable configuration with force balance on each particle and a total stress tensor that satisfies 
\begin{equation}
\Sigma_{xx}^{(N)}=\Sigma_{yy}^{(N)}=\Gamma_N, \quad \Sigma_{xy}^{(N)}=0.  
\end{equation}
For minimization we use the Polak-Ribiere conjugate gradient algorithm \cite{NR}.  We consider the minimization converged when we satisfy the condition $(\tilde U_i - \tilde U_{i+50})/\tilde U_{i+50} < \varepsilon=10^{-10}$, where $\tilde U_i$ is the value at the $i$th step of the minimization.  Tests that this procedure gives numerically well minimized configurations,  with the desired isotropic total stress tensor and force balance on particles, are discussed in the Appendix of Ref.~\cite{WuTeitel-hyper}.

Our results are for a system with $N=8192$ disks, averaged over 10000 isotropic configurations, independently generated at each value of $\Gamma_N$.  We vary the total stress from $\Gamma_N=6.4$ to 18.4, in steps of 0.8.
Since our simulations fix both $N$ and $\Gamma_N$, it is convenient to parameterize our results by the intensive, pressure-like, variable, $\tilde p \equiv \Gamma_N/N = pV/N$, the total stress per particle.

Since our method varies the system volume $L_xL_y$ so as to achieve the desired total stress $\Gamma_N$, the packing fraction,
\begin{equation}
\phi=\frac{N}{L_xL_y}\frac{\pi}{2}\left[\left(\frac{d_s}{2}\right)^2+\left(\frac{d_b}{2}\right)^2\right],
\label{ephi}
\end{equation}
at a fixed $\Gamma_N$ varies slightly from configuration to configuration. 
In Fig.~\ref{GR-hist}a we plot the resulting average $\langle \phi\rangle$ as a function of $\tilde p$ for the range of $\tilde p$ considered in this work. Error bars represent the width of the distribution of $\phi$; the relative width is roughly $0.03-0.04\%$.  The $\Gamma_N$ values we consider here place our systems moderately close above the jamming transition, which for our rapid quench protocol  is $\phi_J\approx 0.842$  \cite{VagbergFSS,WuTeitel-hyper}.

\begin{figure}[h]
\includegraphics[width=3.5in]{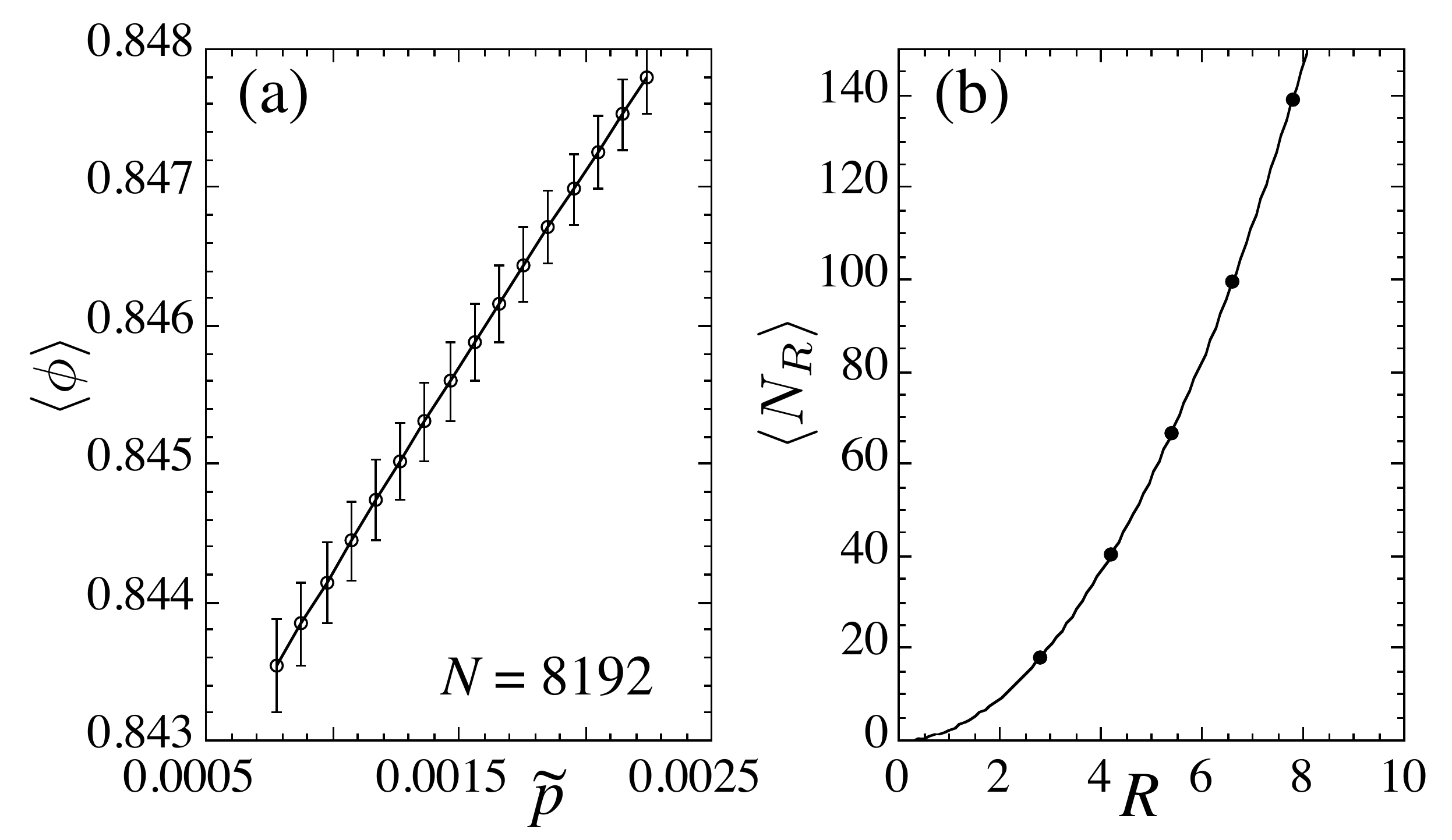}
\caption{(a) Average packing fraction $\langle\phi\rangle$ vs total system stress per particle $\tilde p\equiv\Gamma_N/N$. The error bars represent the width of the distribution of $\phi$, and {\em not} the statistical error of the average.  (b) Average number of particles $\langle N_R\rangle$ in a cluster of radius $R$, at packing fraction $\phi=0.845$.}
\label{phi-NR}
\end{figure}

%%%%%%%%%%%%%%%%%%%%%%
\subsection{Clusters of finite size}
\label{secCluster}

In this work we are interested in the distribution of the stress on finite sized sub-clusters of the system.
To define our particle clusters, we pick a position in the system at random and draw a circle of radius $R$ centered at that point.  All particles whose centers lie within this circle are considered part of the cluster, which we denote as ${\cal C}_R$  \cite{note0}.  The total number of particles $N_R$ in such a cluster will fluctuate from cluster to cluster, but the average $\langle N_R\rangle$ can be obtained from Eq.~(\ref{ephi}) using $\pi R^2$ rather than $L_xL_y$ as the volume on the right hand side.
In Fig.~\ref{phi-NR}b we plot  $\langle N_R\rangle$ vs radius $R$ for a system with packing fraction $\phi=0.845$.  

We can then compute the stress tensor for the cluster ${\cal C}_R$,
\begin{equation}
\Sigma_{\alpha\beta}^{(R)}=\sum_{i\in {\cal C}_R}{\sum_j}^\prime {s}_{ij\alpha}{F}_{ij\beta},\quad \mathbf{F}_{ij}=-{\partial {\cal V}}(r_{ij})/{\partial\mathbf{r}_{j}}.
\end{equation}
The first sum is over all particles $i$ in the cluster ${\cal C}_R$.  The second, primed, sum is over all particles $j$ in contact with $i$, where $\mathbf{s}_{ij}$ is the displacement from the center of particle $i$ to its point of contact with $j$, and $\mathbf{F}_{ij}$ is the force on $j$ due to contact with $i$ \cite{Henkes1}.

Although the total system stress is isotropic, the stress on any particular cluster $\Sigma_{\alpha\beta}^{(R)}$ in general is not.  However the stress averaged over many different clusters will be isotropic.  If we define for each cluster
\begin{equation}
\Gamma_R\equiv \frac{1}{2}\mathrm{Tr}[\Sigma_{\alpha\beta}^{(R)}], 
\end{equation}
then we will have 
\begin{equation}
\langle\Sigma_{\alpha\beta}^{(R)}\rangle=\langle\Gamma_R\rangle\delta_{\alpha\beta}.
\end{equation}
Here and henceforth, we will use $\langle \dots\rangle$ to indicate an average over different clusters.  Our averages in this work are taken over different non-overlapping clusters within a given configuration, and then over the 10000 independently generated configurations at each $\Gamma_N$.

\begin{figure}[h]
\includegraphics[width=3.4in]{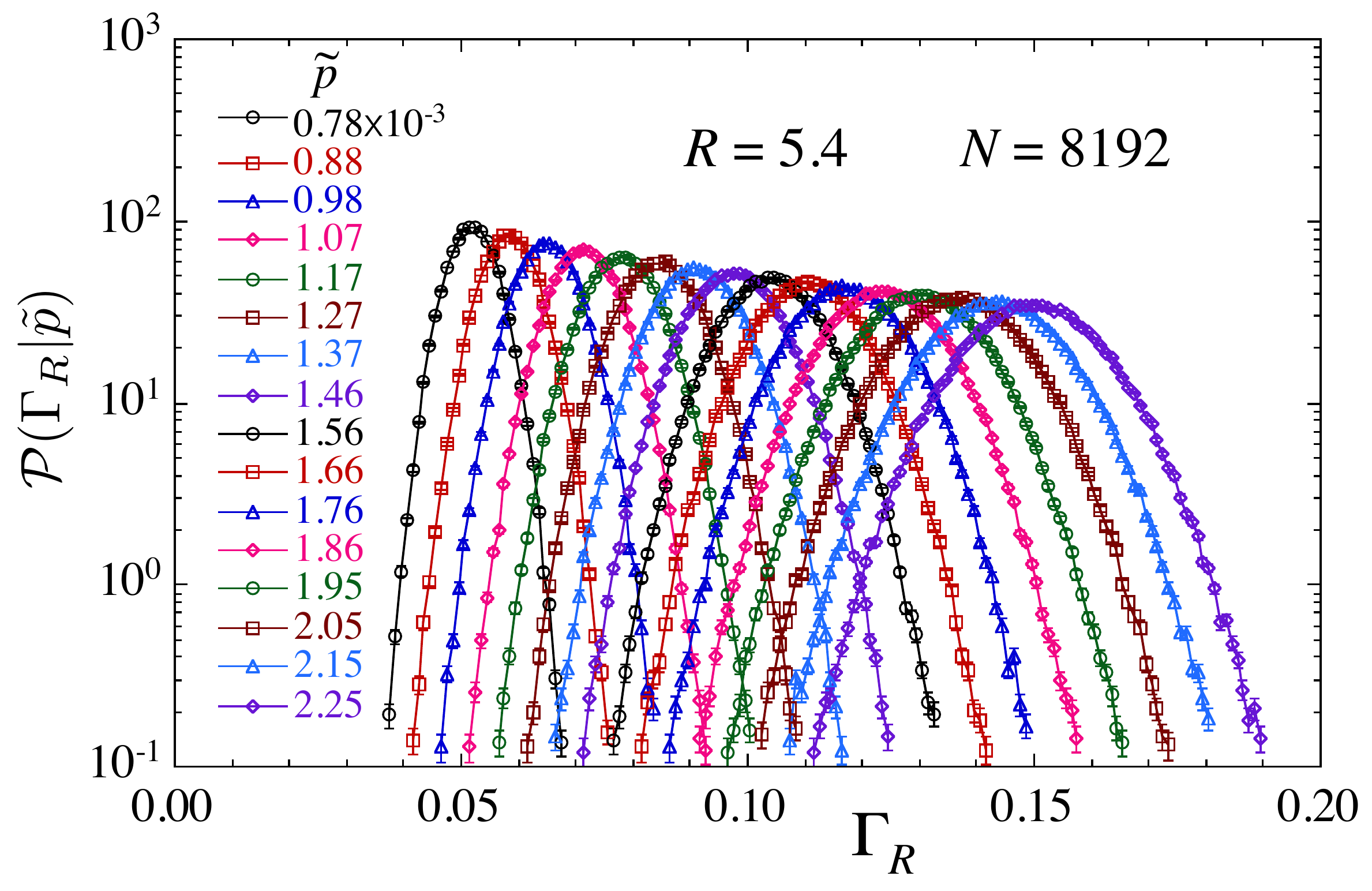}
\caption{(color online) Probability histograms of the stress $\Gamma_R$ on a cluster of radius $R=5.4$ at different values of the total stress per particle $\tilde p=\Gamma_N/N$.}
\label{GR-hist}
\end{figure}

For a system with total stress per particle $\tilde p=\Gamma_N/N$, we will denote the probability that a cluster of radius $R$ has a stress $\Gamma_R$ by ${\cal P}(\Gamma_R|\tilde p)$.
In Fig.~\ref{GR-hist} we show these numerically computed probability histograms ${\cal P}(\Gamma_R|\tilde p)$  over the range of $\tilde p$ we study, for the particular case  of clusters with radius $R=5.4$.  We have chosen our spacing $\Delta\tilde p = \Delta\Gamma_N/N=0.8/8192$ so that the histograms at neighboring values of $\tilde p$ have substantial overlap, as will be needed for our later analysis.  Histograms are normalized so that $\sum_{\Gamma_R}{\cal P}(\Gamma_R|\tilde p)\Delta\Gamma_R=1$, where $\Delta\Gamma_R$ is our bin width; $\Delta\Gamma_R$ is chosen small enough that ${\cal P}(\Gamma_R|\tilde p)$ becomes independent of $\Delta\Gamma_R$.

In this work we will consider a range of cluster sizes from $R=2.8$ to $8.2$, corresponding to clusters with an average number of particles ranging roughly from 18 to 150. Our total system size of $N=8192$ particles was chosen so as to be large enough to explore a moderate range of cluster sizes $R$, while being small enough to generate a large number of independent configurations so as to get good precision for the histograms ${\cal P}(\Gamma_R|\tilde p)$.  The largest cluster size $R$ that we consider is chosen to be  small enough that effects due to the finite size of the total system do not significantly effect the distributions ${\cal P}(\Gamma_R|\tilde p)$.

%%%%%%%%%%%%%%%%%%%

%\section{Results}
\section{Results: The stress ensemble}
\label{sStress}

In an effort to develop a statistical theory for the distribution of stress $\Gamma_R$ on clusters within jammed packings, Henkes et al. \cite{Henkes1,Henkes2} proposed the {\em stress ensemble}.
Noting that the stress tensor $\Sigma_{\alpha\beta}$ is a {\em conserved} quantity, i.e. its global value for the total system is fixed and it is additive over disjoint subsystems, an analogy to the canonical ensemble of statistical mechanics can be made.  For isotropic systems, $\Gamma_R$ plays the role of energy, and the distribution of $\Gamma_R$ was proposed to be,
\begin{equation}
{\cal P}(\Gamma_R|\tilde p) = \Omega_R(\Gamma_R)\dfrac{\mathrm{e}^{-\alpha(\tilde p)\Gamma_R}}{Z_R(\tilde p)}.
\label{ePGR}
\end{equation}
The {\em angoricity} \cite{Henkes2,Blumenfeld} $1/\alpha$ is a temperature-like variable that is set by the total system stress per particle $\tilde p$. The number of available states $\Omega_R(\Gamma_R)$ at a given value of $\Gamma_R$ is presumed independent of $\tilde p$.  The normalizing constant $Z_R$,
\begin{equation}
Z_R(\tilde p)=\int d\Gamma_R\,\Omega_R(\Gamma_R)\mathrm{e}^{-\alpha(\tilde p)\Gamma_R},
\label{eZR}
\end{equation}
is analogous to the partition function, and 
\begin{equation}
{\cal F}_R(\tilde p)\equiv -\ln Z_R(\tilde p)
\label{eFR}
\end{equation}
is analogous to the free energy.

Alternatively, the distribution of Eq.~(\ref{ePGR}) can also be viewed as resulting from a maximum entropy hypothesis \cite{Plischke}, in which all clusters with a given $\Gamma_R$ are presumed equally likely, and the average is constrained to the known value $\langle \Gamma_R\rangle$.  Since the stress is conserved and additive, the average of $\Gamma_R$ is constrained by,
\begin{equation}
\langle\Gamma_R\rangle = \Gamma_N\left(\frac{\pi R^2}{V}\right), 
\label{GammaR}
\end{equation}
a result that we have previously confirmed numerically \cite{WuTeitel}.  The average pressure in the cluster is then equal to the global pressure in the total system,
\begin{equation}
\langle p_R\rangle\equiv \frac{\langle\Gamma_R\rangle}{\pi R^2} = \frac{\Gamma_N}{V}=p.
\label{epR}
\end{equation}

Two particular consequences follow from the distribution of Eq.~(\ref{ePGR}).  The first relates to the fluctuation of stress on the cluster, $\mathrm{var}(\Gamma_R)\equiv\langle\Gamma_R^2\rangle-\langle\Gamma_R\rangle^2$. The second relates to the ratio of distributions at nearby values of $\tilde p$. 

\subsection{Fluctuations} 
\label{G-flucs}

As in an equilibrium thermodynamic system, one can use the free energy of Eq.~(\ref{eFR}) to write,
\begin{equation}
\dfrac{\partial{\cal F}_R}{\partial\alpha} = \langle\Gamma_R\rangle,
\end{equation}
and
\begin{equation}
\dfrac{\partial\langle\Gamma_R\rangle}{\partial\alpha}=\dfrac{\partial^2{\cal F}_R}{\partial\alpha^2} = \langle\Gamma_R\rangle^2-\langle\Gamma_R^2\rangle=-\mathrm{var}(\Gamma_R).
\end{equation}

A change in the inverse angoritcity $\Delta\alpha$ therefore gives a change in the average cluster stress $\langle\Delta\Gamma_R\rangle$,
\begin{equation}
\langle\Delta\Gamma_R\rangle = \dfrac{\partial\langle\Gamma_R\rangle}{\partial \alpha}\Delta\alpha
=-\mathrm{var}(\Gamma_R)\Delta\alpha.
\label{eDalphaBoltz}
\end{equation}
By Eq.~(\ref{epR}) we have $\langle\Delta \Gamma_R\rangle/(\pi R^2)=\langle\Delta p_R\rangle=\Delta p$, hence we conclude that a change in the total system pressure $\Delta p$ induces a change in the inverse angoricity $\Delta\alpha$, given by,
\begin{equation}
\Delta\alpha=-\left[\dfrac{\pi R^2}{\mathrm{var}(\Gamma_R)}\right]\Delta p. 
%= -\left[\dfrac{\pi R^2}{\mathrm{var}(\Gamma_R)}\right]\left[\dfrac{N}{V}\right]\Delta \tilde p 
\label{eDalpha}
\end{equation}
Taking the limit $\Delta p\to 0$ we then get,
\begin{equation}
\dfrac{d\alpha}{dp}=-\left[\dfrac{\pi R^2}{\mathrm{var}(\Gamma_R)}\right].
\label{edalphadp}
\end{equation}

In Ref.~\cite{WuTeitel} we showed that, for the range of cluster sizes and pressures considered here, the dependence of $\mathrm{var}(\Gamma_R)$ on cluster size $R$ was well fit by the form $\mathrm{var}(\Gamma_R)/(\pi R^2) =c_1+c_2/R$.
Thus from Eq.~(\ref{edalphadp}) we might expect to see $1/R$ corrections to $\alpha(\tilde p)$ arising from the finite sizes of our clusters.

\subsection{Histogram ratio}
\label{s1DRatio}

The results of the previous subsection, in particular Eq.~(\ref{edalphadp}), hold {\em if} the distribution of stress $\Gamma_R$ obeys the form of Eq.~(\ref{ePGR}).  However it is necessary to first demonstrate that this form does indeed hold.  A direct test of whether or not the distributions ${\cal P}(\Gamma_R|\tilde p)$ obey Eq.~(\ref{ePGR}) is given by considering the ratio of numerically measured histograms at two neighboring values of $\tilde p$  \cite{Dean}.

%%%%%%%%%%%%%%%
Denoting quantities at a given $\tilde p_1$ or $\tilde p_2$ by the subscript 1 or 2, the log ratio of histograms at two neighboring values of $\tilde p_1<\tilde p_2$ is given by,
\begin{equation}
 \ln\left[\frac{{\cal P}_1}{{\cal P}_2}\right]
= \ln\left[\frac{Z_{R,2}}{Z_{R,1}}\right] +(\alpha_2-\alpha_1)\Gamma_R
=-\Delta{\cal F}_R+\Delta\alpha\Gamma_R,
\label{e5}
\end{equation}
where $\Delta{\cal F}_R\equiv{\cal F}_{R,2}-{\cal F}_{R,1}$ and $\Delta\alpha\equiv\alpha_2-\alpha_1$.

Expecting that the right hand side of Eq.~(\ref{e5}) scales proportional to the cluster area $\pi R^2$, we define an {\em intensive} log ratio,
\begin{equation}
{\cal R}\equiv \frac{1}{\pi R^2} \ln\left[\frac{{\cal P}_1}{{\cal P}_2}\right]
 =-\Delta f+\Delta \alpha\, p_R.
\label{Rslinear}
\end{equation}
where $p_R\equiv\Gamma_R/(\pi R^2)$, and $f\equiv {\cal F}_R/(\pi R^2)$.  The condition ${\cal R}=0$ locates the point of greatest overlap between neighboring histograms, where ${\cal P}_1={\cal P}_2$.

\begin{figure}[h]
\includegraphics[width=3.5in]{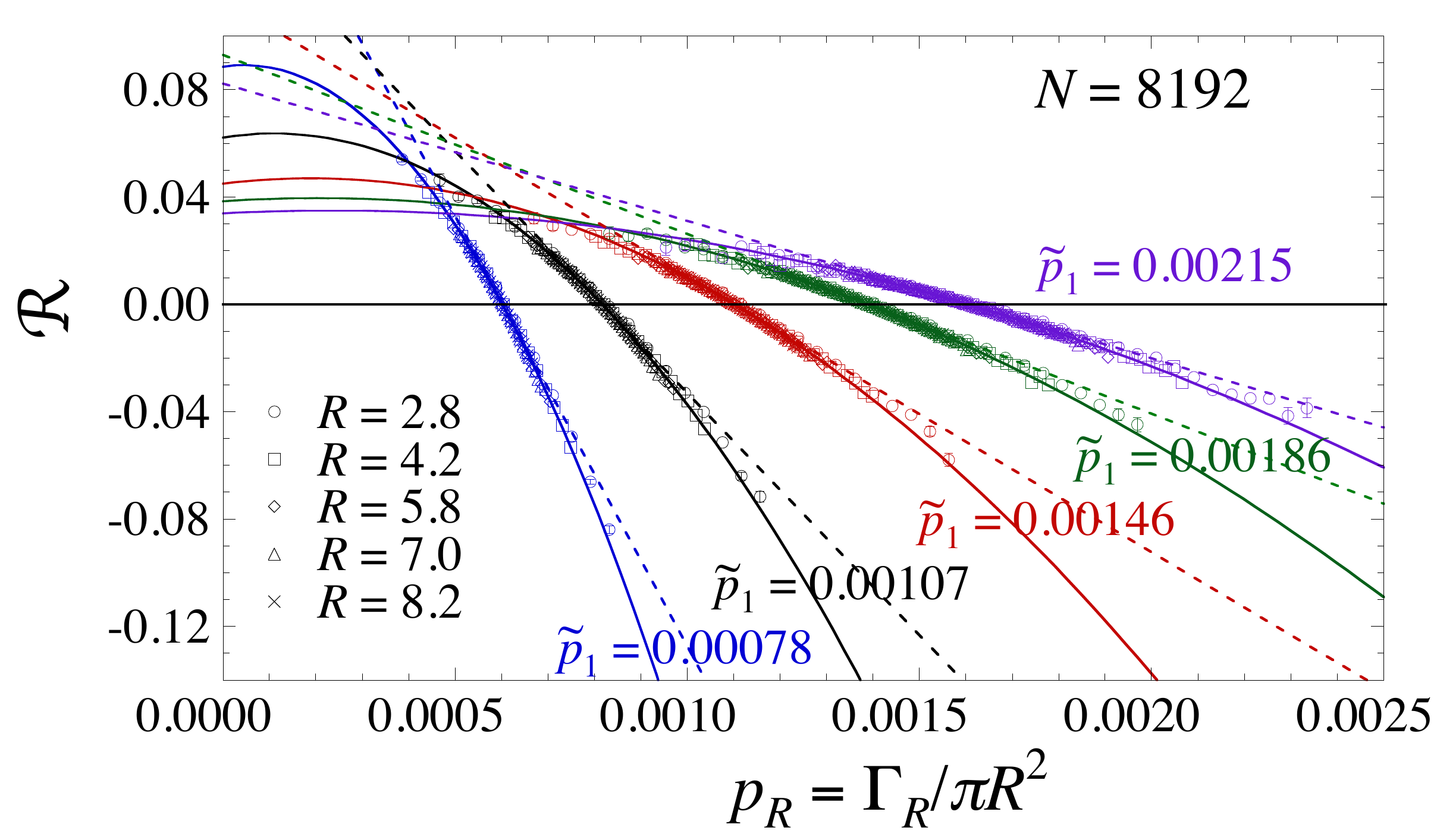}
\caption{ 
(color online) Log ratio ${\cal R}\equiv(1/\pi R^2)\ln[{\cal P}_1/{\cal P}_2]$  of histograms ${\cal P}_1$ and ${\cal P}_2$ at total system stresses per particle $\tilde p_1$ and $\tilde p_2=\tilde p_1+\Delta\tilde p$, 
vs cluster pressure $p_R=\Gamma_R/\pi R^2$. 
Data for different cluster sizes $R$ (denoted by different symbol shapes) but the same $\tilde p_1,\tilde p_2$ collapse to a common curve that is well fit by a parabola (solid curves); dashed lines are the tangents at the point of greatest overlap between the  histograms ${\cal P}_1$ and ${\cal P}_2$, given by the condition ${\cal R}=0$. We show results for stress per particle $\tilde p = 0.00078$ to $0.00215$.  Representative error bars are shown at the tail ends of $p_R$.
}
\label{rescaled-Ratios}
\end{figure}

In Fig.~\ref{rescaled-Ratios} we plot ${\cal R}$ vs $p_R$ for several different pairs of $\tilde p_1$ and $\tilde p_2=\tilde p_1+\Delta\tilde p$, for cluster sizes $R=2.8$ to $8.2$. We find a fairly good looking collapse of the data for different cluster radii $R$.  This suggests that, to leading order in $1/R$, ${\cal R}$, and hence $\Delta f$ and $\Delta\alpha$, are intensive quantities independent of the cluster size.
However we find that the data for ${\cal R}$ show a clear {\em curvature}, not the linear dependence on $p_R$ predicted by Eq.~(\ref{Rslinear}).

%Note, that since $\mathrm{var}(p_R)=\mathrm{var}(\Gamma_R)/(\pi R^2)^2\sim 1/R^2$, the relative width of the histograms ${\cal P}(\Gamma_R|\tilde p)$ decrease as $R$ increases, hence the measurable region of overlap of neighboring histograms ${\cal P}_1$ and ${\cal P}_2$ decreases.  Thus, as $R$ increases, the data in Fig.~\ref{rescaled-Ratios} gets confined to a narrower region of $p_R$, and the curvature becomes less evident as $R$ increases.

Instead of using Eq.~(\ref{Rslinear}) we may empirically fit our data in Fig.~\ref{rescaled-Ratios} to a quadratic form,
\begin{equation}
{\cal R}= -\Delta f+ \Delta\alpha\, p_R + \Delta\lambda\, p_R^2,
\label{RspR}
\end{equation}
where $\Delta f$, $\Delta\alpha$ and $\Delta\lambda$ vary with the stress $\tilde p_1$, but are independent of the cluster radius $R$.  Such fits give the solid curves in Fig.~\ref{rescaled-Ratios}.

{\bf Linear approximation:} If we for the moment ignore the curvature in the data of Fig.~\ref{rescaled-Ratios}, we can approximate ${\cal R}$ by its tangent line at the value $p_R^*$ where ${\cal R}=0$. This is the point where the two distributions ${\cal P}_1$ and ${\cal P}_2$ have their largest overlap.  Such tangents are shown as the dashed lines in Fig.~\ref{rescaled-Ratios}, and have slopes,
\begin{equation}
\Delta\bar\alpha=\Delta\alpha+2\Delta\lambda p_R^*.
\label{ebarDalpha}
\end{equation}

\begin{figure}[h]
\includegraphics[width=3.5in]{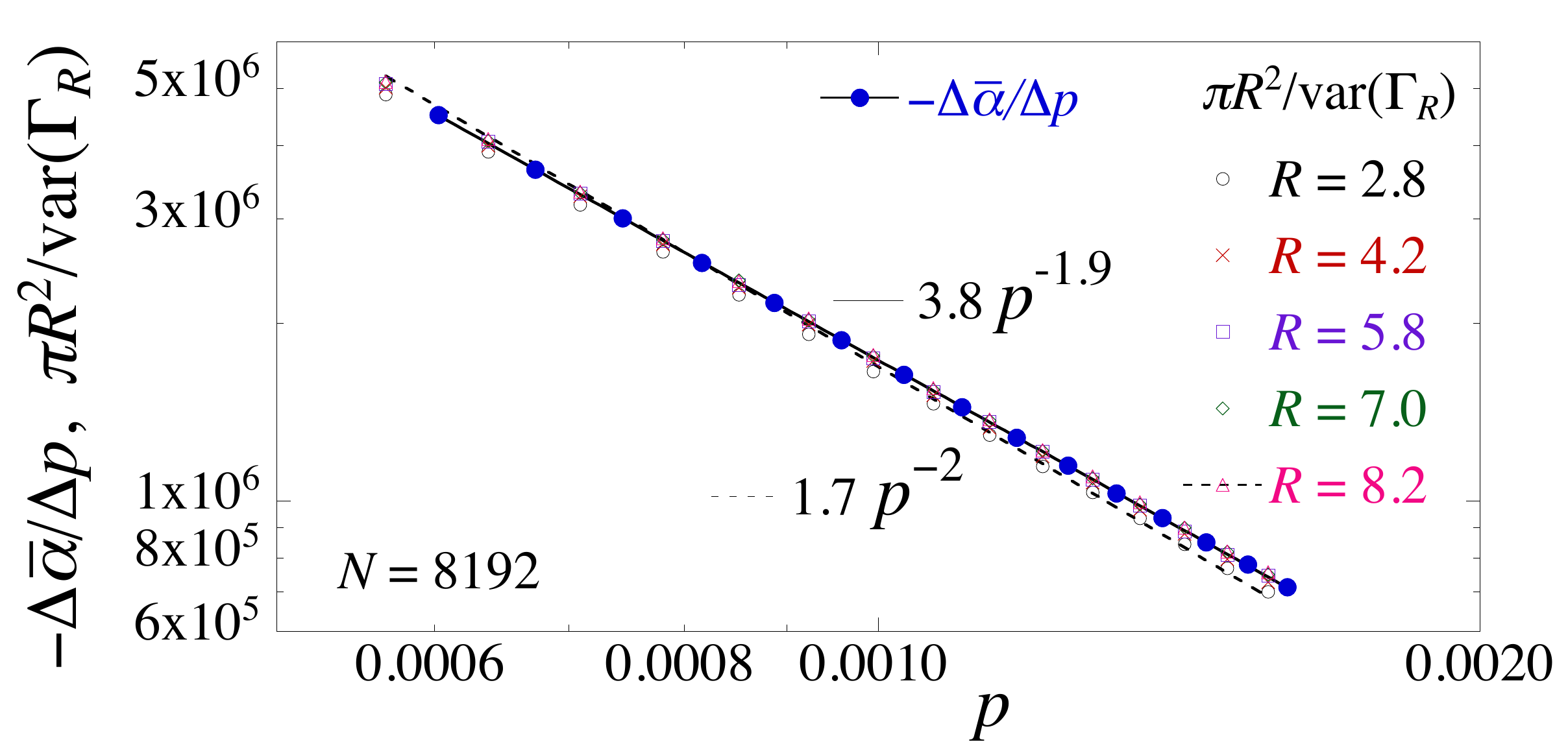}
\caption{(color online) Comparison of (i) $-\Delta\bar\alpha/\Delta p$ vs $p=(p_1+ p_2)/2$ (solid circles) as computed from the linear approximation to the log histogram ratio ${\cal R}$, given by the slopes Eq.~(\ref{ebarDalpha}) of the tangent lines in Fig.~\ref{rescaled-Ratios}, with (ii) $\pi R^2/\mathrm{var}(\Gamma_R)$ vs $p$, for several different cluster radii $R$ (open symbols), which by the fluctuation expression of Eq.~(\ref{edalphadp}) gives $-d\alpha/dp$ in the stress ensemble approximation.  The solid line is a fit to an arbitrary power-law, and finds $-\Delta\bar\alpha/\Delta p\approx 3.8  p^{-1.9}$.  The dashed line is a fit to the the power-law $ p^{-2}$.  
}
\label{barDalpha-Dp-1Dhist}
\end{figure}

In Fig.~\ref{barDalpha-Dp-1Dhist} we plot $-\Delta\bar\alpha/\Delta p$ vs $p\equiv (p_1+p_2)/2$, where $p_{1,2}=\tilde p_{1,2}(N/V_{1,2})$ gives the corresponding total system pressure of the two overlapping histograms.  We find an excellent fit to a power-law, $-\Delta\bar\alpha/\Delta p\approx 3.8 p^{-1.9}$.  Taking $\Delta\bar\alpha/\Delta p$ as an approximation to the derivative, we  can integrate to get $\bar\alpha(p) \approx 4.2 p^{-0.9}$.  Given the rather limited range of our data, however, it is unclear how much significance should be given to the specific numerical value of this fitted exponent; the data is also well fit by the expression $-\Delta\bar\alpha/\Delta p\approx(1.9/p^2)(1-0.000094/p)$.

Viewing the stress ensemble of Eq.~(\ref{ePGR}) as an approximation to the true distribution, we can test whether $-\Delta\bar\alpha/\Delta p$ from the above linear approximation to the histogram ratio is in agreement with the $-d\alpha/dp$ one would expect from the fluctuation expression of Eq.~(\ref{edalphadp}).  We therefore also plot in Fig.~\ref{barDalpha-Dp-1Dhist} the quantity $\pi R^2/\mathrm{var}(\Gamma_R)$ vs $p$, showing results for several different cluster sizes $R$.
We see an excellent agreement.  

The agreement shown in Fig.~\ref{barDalpha-Dp-1Dhist} might naively be taken as evidence that the stress ensemble, while failing to give a strictly linear log ratio ${\cal R}$ as predicted, is nevertheless not a bad approximation to the stress distribution. However, as we will show in the next section, Eq.~(\ref{edalphadp}) also results  from the assumption that the distribution ${\cal P}(\Gamma_R|\tilde p)$ is a simple Gaussian, provided that the spacing $\Delta p$ between the overlaping distributions is not too great \cite{McNamara}.  Moreover, such a Gaussian model also provides a simple mechanism for producing the curvature in ${\cal R}$ that is evident in Fig.~\ref{rescaled-Ratios}.  We will discuss the extent to which a Gaussian approximation can explain the data of Fig.~\ref{rescaled-Ratios} in Sec.~\ref{gauss-1D}.

{\bf Quadratic fit:} The quadratic form for the log ratio ${\cal R}$, given by Eq.~(\ref{RspR}),  clearly describes the data better than the linear expression of Eq.~(\ref{Rslinear}).  However, while the quadratic fits in Fig.~\ref{rescaled-Ratios} look reasonable, a quantitative test shows that they are not particularly accurate, given the high precision of our data.  As a measure of the goodness of our fits we will use the chi squared per degree of freedom $\chi^2/\nu$,
\begin{equation}
\chi^2/\nu \equiv \frac{1}{M_d-M_f}\sum_{i=1}^{M_d}\left[\dfrac{y_i-y(\mathbf{x}_i)}{\delta y_i}\right]^2,
\label{echisq}
\end{equation}
where $M_d$ is the number of data points, $M_f$ the number of fit parameters, $\mathbf{x}_i$ the independent variables, $y_i$ the measured dependent variable at $\mathbf{x}_i$, $\delta y_i$ the estimated statistical error in $y_i$, and $y(\mathbf{x}_i)$ the fitting function.  A good fit is usually indicated by $\chi^2/\nu\lesssim O(1)$.

\begin{figure}[h]
\includegraphics[width=3.5in]{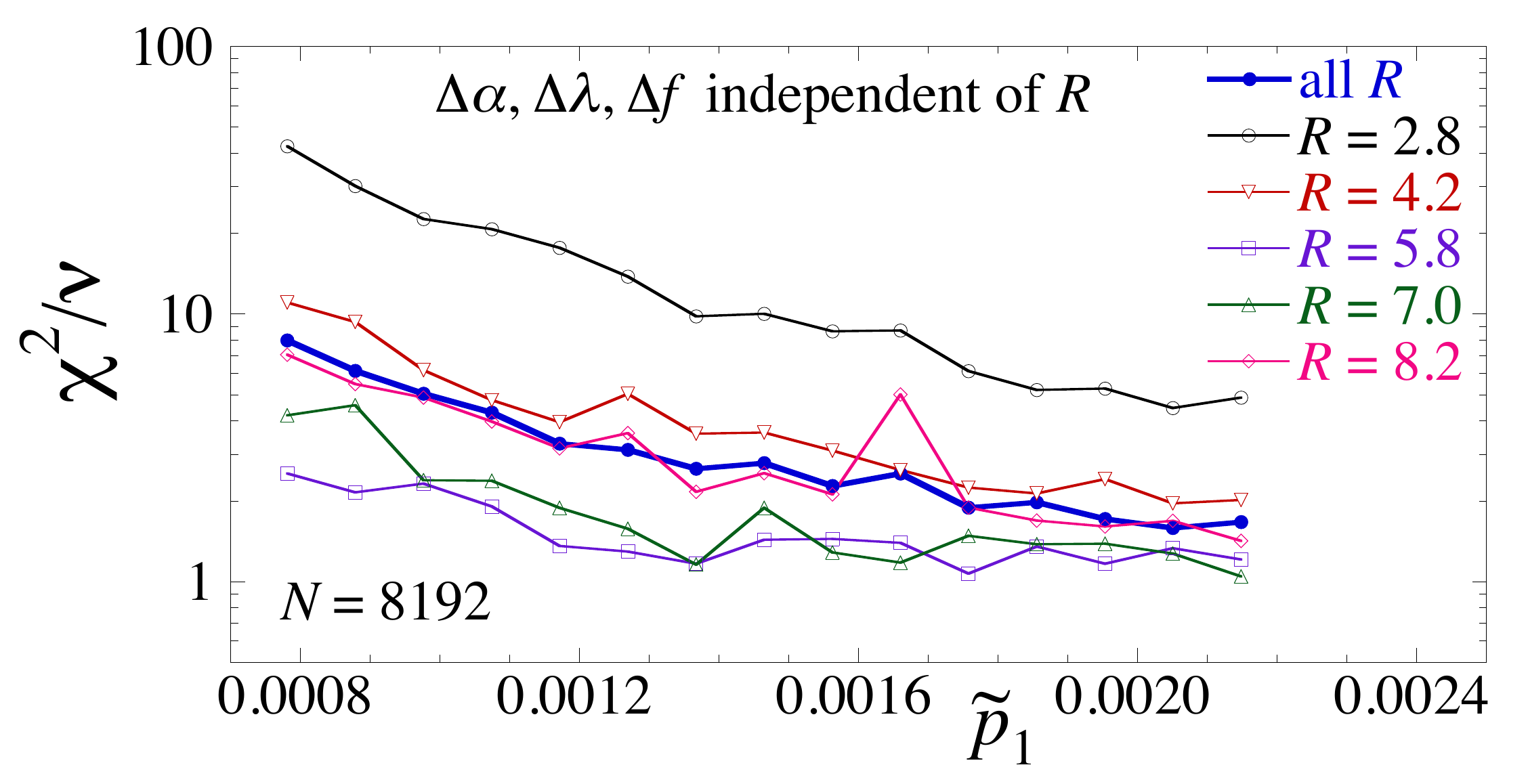}
\caption{(color online) Chi squared per degree of freedom, $\chi^2/\nu$, for the fit of the histogram ratio ${\cal R}$ to the quadratic form of Eq.~(\ref{RspR}), where the fitting parameters $\Delta f$, $\Delta\alpha$ and $\Delta\lambda$ are assumed to be independent of the cluster radius $R$.  Results are plotted vs $\tilde p_1$, the stress per particle at the lower of the two stresses $\tilde p_1$, $\tilde p_2$ used to define ${\cal R}$ (see Eq.~(\ref{Rslinear})).  ``all $R$" denotes the $\chi^2/\nu$ of the fit to the entire data set including all cluster sizes $R$, while the other symbols denote the $\chi^2/\nu$ of the same fit, but  restricted to data at a given fixed cluster size $R$.
}
\label{chisq-indepR}
\end{figure}

In Fig.~\ref{chisq-indepR} we plot the $\chi^2/\nu$ of the fit to ${\cal R}$ using the quadratic form of Eq.~(\ref{RspR}), where the fitting parameters $\Delta f$, $\Delta\alpha$ and $\Delta\lambda$ are assumed to be independent of the cluster radius $R$.  Our results are plotted vs $\tilde p_1$, the stress per particle at the lower of the two stresses $\tilde p_1$, $\tilde p_2$ used to define the histogram ratio.  We show  results (solid circles) for the fit to the entire data set including all cluster sizes $R$, as well as the $\chi^2/\nu$ (open symbols) for the data set restricted to clusters of a given fixed radius $R$ (we keep $\Delta f$, $\Delta\alpha$ and $\Delta\lambda$ the same, but sum Eq.~(\ref{echisq}) over only the data for a given cluster size $R$, with $M_d$ now being the number of data points at radius $R$, and $M_f$ the number of fit parameters divided by the number of different cluster radii).
We see that the $\chi^2/\nu$ becomes $\sim O(1)$ only as $\tilde p_1$ increases, and only for the larger cluster sizes; as $\tilde p_1$ decreases, the $\chi^2/\nu$ steadily increases and becomes $\sim O(10)$ at our smallest $\tilde p_1$, indicating a poor fit.

The fits discussed above in connection with Figs.~\ref{rescaled-Ratios}-\ref{chisq-indepR} assumed that the fitting parameters $\Delta f$, $\Delta\alpha$ and $\Delta\lambda$ were independent of the cluster radius $R$.  However the discussion at the end of Sec.~\ref{G-flucs} leads one to suspect that these parameters may have $1/R$ corrections arising from the finite size of the clusters.  We therefore extend our analysis to include this possibility by using,
\begin{align}
\Delta\alpha(R)=A(1+a/R),&\quad\Delta\lambda(R)=B(1+b/R),
\label{ealpha-1oR}
\\\nonumber
\Delta f(R)&=C(1+c/R),
\end{align}
in the fit to Eq.~(\ref{RspR}), where $A$, $B$, $C$, $a$, $b$ and $c$ are taken to be independent of $R$.  The values of $A$, $B$ and $C$ thus represent the limiting $R\to\infty$ values of $\Delta\alpha$, $\Delta\lambda$ and $\Delta f$.

\begin{figure}[h!]
\includegraphics[width=3.5in]{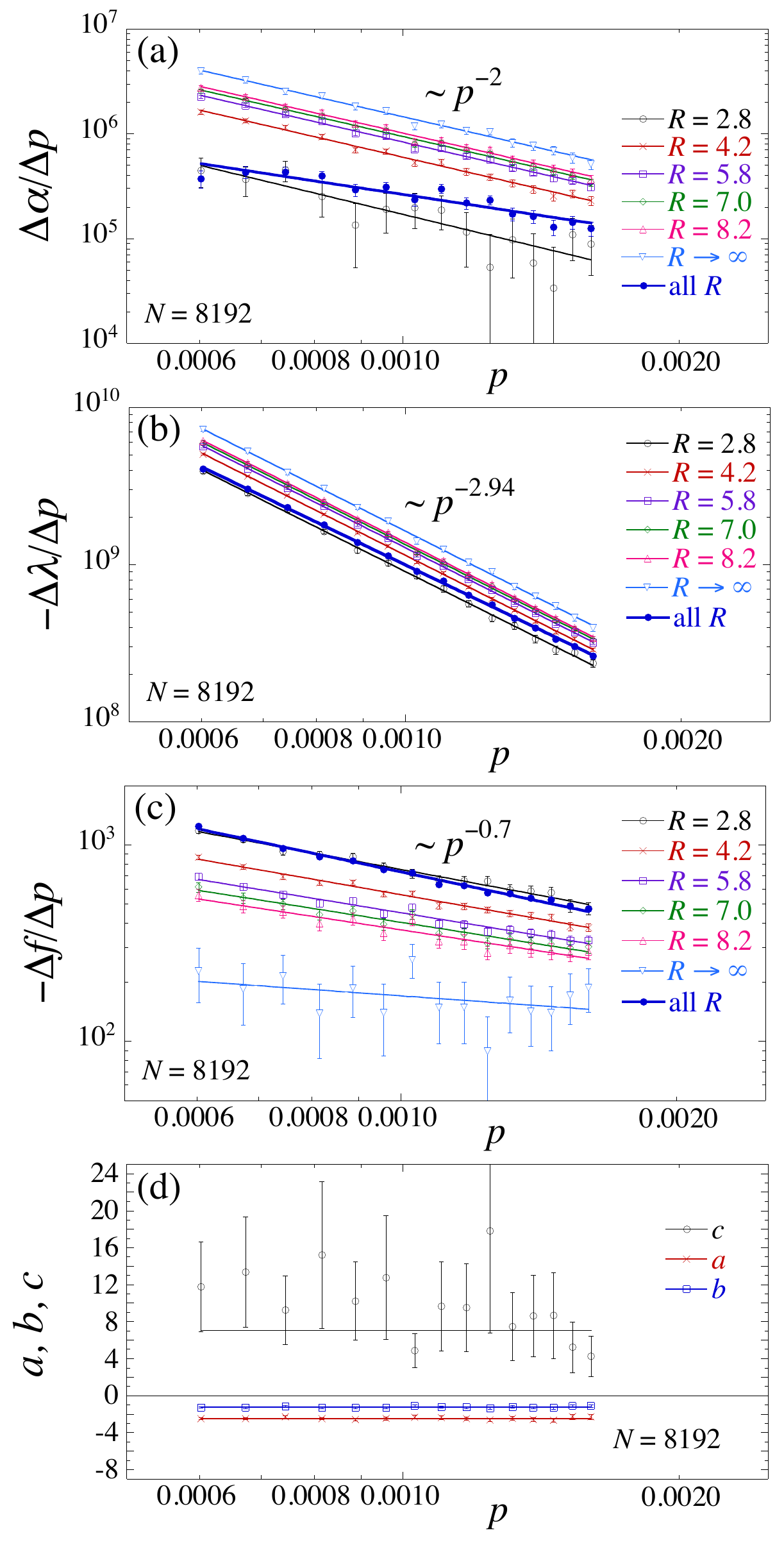}
\caption{(color online) (a) $\Delta\alpha/\Delta p$, (b) $-\Delta\lambda/\Delta p$, (c) $-\Delta f/\Delta p$ vs pressure $p=(p_1+p_2)/2$ from quadratic fits to the histogram ratio ${\cal R}$ with $1/R$ corrections as in Eq.~(\ref{ealpha-1oR}), for clusters of different radii $R$.  Also shown are the $R\to\infty$ limiting values $A$, $B$, $C$ of Eq.~(\ref{ealpha-1oR}), as well as the values from fits keeping $\Delta\alpha$, $\Delta\lambda$ and $\Delta f$ constant for all $R$ (labeled as ``all $R$").  Solid lines are power-law fits, with the power indicated for the fit to the largest value of $R$. (d) Length scale parameters $a$, $b$ and $c$ of Eq.~(\ref{ealpha-1oR}) that determine the strength of the $1/R$ corrections, vs $p$; the solid lines are the best fit to a constant, indicating no systematic dependence on pressure.  
}
\label{quadFits-1oR}
\end{figure}

In Fig.~\ref{quadFits-1oR} we plot the results of such fits with $1/R$ corrections, showing in panels a,b,c $\Delta\alpha(R)/\Delta p$, $\Delta\lambda(R)/\Delta p$ and $\Delta f(R)/\Delta p$ vs the average histogram pressure $p=(p_1+p_2)/2$ for several different cluster radii $R$, as well as the limiting $R\to\infty$ values $A$, $B$, and $C$.  We see that as $R$ increases, all parameters are approaching finite values.
We also show in these figures the results from our earlier fit keeping $\Delta\alpha$, $\Delta\lambda$, and $\Delta f$ as constants independent of $R$; these are labeled in the figures as ``all $R$."  The power-law behavior of the data for the largest $R$ is indicated in the figures, where we find $\Delta\alpha/\Delta p\sim p^{-2}$, $\Delta\lambda/\Delta p\sim p^{-2.94}$, and $\Delta f/\Delta p\sim p^{-0.7}$.  Given the limited range of our data, it is unclear how much significance should be given to the specific numerical values of these exponents. We see that the $1/R$ corrections are quite noticeable for our finite cluster sizes, and that the results we get when ignoring these corrections (the results labeled ``all  $R$") tend to roughly agree with the values found for the smallest $R$ when the $1/R$ corrections are included.

The parameters $a$, $b$ and $c$ of Eq.~(\ref{ealpha-1oR}) represent length scales that determine the strength of the $1/R$ corrections.  We plot these vs $p$ in Fig.~\ref{quadFits-1oR}d and find that these are consistent with being {\em constant}, independent of the pressure.  The lengths $a\approx -2.5$, $b\approx -1.3$ and $c\approx 7.0$ are large enough compared to the range of our cluster sizes $R=2.8-8.2$, so as to explain the noticeable finite size effects we see in Figs.~\ref{quadFits-1oR}a,b,c. 

\begin{figure}[h!]
\includegraphics[width=3.5in]{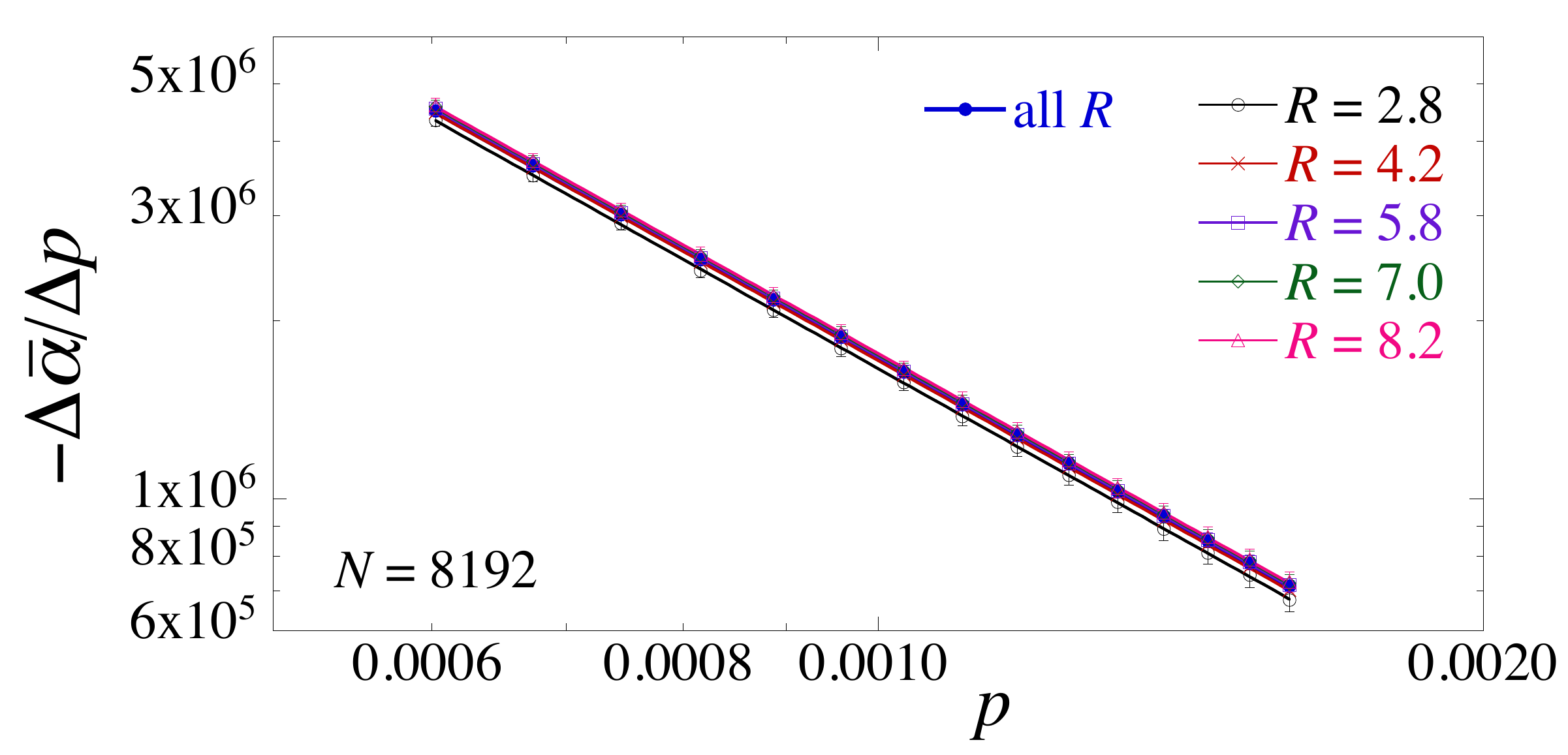}
\caption{(color online) Comparison of
$-\Delta\bar\alpha/\Delta p$ vs $p=(p_1+p_2)/2$ as computed from Eq.~(\ref{ebarDalpha}) using the $\Delta\alpha(R)$ and $\Delta\lambda(R)$  determined from the fits to the histogram ratio ${\cal R}$ with the $1/R$ corrections of Eq.~(\ref{ealpha-1oR}) (open symbols for different $R$), vs from  fits to ${\cal R}$ using $\Delta\alpha$, $\Delta\lambda$ and $\Delta f$ taken to be independent of $R$ (solid circles, previously shown in Fig.~\ref{barDalpha-Dp-1Dhist} and denoted here as ``all $R$").
} 
\label{barDalpha-1oR}
\end{figure}

The parameters $\Delta\alpha$, $\Delta\lambda$ and $\Delta f$, that describe the quadratic shape of the histogram ratio ${\cal R}$, thus show a clear dependence on the cluster size $R$.  However, if we use the $\Delta\alpha(R)$ and $\Delta\lambda(R)$ from Fig.~\ref{quadFits-1oR} in Eq.~(\ref{ebarDalpha}) to compute $\Delta\bar\alpha(R)$, the slope of ${\cal R}$ at the point of maximum histogram overlap, we find that this shows essentially no dependence on the cluster size $R$.  In Fig.~\ref{barDalpha-1oR} we plot this $\Delta\bar\alpha(R)/\Delta p$ vs the average histogram pressure $p=(p_1+p_2)/2$ for several different $R$.  For comparison we also plot the $\Delta\bar\alpha/Dp$, previously shown in Fig.~\ref{barDalpha-Dp-1Dhist}, obtained from fits assuming $\Delta\alpha$, $\Delta\lambda$ and $\Delta f$ independent of $R$.  We see that there is essentially no difference between the two fits, nor between any of the cluster sizes $R$, except for the smallest size $R=2.8$.  Since $\Delta\bar\alpha$ is a measure of behavior at the point of greatest overlap of the two histograms, and this point lies near the peaks of the distributions, the insensitivity of $\Delta\bar\alpha$ to the cluster size $R$ illustrates, not surprisingly, that the dependence on the cluster size $R$ which is observed for the parameters $\Delta\alpha$, $\Delta\lambda$ and $\Delta f$   in Fig.~\ref{quadFits-1oR} is due to the dependence on $R$ of the {\em tails} of the distributions ${\cal P}(\Gamma_R|\tilde p)$.

It is interesting to note that, while the $\bar\alpha(p)$ associated with the linear approximation to ${\cal R}$ at the point of greatest histogram overlap is positive, the $\alpha(p)$ obtained from the quadratic fit to Eq.~(\ref{RspR}) is {\em negative}.  We can see this from Fig.~\ref{barDalpha-Dp-1Dhist} where we find $\Delta\bar\alpha/\Delta p \sim -p^{-1.9}$, and so $\bar\alpha(p) \sim p^{-0.9}$, compared to Fig.~\ref{quadFits-1oR}a where we find that $\Delta\alpha/\Delta p\sim p^{-2}$, and so $\alpha(p)\sim -p^{-1}$.

\begin{figure}[h!]
\includegraphics[width=3.5in]{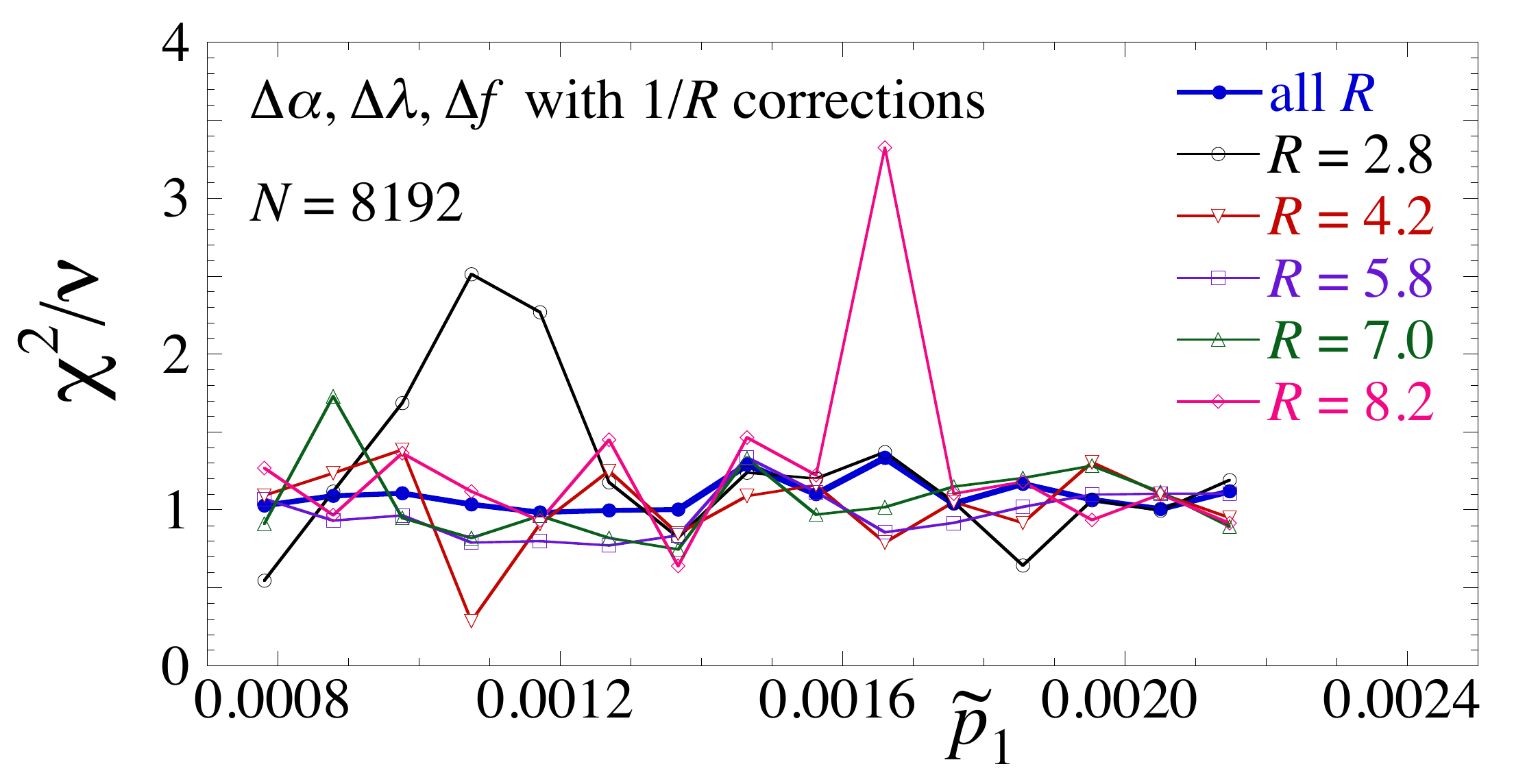}
\caption{(color online)  Chi squared per degree of freedom, $\chi^2/\nu$, for the fit of the histogram ratio ${\cal R}$ to the quadratic form of Eq.~(\ref{RspR}), where the fitting parameters $\Delta f$, $\Delta\alpha$ and $\Delta\lambda$ have the $1/R$ corrections of Eq.~(\ref{ealpha-1oR}).  Results are plotted vs $\tilde p_1$, the stress per particle at the lower of the two stresses $\tilde p_1$, $\tilde p_2$ used to define ${\cal R}$ (see Eq.~(\ref{Rslinear})).  ``all $R$" denotes the $\chi^2/\nu$ of the fit to the entire data set including all cluster sizes $R$, while the other symbols denote the $\chi^2/\nu$ of the fit restricted to data at a given fixed cluster size.
} 
\label{chisq-1oR}
\end{figure}

Finally, we  test the accuracy of our model with $1/R$ corrections by computing the $\chi^2/\nu$ of the fit.  In Fig.~\ref{chisq-1oR} we show $\chi^2/\nu$ as computed for the entire set of data including all cluster sizes $R$, as well as the $\chi^2/\nu$ restricted to data for specific cluster sizes $R$.
We now see, in contrast to the results in Fig.~\ref{chisq-indepR}, that in essentially all cases $\chi^2/\nu\sim O(1)$.  Including such $1/R$ corrections to $\Delta\alpha$, $\Delta\lambda$ and $\Delta f$ thus significantly improves the quality of the fit.

\subsection{Gaussian approximation}
\label{gauss-1D}

In this section we consider an alternative possibility, that the distribution of stress on clusters is given by a simple Gaussian distribution.  We will show that such a Gaussian approximation gives  both (i) a simple mechanism for producing a histogram ratio ${\cal R}$ that is quadratic in the cluster pressure, as in Eq.~(\ref{RspR}), and (ii) a variation of an effective inverse angoricity  (defined by the histogram ratio) with pressure, $d\alpha/dp$, that is the same as found in Eq.~(\ref{edalphadp}) for the Boltzmann distribution, provided the spacing $\Delta p =p_2-p_2$ between the histograms used in computing ${\cal R}$ is sufficiently small.  Similar results have been presented earlier by McNamara et al. \cite{McNamara} in the context of the volume distribution of granular packings.  However, we will show that this Gaussian approximation gives a poorer description of our data than does the quadratic fit of the previous section.

We will here assume that the distribution of stress $\Gamma_R$ on a cluster of radius $R$ is given by the Gaussian,
\begin{equation}
{\cal P}(\Gamma_R|\tilde p)=\dfrac{1}{\sqrt{2\pi\sigma^2}}\mathrm{e}^{-\frac{1}{2}\delta\Gamma_R^2/\sigma^2}
\label{eGauss}
\end{equation}
where $\delta\Gamma_R\equiv \Gamma_R-\langle\Gamma_R\rangle$ is the fluctuation of $\Gamma_R$ away from its ensemble average, and $\sigma^2\equiv\mathrm{var}(\Gamma_R)=\langle\delta\Gamma_R^2\rangle$ is the variance of $\Gamma_R$.  Both $\langle\Gamma_R\rangle$ and $\sigma^2$ are functions of the total system stress per particle, $\tilde p=\Gamma_N/N$.

Using the above Gaussian distribution, it is straightforward to compute the histogram ratio ${\cal R}$ at two neighboring values $\tilde p_1$ and $\tilde p_2$. Doing so, one find a quadratic form as in Eq.~(\ref{RspR}).  We use the coefficients of this quadratic form to define effective parameters $\Delta\alpha_g$, $\Delta\lambda_g$ and $\Delta f_g$, so that,
\begin{equation}
{\cal R}\equiv\dfrac{1}{\pi R^2}\ln({\cal P}_1/{\cal P}_2)=-\Delta f_g +\Delta\alpha_g p_R+\Delta\lambda_g p_R^2
\label{eRGauss}
\end{equation}
where $p_R\equiv \Gamma_R/(\pi R^2)$, and
\begin{align}
\Delta f_g&= \frac{1}{\pi R^2}\left(\ln\left[\frac{\sigma_1}{\sigma_2}\right]+\dfrac{\langle\Gamma_R\rangle^2_2}{2\sigma_2^2}-\dfrac{\langle\Gamma_R\rangle_1^2}{2\sigma_1^2}\right)\nonumber
\\
\Delta\alpha_g&=\dfrac{\langle\Gamma_R\rangle_1}{\sigma_1^2}-\dfrac{\langle\Gamma_R\rangle_2}{\sigma_2^2}
\label{eGaussParams}
\\
\Delta\lambda_g&=\pi R^2\left(\dfrac{1}{2\sigma_2^2}-\dfrac{1}{2\sigma_1^2}\right),
\nonumber
\end{align}
where the subscripts 1,2 refer to values at $\tilde p_{1,2}$.

\begin{figure}[h]
\includegraphics[width=3.5in]{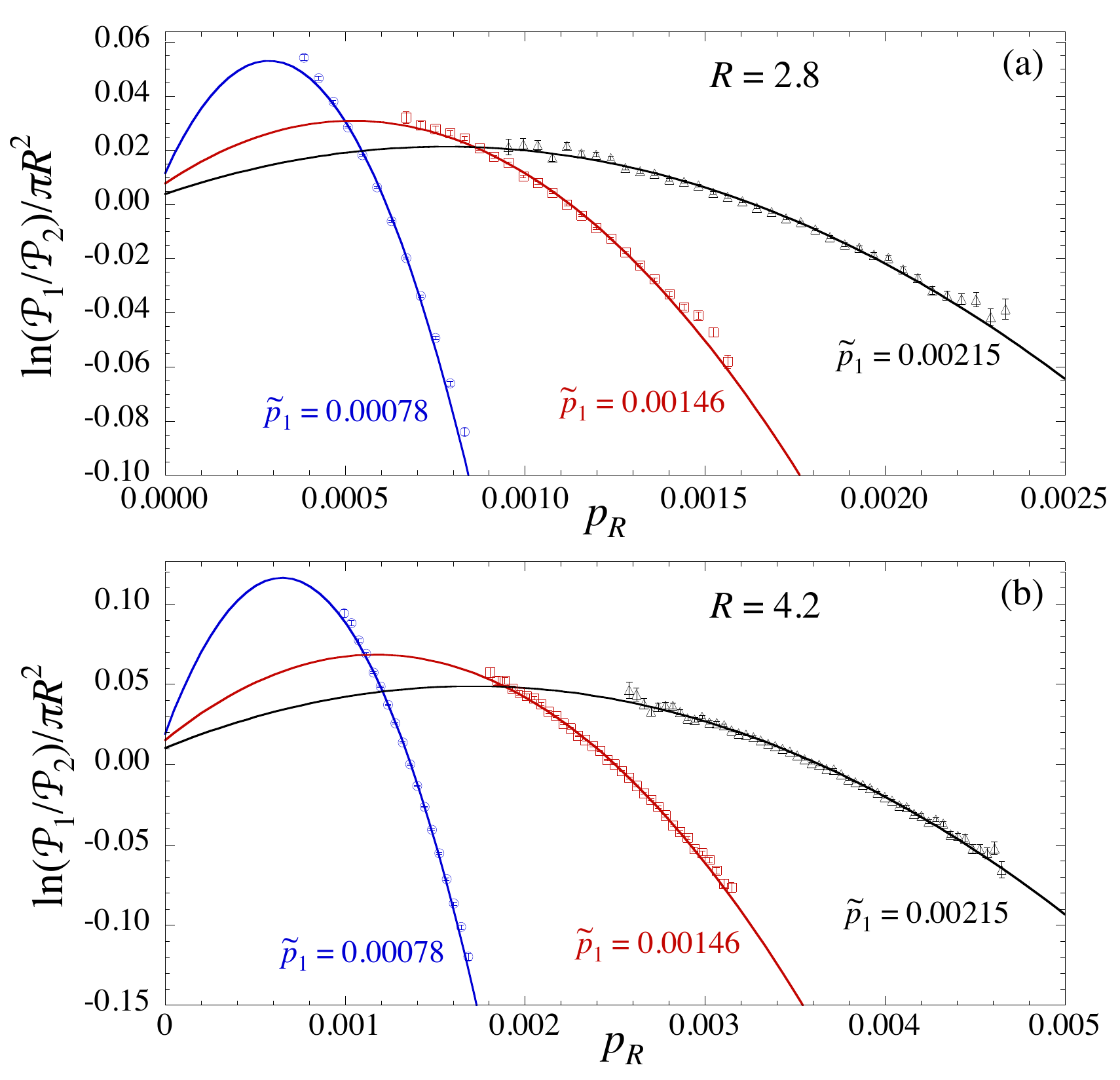}
\caption{(color online) Histogram ratio ${\cal R}$ at neighboring values of the total system stress per particle $\tilde p_1$ and $\tilde p_2$, vs cluster stress per area $p_R=\Gamma_R/(\pi R^2)$.  Data are shown for three different values of $\tilde p_1$, for clusters of radius (a) $R=2.8$ and (b) $R=4.2$.  Solid lines give the prediction of the Gaussian approximation of Eqs.~(\ref{eRGauss}) and (\ref{eGaussParams}).
}
\label{RGauss}
\end{figure}

Since we can easily compute averages and variances of $\Gamma_R$ \cite{WuTeitel}, the result of Eq.~(\ref{eGaussParams}) involves no adjustable parameters, and we can directly see how well it agrees with our numerically computed values for the histogram ratio.  In Fig.~\ref{RGauss} we plot our data together with the prediction of Eq.~(\ref{eGaussParams}) (solid lines) for two different cluster radii, $R=2.8$ and $R=4.2$, at three different values of the total stress per particle $\tilde p_1$.  We see that the agreement is not bad, although the prediction of Eq.~(\ref{eGaussParams}) noticeably curves away from the data at both the high and low ends, particularly for the smaller value of $R$.

\begin{figure}[h]
\includegraphics[width=3.5in]{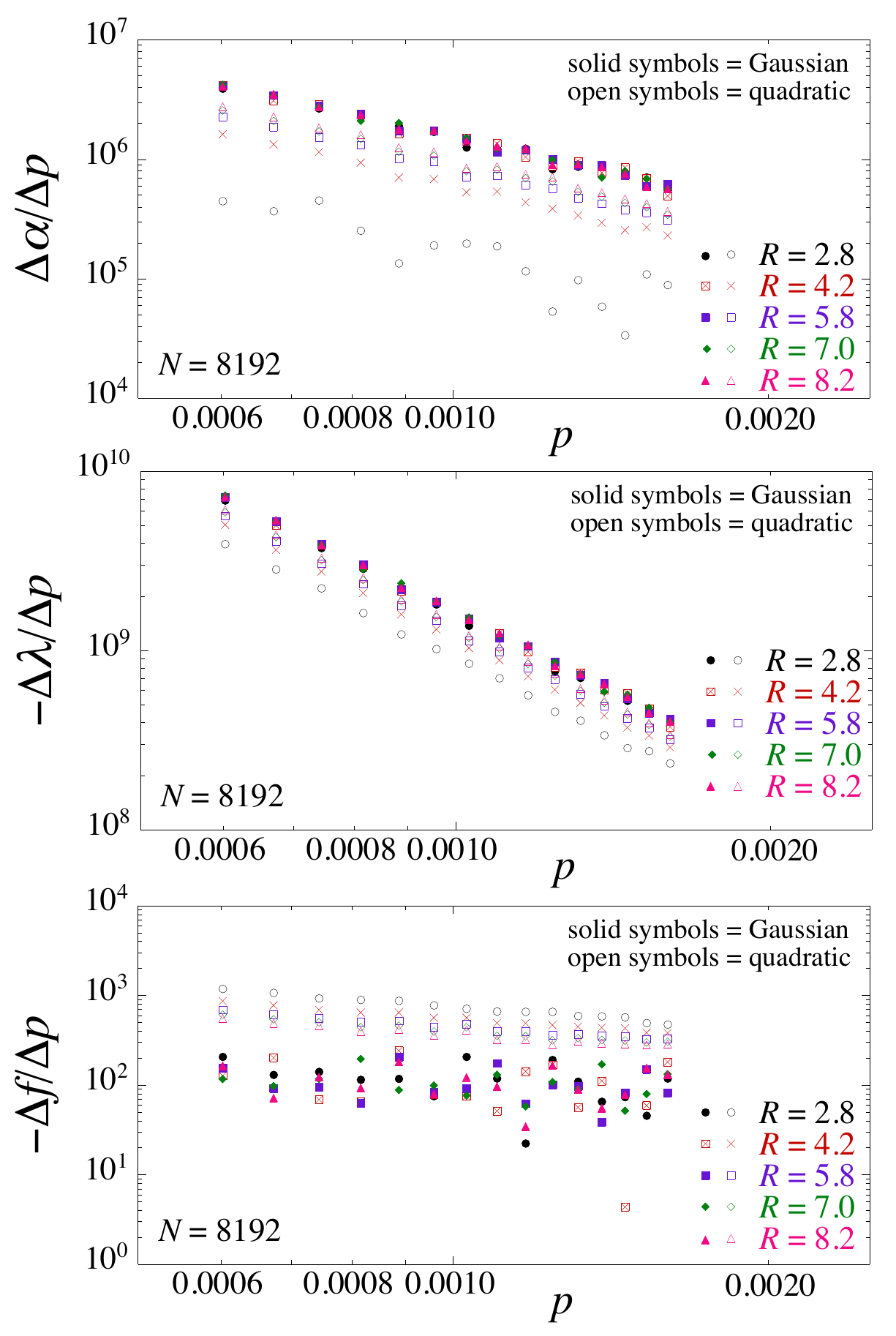}
\caption{(color online) Comparison of parameters $\Delta\alpha/\Delta p$, $\Delta\lambda/\Delta p$ and $\Delta f/\Delta p$ from the Gaussian approximation of Eq.~(\ref{eGaussParams}) with those obtained from the quadratic fit with $1/R$ corrections shown previously in Figs.~\ref{quadFits-1oR}a,b,c.  Solid symbols are for the Gaussian approximation, while open symbols are for the quadratic fit.  Results are plotted vs the system pressure $p$ for several different cluster radii $R$.  
}
\label{gaussCompare}
\end{figure}

In Fig.~\ref{gaussCompare} we compare the values of $\Delta\alpha_g$, $\Delta\lambda_g$ and $\Delta f_g$ from the Gaussian approximation of Eq.~(\ref{eGaussParams}) with the values of $\Delta\alpha$, $\Delta\lambda$ and $\Delta f$ obtained previously by the quadratic fit to ${\cal R}$ with $1/R$ corrections.  We see that the two sets of parameters are noticeably different.  However, if we consider the slope $\Delta \bar\alpha$ of ${\cal R}$ at the point of greatest histogram overlap, one can show that the Gaussian approximation gives results  essentially identical to that predicted for the Boltzmann distribution of Eq.~(\ref{edalphadp}) and so also identical to that found from the quadratic fit to ${\cal R}$, as shown in Figs.~\ref{barDalpha-Dp-1Dhist} and \ref{barDalpha-1oR}.  

Defining $\Delta\bar\alpha_g=\Delta\alpha_g+\Delta\lambda_g p_R^*$, and assuming the point of greatest overlap between the two histograms is at $p_R^*=(p_1+p_2)/2$, we find from Eq.~(\ref{eGaussParams}),
\begin{equation}
\Delta\bar\alpha_g= -\frac{\pi R^2}{2}\left[\dfrac{1}{\sigma_1^2}+\dfrac{1}{\sigma_2^2}\right]\Delta p,
\label{eDbaralphag}
\end{equation}
where $\Delta p=p_2-p_1$.  For $\Delta p$ sufficiently small, we can take to leading order $\sigma_1^2\approx\sigma_2^2$ in Eq.~(\ref{eDbaralphag}) and hence the above becomes equal to Eq.~(\ref{eDalpha}) found for the Boltzmann distribution.  Hence the agreement of $\Delta\bar\alpha$ between the numerically computed histogram ratio ${\cal R}$ and the value found via the fluctuations of $\Gamma_R$ as in Eq.~(\ref{eDalpha}) cannot in itself be taken as  evidence for the correctness of the Boltzmann distribution of Eq.~(\ref{ePGR}); the same relation holds just as well for a Gaussian distribution, provided $\Delta p$ is not too big.  The true test for the Boltzmann distribution of Eq.~(\ref{ePGR}) is therfore the linearity of the histogram ratio ${\cal R}$ in the cluster pressure $p_R$.

\begin{figure}[h]
\includegraphics[width=3.5in]{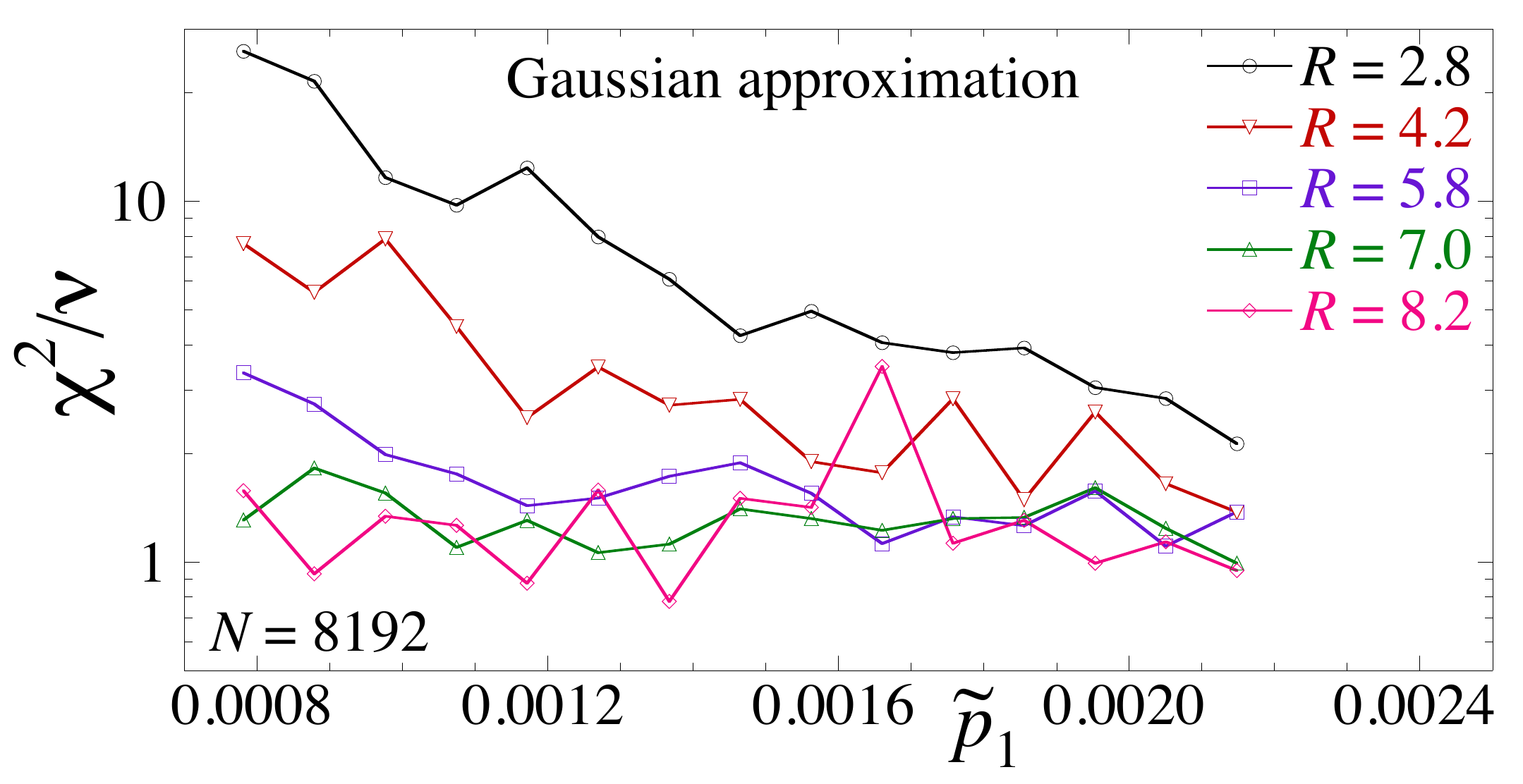}
\caption{(color online) Chi squared per degree of freedom, $\chi^2/\nu$, of the fit of the histogram ratio ${\cal R}$ to the Gaussian approximation of Eqs.~(\ref{eRGauss}) and (\ref{eGaussParams}).  Results are plotted vs $\tilde p_1$, the stress per particle at the lower of the two stresses $\tilde p_1$, $\tilde p_2$ used to define ${\cal R}$ (see Eq.~(\ref{Rslinear})), for several different cluster radii $R$.  }
\label{chisqGauss}
\end{figure}

Finally, to check quantitatively how well the Gaussian approximation is describing the histogram ratio data, we can compute the $\chi^2/\nu$ of the fit of the Gaussian results of Eq.~(\ref{eGaussParams}) to the measured data for ${\cal R}$.  In Fig.~\ref{chisqGauss} we plot this $\chi^2/\nu$ vs $\tilde p_1$, the stress per particle at the lower of the two stresses $\tilde p_1$, $\tilde p_2$ used to define ${\cal R}$, for several different cluster radii $R$.  We see that the Gaussian approximation is quite noticeably worse than the quadratic fits to ${\cal R}$ with $1/R$ corrections in the fitting parameters, as shown earlier in Fig.~\ref{chisq-1oR}.  Only for the largest cluster sizes $R$ is the Gaussian approximation reasonable, with $\chi^2/\nu\sim O(1)$.  This is because as $R$  increases at fixed $\Delta p$, our finite data sampling for ${\cal R}$ gets confined to an ever smaller region of $p_R$ about the point of greatest histogram overlap $p_R^*$, and so the data is decreasingly sensitive to the curvature in ${\cal R}$.

\subsection{Relation to previous work}
\label{secRelation}

A similar analysis of the same bidisperse two dimensional model has previously been carried out by Henkes et al. \cite{Henkes1}.  They used configurations quenched at constant packing fraction $\phi$ in a square box, rather than constant isotropic stress $\Gamma_N$.  They also considered a somewhat different histogram ratio than that considered in the present work.  They used,
\begin{equation}
\tilde{\cal R}_H\equiv\ln\left[\dfrac{{\cal P}(\Gamma_R|\tilde p_1)}{{\cal P}(\Gamma_R|\tilde p_2)}\dfrac{{\cal P}(\Gamma^\prime_R|\tilde p_2)}{{\cal P}(\Gamma^\prime_R|\tilde p_1)}\right].
\label{etildeR}
\end{equation}
Plotting $\tilde{\cal R}_H$ vs $\Gamma_R-\Gamma_R^\prime$, they found a linear relation, in agreement with expectations from the stress ensemble of Eq.~(\ref{ePGR}).

\begin{figure}[h]
\includegraphics[width=3.5in]{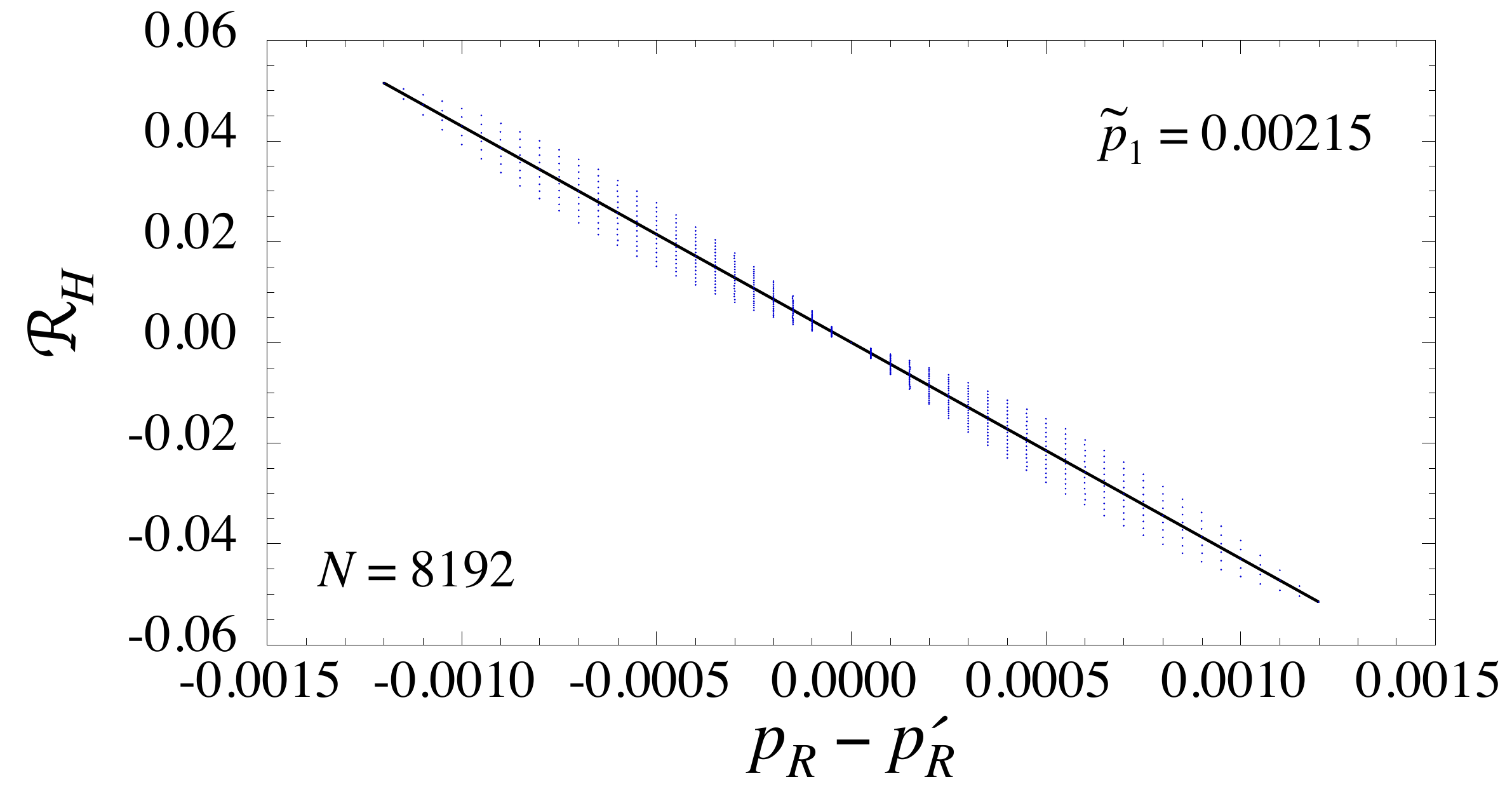}
\caption{(color online) Histogram ratio  of Eq.~(\ref{etildeR}) used by Henkes et al. in Ref.~\cite{Henkes1}, normalized by the cluster volume ${\cal R}_H=\tilde{\cal R}/(\pi R^2)$, vs $\tilde p_R-\tilde p_R^\prime$.  Data is computed using Eq.~(\ref{HenkesR1}) and previously determined values of $\Delta\alpha$ and $\Delta\lambda$ for the case of a stress per particle of $\tilde p=0.00215$.  Solid line is a linear fit to the data.
}
\label{RHenkes}
\end{figure}

However, our result of Eq.~(\ref{RspR}) for ${\cal R}$ leads to the conclusion that the ratio used by Henkes et al., when scaled by the cluster volume to be an intensive quantity, ${\cal R}_H\equiv\tilde{\cal R}_H/(\pi R^2)$, should obey,
\begin{align}
{\cal R}_H&= \Delta\alpha(p_R-p_R^\prime)+\Delta\lambda(p_R^2-p_R^{\prime 2})
\label{HenkesR1}\\
&=\left[\Delta\alpha+\Delta\lambda(p_R+p_R^\prime)\right](p_R-p_R^\prime).
\label{HenkesR2}
\end{align}
To check the behavior of ${\cal R}_H$, we consider the case with a stress per particle $\tilde p_1=0.00215$.  Generating a discrete set of evenly spaced values of $p_R$ that span the range of the data for this $\tilde p_1$ in Fig.~\ref{rescaled-Ratios}, and applying Eq.~(\ref{HenkesR1}) using the values of $\Delta\alpha$ and $\Delta\lambda$ obtained from the fit to Eq.~(\ref{RspR}) for this $\tilde p_1$, we plot  $R_H$ vs $p_R-p_R^\prime$ in Fig.~\ref{RHenkes}.

At each of the discrete values of $p_R-p_R^\prime$ there is a range of values of ${\cal R}_H$ corresponding to the different possible values of $p_R+p_R^\prime$, as seen from Eq.~(\ref{HenkesR2}).  But the average about these values is a straight line (solid line in Fig.~\ref{RHenkes}) of slope $\Delta\alpha +\Delta\lambda(p_1+p_2)$, where $p_{1,2}=\tilde p_{1,2}N/V_{1,2}$; $(p_1+p_2)/2$ locates the pressure at the point of greatest overlap between the two distributions ${\cal P}_{1,2}$ at $\tilde p_1$ and $\tilde p_2$.  This slope is thus exactly equal to the slope of the linear approximation to our ${\cal R}$ given by Eq.~(\ref{ebarDalpha}), and hence the results of Henkes et al. should be equivalent to the results shown in our Fig.~\ref{barDalpha-Dp-1Dhist}.  The straight line relation Henkes et al. observed between $\tilde{\cal R}_H$ and $\Gamma_R-\Gamma_R^\prime$, as opposed to the quadratic relation we find for our simpler ${\cal R}$, is therefore just an artifact of their having used the ratio of Eq.~(\ref{etildeR}), which upon averaging data at fixed $p_R-p_R^\prime$ averages away the non-linear behavior.

%Using their method, Henkes et al. then found $\alpha(\tilde p)=2\tilde p^{-1}$, and so $-d\alpha/d\tilde p = 2\tilde p^{-2}$.  If we fit our data in Fig.~\ref{barDalpha-Dp-1Dhist} to the power-law $\tilde p^{-2}$ we find (dashed line in Fig.~\ref{barDalpha-Dp-1Dhist}) $-\Delta\alpha/\Delta\tilde p\approx 2.3\tilde p^{-2}$, in reasonable agreement.

%%%%%%%%%%%%%%%
\section{Results: The stress -- force-tile ensemble}
\label{sResultsForceTile}

The results discussed in the previous section thus provide no compelling evidence that the stress distribution ${\cal P}(\Gamma_R|\tilde p)$ in our two dimensional system is indeed given by the simple stress ensemble form of Eq.~(\ref{ePGR}).  The Gaussian approximation also seems to be a poor representation of the distribution.  The good fit of the histogram ratio ${\cal R}$ to the quadratic form 
of Eq.~(\ref{RspR}) suggests instead that the distribution ${\cal P}(\Gamma_R|\tilde p)$ involves a Boltzmann factor with a quadratic term in the stress,
\begin{equation}
\mathrm{exp}\left[-\alpha\Gamma_R-\dfrac{\lambda}{\pi R^2}\Gamma_R^2\right],
\label{eBoltz}
\end{equation}
with $1/\alpha$ and $1/\lambda$ as intensive temperature-like variables that vary with the total system pressure, and that approach well defined values (with $1/R$ corrections) as the cluster size $R$ increases.   In this section we discuss and test one proposed mechanism for generating the above Boltzmann factor.

As mentioned earlier, the stress ensemble of Eq.~(\ref{ePGR}) may be viewed as resulting from a maximum entropy hypothesis, given that the average stress on the cluster $\langle\Gamma_R\rangle$ is constrained by the total system stress $\Gamma_N$, according to Eq.~(\ref{GammaR}).  However, if the system possesses other constrained observables, these too can effect the cluster stress distribution.  As pointed out by the work of Tighe et al. \cite{Tighe1,Tighe3,Tighe4}, in two dimensions the Maxwell-Cremona force-tile area \cite{Maxwell} is another such constraining quantity.  Moreover, they showed that this force-tile area leads naturally to a stress distribution with a Boltmann factor such as in Eq.~(\ref{eBoltz}).

\subsection{The Maxwell-Cremona force-tile area}
\label{secForceTile}

\begin{figure}
\includegraphics[width=3.2in]{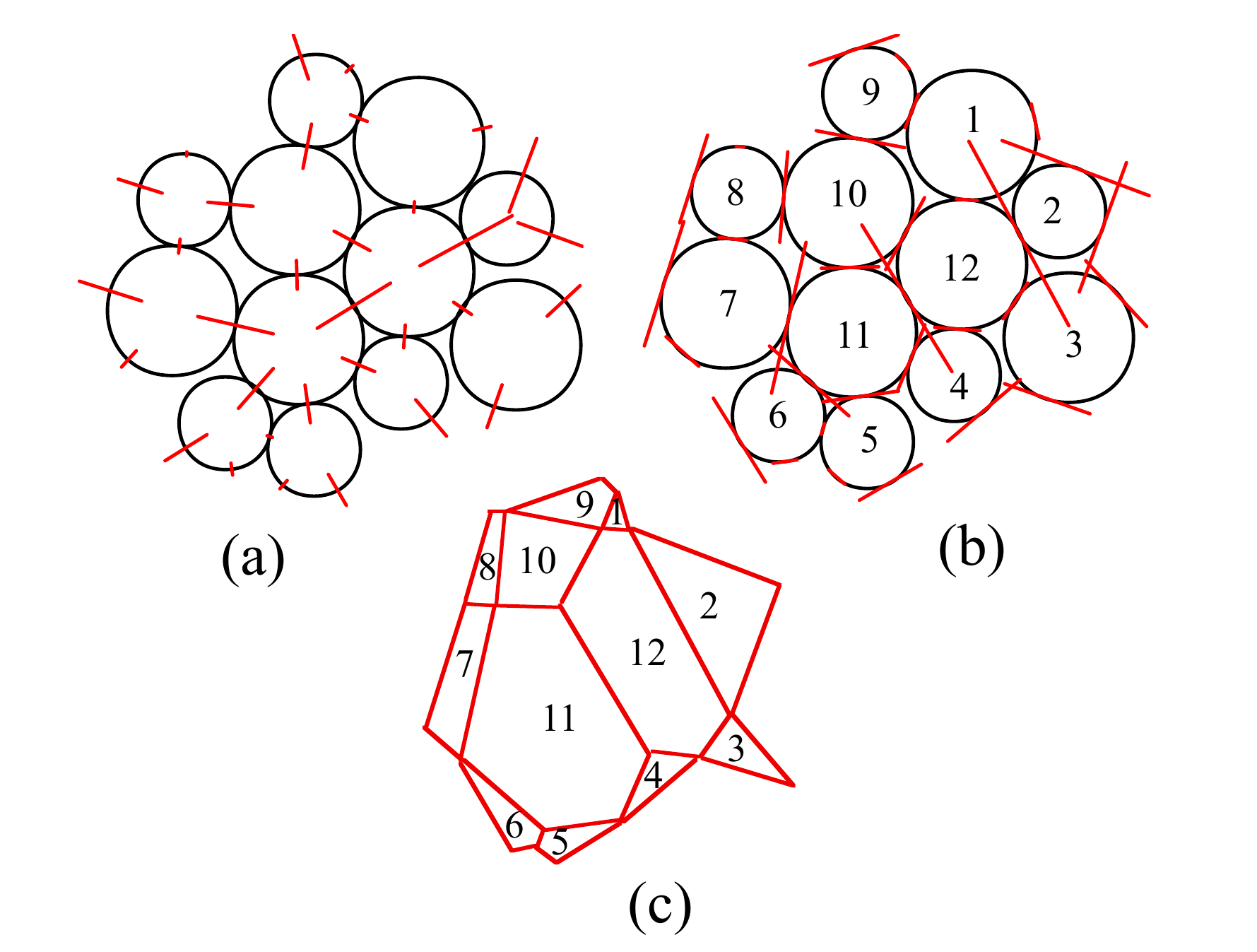}
\caption{(color online) Construction of the Maxwell-Cremona force-tiles for a sub cluster of our system: (a) red lines represent contact forces between the particles; the magnitude of the force is proportional to the length of the line; (b) force lines are rotated by $90^\circ$; (c) rotated force lines are translated to lie tip-to-tail forming closed loops that are the force tiles.  In (b) and (c), numbers denote particular particles and their corresponding force-tiles.
}
\label{tiles-abc}
\end{figure}

The Maxwell-Cremona force-tiles were introduced by Maxwell in 1864 \cite{Maxwell}.
We illustrate the construction of the force-tiles, a concept which applies only to two dimensional packings, in Fig.~\ref{tiles-abc}.  Panel a shows a sub cluster of particles within in a mechanically stable packing.  The red lines indicate the elastic forces between particles in contact; the length of each line is proportional to the magnitude of the contact force.  For our frictionless particles, these forces always point normal to the surface at the point of contact.  In panel b, the force lines of panel a are rotated $90^\circ$ so that they are now tangential to the particle surface.  In panel c, these rotated force lines are translated so as to place the force lines from each particle tip-to-tail going counterclockwise around each particle. 
Since the net force on each particle vanishes, the force lines for each particle must form a closed loop \cite{Ball}.  The area of the loop for particle $i$ is the particle's force-tile area $A_i$.  For frictionless particles, such as studied here, the force-tiles always have convex surfaces.  In panels b and c we number the particles and their corresponding force-tiles.

Because the contact force that defines a given edge of the force-tile of a particle $i$ must also be an edge of the force-tile of the particle $j$ that shares that contact, one can show
that the force-tiles tile space with no gaps or overlaps \cite{Tighe3}.  
The force-tile area of a cluster of particles ${\cal C}$ is then just the sum of force-tile areas for each member particle, $A_{\cal C}=\sum_{i\in{\cal C}}A_i$.  

\begin{figure}[h]
\includegraphics[width=3.4in]{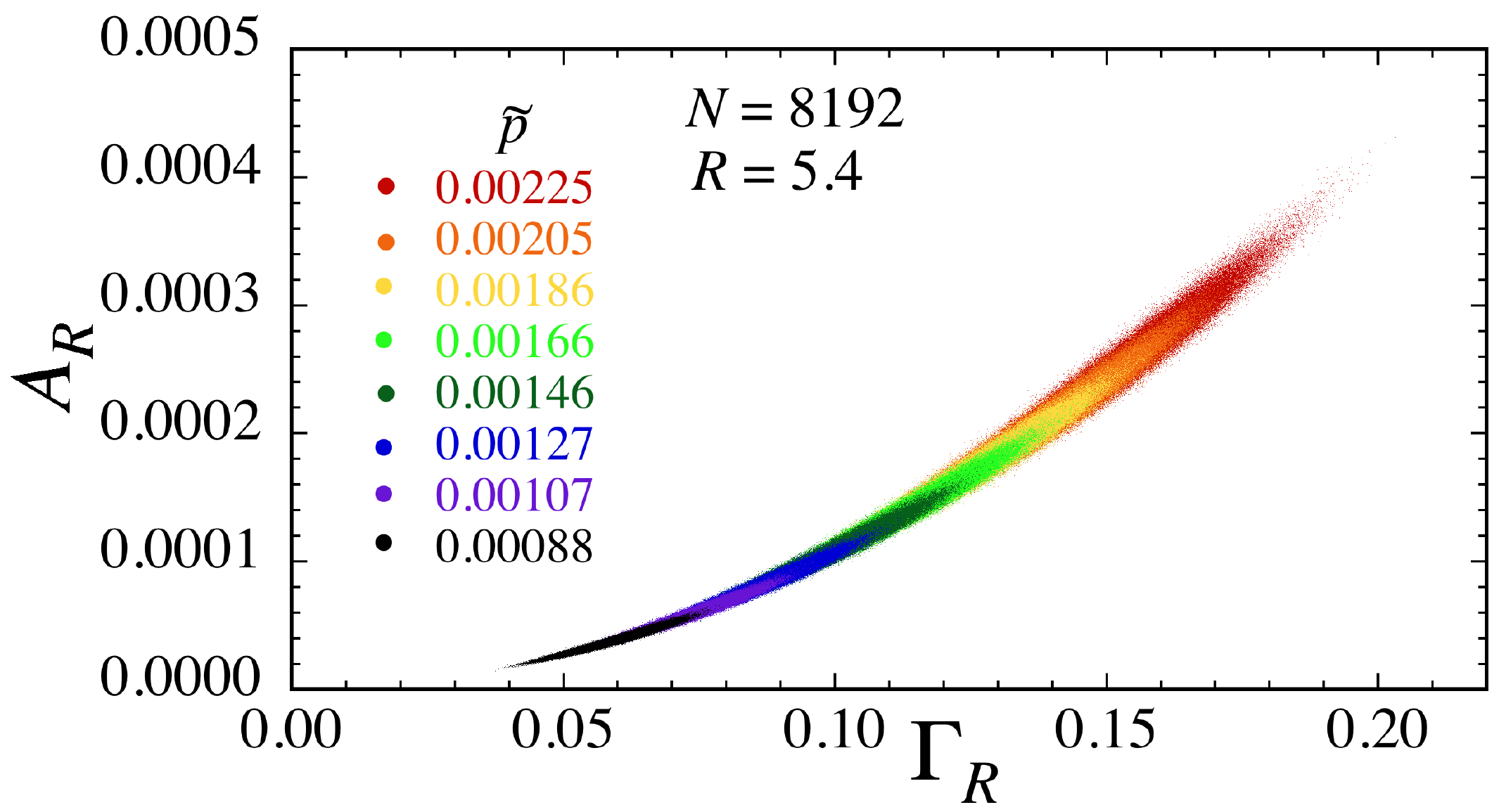}
\caption{(color online) Scatter plot of values of stress $\Gamma_R$ and force-tile area $A_R$, for clusters of radius $R=5.4$, for different values of total system stress per particle $\tilde p$ ranging from $0.00088$ to $0.00225$.  The smaller the value of $\tilde p$, the more compact is the distribution.  }
\label{2D-hist}
\end{figure}

For a packing with periodic Lees-Edwards boundary conditions, the force-tiling is similarly periodic, and the force-tile area for the total system $A_N$ is determined uniquely by the total system stress tensor, $A_N=\mathrm{det}[\Sigma^{(N)}_{\alpha\beta}]/V$  \cite{Tighe3}.  For our system with isotropic stress $\Sigma_{\alpha\beta}^{(N)}=\Gamma_N\delta_{\alpha\beta}$, and so 
\begin{equation}
A_N=\Gamma_N^2/V.  
\label{ANGN}
\end{equation}
For finite clusters of radius $R$, however, since the boundary is not fixed, $A_R$  may take a distribution of values for each given value of $\Gamma_R$.  We illustrate this in Fig.~\ref{2D-hist} where we show a scatter plot of the values of $A_R$ and $\Gamma_R$ found in individual clusters, for the particular cluster size $R=5.4$, at several different values of the total system stress per particle $\tilde p$. The distributions for neighboring values of $\tilde p$ overlap each other, similar to the distributions of $\Gamma_R$ in Fig.~\ref{GR-hist}.

Since  the force-tile area is conserved (i.e. the total system value $A_N$ is fixed and $A$ is additive over disjoint subsystems) the average on clusters of radius $R$ is constrained by,
\begin{equation}
\langle A_R\rangle = A_N\left(\frac{\pi R^2}{V}\right),
\label{AR}
\end{equation}
a result which we have numerically confirmed elsewhere  \cite{WuTeitel}.
Combining the above with Eq.~(\ref{GammaR}), and using the fixed relation between $A_N$ and $\Gamma_N$ given by Eq.~(\ref{ANGN}), then yields the relation between the average cluster force-tile area and the average cluster stress, 
\begin{equation}
\langle A_R\rangle=\dfrac{\langle\Gamma_R\rangle^2}{\pi R^2}.  
\label{eAG2}
\end{equation}
Defining an {\em intensive} force-tile area, $a_R\equiv A_R/(\pi R^2)$, and recalling $p_R\equiv \Gamma_R/(\pi R^2)$, the above becomes simply,
\begin{equation}
\langle a_R\rangle=\langle p_R\rangle^2=p^2.
\label{eap2}
\end{equation}
%It is this relation which lies behind the quadratic curvature in the data of Fig.~\ref{f2}a.
Thus a maximum entropy formulation should consider the joint distribution of both $\Gamma_R$ and $A_R$, treating both as constrained variables whose averages are known.  Assuming that all configurations with a given pair of $(\Gamma_R,A_R)$ are equally likely, one gets,
\begin{equation}
{\cal P}(\Gamma_R, A_R|\tilde p)=\Omega_R(\Gamma_R,A_R)\dfrac{\mathrm{e}^{-\alpha(\tilde p)\Gamma_R-\lambda(\tilde p) A_R}}{Z_R(\tilde p)},
\label{PGA}
\end{equation}
with 
\begin{equation}
Z_R(\tilde p)\equiv\int d\Gamma_R\int d A_R\Omega_R(\Gamma_R,A_R)\mathrm{e}^{-\alpha(\tilde p)\Gamma_R-\lambda(\tilde p) A_R}.
\label{eZGA}
\end{equation}

\subsection{Histogram ratio}
\label{secHist2D}

Considering the joint distribution of $\Gamma_R$ and $A_R$ at two neighboring values of the total system stress per particle, $\tilde p_1$ and $\tilde p_2$, we can again construct the log histogram ratio ${\cal R}$.  From Eq.~(\ref{PGA}) we get,
\begin{equation}
{\cal R}\equiv\frac{1}{\pi R^2}\ln\left[\frac{{\cal P}_1}{{\cal P}_2}\right]=-\Delta f+\Delta\alpha\, p_R+\Delta\lambda\, a_R
\label{pRaR}
\end{equation}
where $\Delta f\equiv -\ln[Z_{R,2}/Z_{R,1}]/(\pi R^2)$, 
$\Delta\alpha\equiv \alpha_2-\alpha_1$ and $\Delta\lambda\equiv\lambda_2-\lambda_1$.  If the parameters $\Delta f$, $\Delta\alpha$ and $\Delta\lambda$ are intensive, with only a weak dependence on the cluster size $R$, then plotting ${\cal R}$ vs the intensive quantities $p_R$ and $a_R$, data for different cluster sizes $R$ should all collapse to a single flat plane for a given pair $\tilde p_1,\tilde p_2$.  The slopes of the plane in directions $p_R$ and $a_R$ determine the values of $\Delta\alpha$ and $\Delta\lambda$.

\begin{figure}[h]
\includegraphics[width=3.5in]{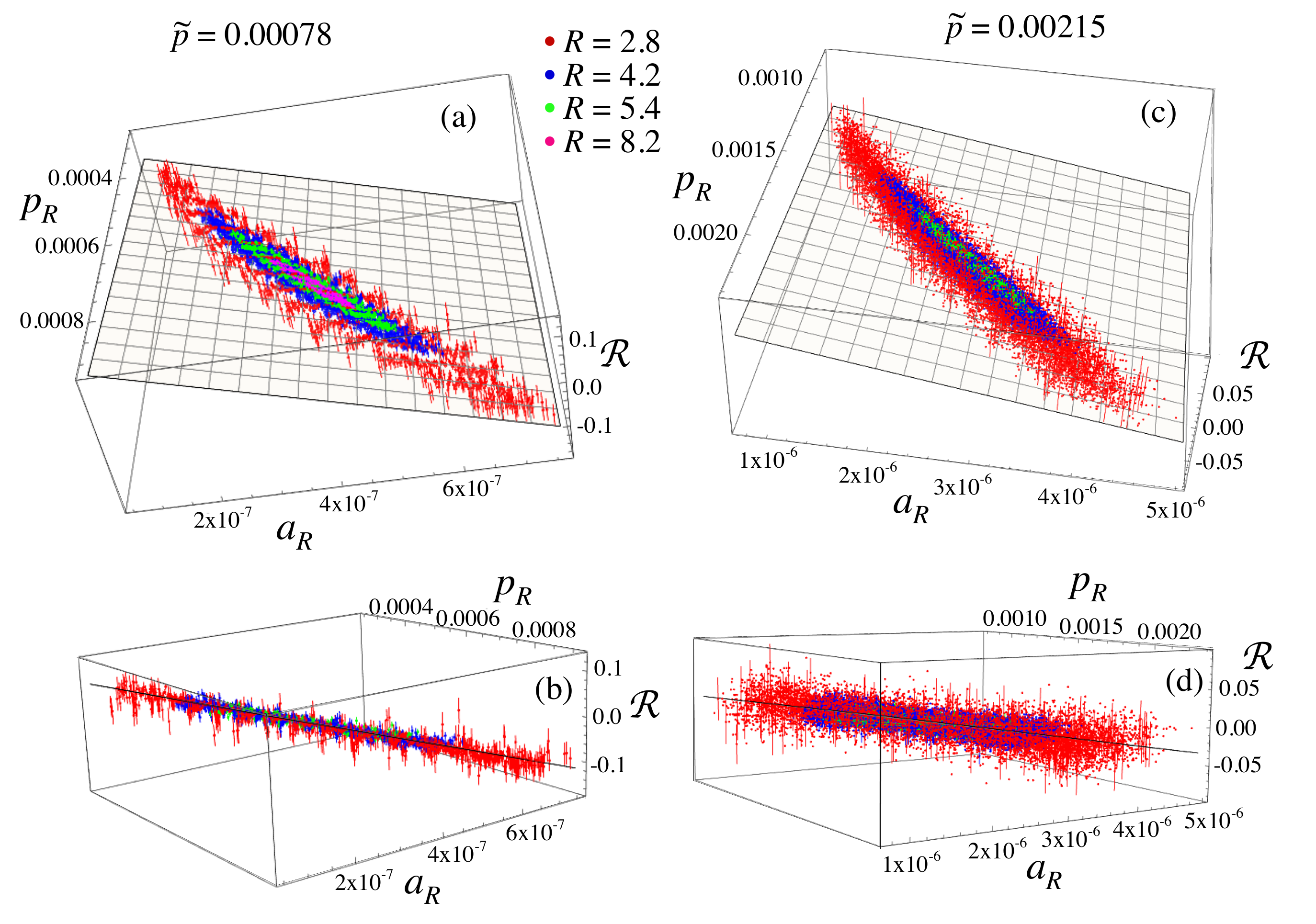}
\caption{(color online) (a) Plot of log histogram ratio ${\cal R}$ vs cluster pressure $p_R$ and force-tile area per volume $a_R$ for different cluster radii $R$ at the total system stress per particle $\tilde p=0.00078$.  The data cluster into more compact regions as $R$ increases.  Shaded region shows the best planar fit to the data, where all fit parameters are taken to be independent of $R$. (b) Same as (a) but looking edge on at the fitted plane, confirming that all data lies on a common flat plane. Panels (c) and (d) are the same as (a) and (b) but at the total system stress per particle $\tilde p=0.00215$.  To increase the clarity of the figure, in panels (c) and (d) error bars are shown on only a randomly selected 5\% of the data points.}
\label{2D-ratio}
\end{figure}

Computing ${\cal R}$ from our numerically determined joint histograms, we find that our data for ${\cal R}$ do indeed  collapse quite well onto a single flat plane for all $R$.  
In Fig.~\ref{2D-ratio} we show  ${\cal R}$ vs $p_R$ and $a_R$ for several different cluster radii $R$.    Panels a and b show results for our lowest system stress, $\tilde p_1=0.00078$.  Panel a shows a side view looking down upon this plane from the side; the data cluster into  more compact regions as $R$ increases.   Panel b shows a view looking edge on at the plane, thus confirming that the surface defined by our data is indeed a flat common plane for all $R$.  Panels c and d show similar results for our largest system stress, $\tilde p_1=0.00215$.

\begin{figure}[h]
\includegraphics[width=3.5in]{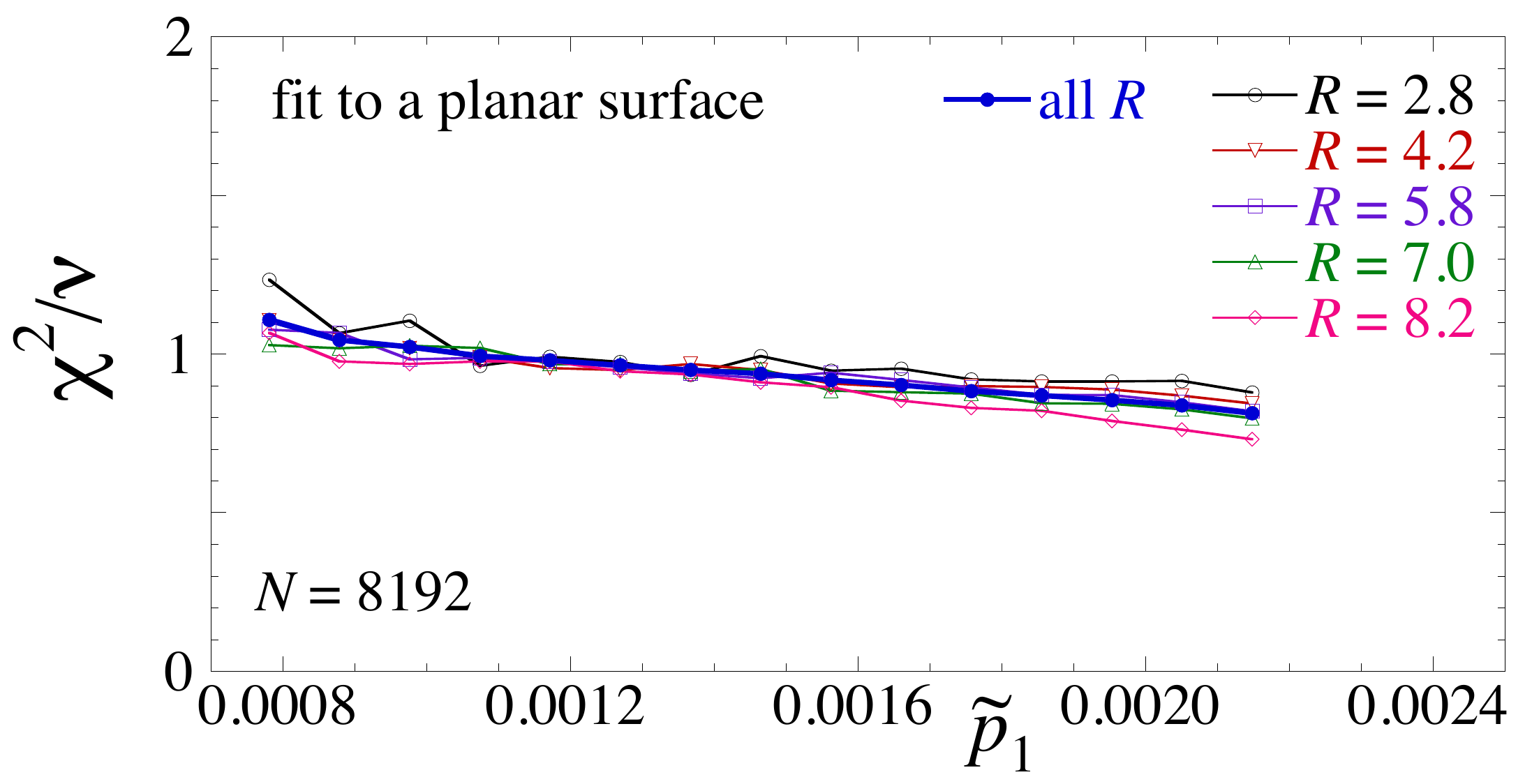}
\caption{(color online) Chi squared per degree of freedom, $\chi^2/\nu$, of fits of the histogram ration ${\cal R}$ to the planar form of Eq.~(\ref{pRaR}).  Results labeled ``all R" (solid circles) are fits keeping the parameters $\Delta f$, $\Delta\alpha$ and $\Delta\lambda$ the same for all cluster sizes $R$.  Other results are for fits specifically to the indicated cluster size $R$ alone. Results are plotted vs $\tilde p_1$, the stress per particle at the lower of the two stresses $\tilde p_1$, $\tilde p_2$ used to define ${\cal R}$ (see Eq.~(\ref{pRaR})).  $\chi^2/\nu\approx 1$ indicates an excellent fit.
}
\label{2DchisqPlanar}
\end{figure}

Fitting our data to the planar form of Eq.~(\ref{pRaR}), and taking the fit parameters $\Delta f$, $\Delta\alpha$ and $\Delta\lambda$ as constants independent of the cluster radius $R$, our fit gives the shaded planes shown in Fig.~\ref{2D-ratio}.  In Fig.~\ref{2DchisqPlanar} we show the $\chi^2/\nu$ of this fit (solid circles) to the entire data set of all cluster sizes $R$; we see that the fit is excellent with $\chi^2/\nu\approx 1$ for all stresses $\tilde p_1$. We have also tried fits where we allow the parameters $\Delta f$, $\Delta\alpha$ and $\Delta\lambda$ to have $1/R$ corrections, as in Eq.~(\ref{ealpha-1oR}).  We find little change in our results, with $\chi^2/\nu\approx 1$ remaining for all $\tilde p_1$, essentially no change in $\Delta\alpha$ and $\Delta\lambda$, and only a small shift in $\Delta f$.  Finally, we have also done planar fits to each cluster size $R$ independently, so that $\Delta f$, $\Delta\alpha$ and $\Delta\lambda$ may depend on $R$ in any arbitrary way.  The resulting $\chi^2/\nu$ from such fits are shown in Fig.~\ref{2DchisqPlanar} for several different $R$ (open symbols), and we see again that $\chi^2/\nu\approx 1$ everywhere.

\begin{figure}[h]
\includegraphics[width=3.5in]{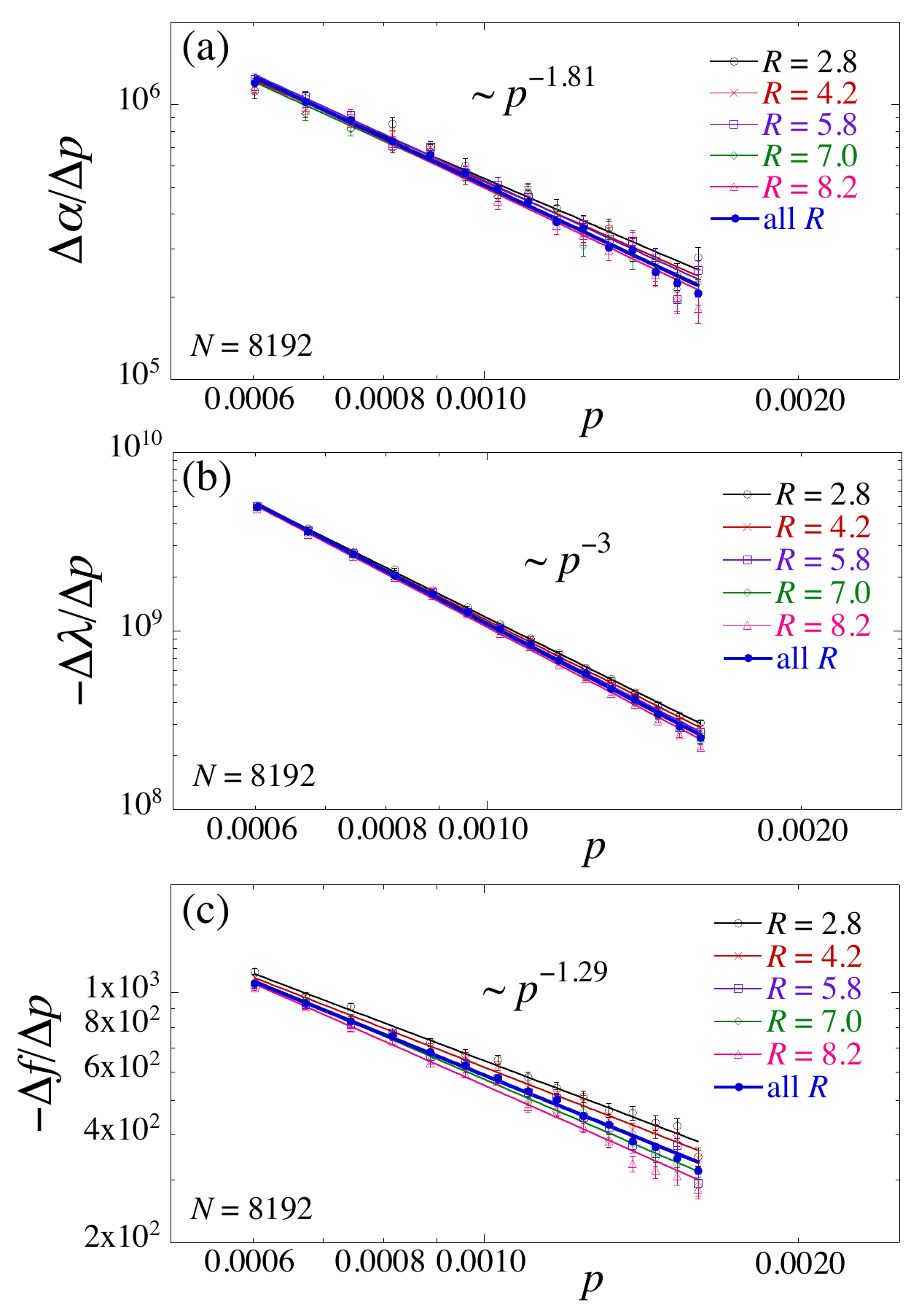}
\caption{(color online) Comparison of parameters (a) $\Delta\alpha/\Delta p$, (b) $-\Delta\lambda/\Delta p$ and (c) $-\Delta f/\Delta p$ obtained from fits of the histogram ratio ${\cal R}$ to the planar form of Eq.~(\ref{pRaR}).  Results labeled ``all $R$" (solid circles) are from fits where $\Delta\alpha$, $\Delta\lambda$ and $\Delta f$ are taken the same for all cluster sizes $R$.  Other results (open symbols) are from fits specifically to the indicated cluster size $R$ alone, with $\Delta\alpha$, $\Delta\lambda$ and $\Delta f$ chosen independently for each $R$.  Results are plotted vs the pressure $p=(p_1+p_2)/2$.  Solid lines are fits to a power-law, with the indicated power-law being the result from the fit to the largest value of $R$.
}
\label{2Dplanar-params}
\end{figure}

In Fig.~\ref{2Dplanar-params} we plot the resulting fit parameters as $\Delta\alpha/\Delta p$, $-\Delta\lambda/\Delta p$ and $-\Delta f/\Delta p$ vs the pressure $p=(p_1+p_2)/2$.  We show results for the case where we take $\Delta\alpha$, $\Delta\lambda$ and $\Delta f$ to be the same for all cluster radii $R$ (solid circles), as well the case where we fit separately to clusters of a specific $R$ (open symbols).  For $\Delta\alpha/\Delta p$ and $\Delta\lambda/\Delta p$ the results show little sensitivity to which case is used, or to the cluster size $R$ in the second case;  $\Delta f/\Delta p$ shows a somewhat greater sensitivity at the larger values of $p$,  suggesting that some $R$-dependence does exist for $\Delta f$.

\subsection{Fluctuations}
\label{sfluc2D}

Similar to the discussion for the stress ensemble in Sec.~\ref{G-flucs}, in the stress -- force-tile ensemble we can relate the parameters $\alpha$ and $\lambda$ to the fluctuations of stress $\Gamma_R$ and force-tile area $A_R$.  For the ensemble of Eq.~(\ref{PGA}), and with ${\cal F}_R\equiv -\ln Z_R$, we have,
\begin{equation}
\left(\dfrac{\partial{\cal F}_R}{\partial \alpha}\right)_\lambda = \langle\Gamma_R\rangle,\quad
\left(\dfrac{\partial{\cal F}_R}{\partial \lambda}\right)_\alpha = \langle A_R\rangle,
\end{equation}
and
\begin{align}
\left(\dfrac{\partial^2{\cal F}_R}{\partial \alpha^2}\right)_\lambda &= \left(\dfrac{\partial\langle\Gamma_R\rangle}{\partial\alpha}\right)_\lambda=-\mathrm{var}(\Gamma_R)\\[10pt]
\left(\dfrac{\partial^2{\cal F}_R}{\partial \lambda^2}\right)_\alpha &= \left(\dfrac{\partial\langle A_R\rangle}{\partial\lambda}\right)_\alpha=-\mathrm{var}(A_R)\\[10pt]
\left(\dfrac{\partial^2{\cal F}_R}{\partial \alpha\partial\lambda}\right) &=\left(\dfrac{\partial\langle\Gamma_R\rangle}{\partial\lambda}\right)_\alpha = \left(\dfrac{\partial\langle A_R\rangle}{\partial\alpha}\right)_\lambda=-\mathrm{cov}(\Gamma_R,A_R),
\end{align}
where $\mathrm{cov}(\Gamma_R,A_R)=\langle\Gamma_R A_R\rangle-\langle\Gamma_R\rangle\langle A_R\rangle$ is the covariance.

Defining the covariance matrix
\begin{equation}
\mathbb{C}\equiv\left[
\begin{array}{ll}
\mathrm{var}(\Gamma_R)&\mathrm{cov}(\Gamma_R,A_R)\\
\mathrm{cov}(\Gamma_R,A_R)&\mathrm{var}(A_R)
\end{array}
\right],
\label{eCovar}
\end{equation}
the changes in the average cluster stress and average cluster force-tile area in response to changes $\Delta\alpha$ and $\Delta\lambda$ in the parameters $\alpha$ and $\lambda$, are given by,
\begin{equation}
\left[
\begin{array}{l}
\langle\Delta\Gamma_R\rangle\\ \langle\Delta A_R\rangle
\end{array}
\right]
=-\mathbb{C}\cdot\left[
\begin{array}{l}
\Delta\alpha\\ \Delta\lambda
\end{array}
\right].
\label{eC1}
\end{equation}

Consider now our global system with periodic boundary conditions.  If we vary the total system pressure an amount $\Delta p$ from $p_1=\Gamma_{N1}/V_1$ to $p_2=\Gamma_{N2}/V_2$, then by Eq.~(\ref{epR}) the average stress on the cluster will vary as $\langle\Delta\Gamma_R\rangle/(\pi R^2)=\langle \Delta p_R\rangle =  \Delta p$.  By Eq.~(\ref{eap2}), the average force-tile area of the cluster will vary as
$\langle\Delta A_R\rangle /(\pi R^2)= \langle \Delta a_R\rangle =\langle p_R\rangle_2^2 -\langle p_R\rangle^2_1=p_2^2-p_1^2=(p_1+p_2)\Delta p$.  Taking the limit $\Delta p\to 0$ and inverting Eq.~(\ref{eC1}) we then get,
\begin{equation}
\left[
\begin{array}{l}
d\alpha/dp\\d\lambda/dp
\end{array}
\right]=-\pi R^2\mathbb{C}^{-1}\cdot\left[
\begin{array}{c}
1\\ 2p
\end{array}
\right]
\label{eCinv}
\end{equation}
where $\mathbb{C}^{-1}$ is the inverse of the covariance matrix.  
Thus the use of a global system with periodic boundary conditions, which by Eq.~(\ref{eap2}) restricts the average cluster behavior to lie on  the specific curve $\langle a_R\rangle =\langle p_R\rangle^2$ in $(p_R,a_R)$ space, similarly requires that $\alpha$ and $\lambda$ for the periodic system can not be chosen as independent parameters, but must be related to each other parametrically via the global pressure $p$ so as to satisfy Eq.~(\ref{eCinv}).  Or to put it another way, the use of a global system with periodic boundary conditions restricts the Boltzmann distribution of Eq.~(\ref{PGA}) to parameters that lie on a specific parametric curve $(\alpha(p),\lambda(p))$ in the more general $(\alpha,\lambda)$ space.

\begin{figure}[h]
\includegraphics[width=3.5in]{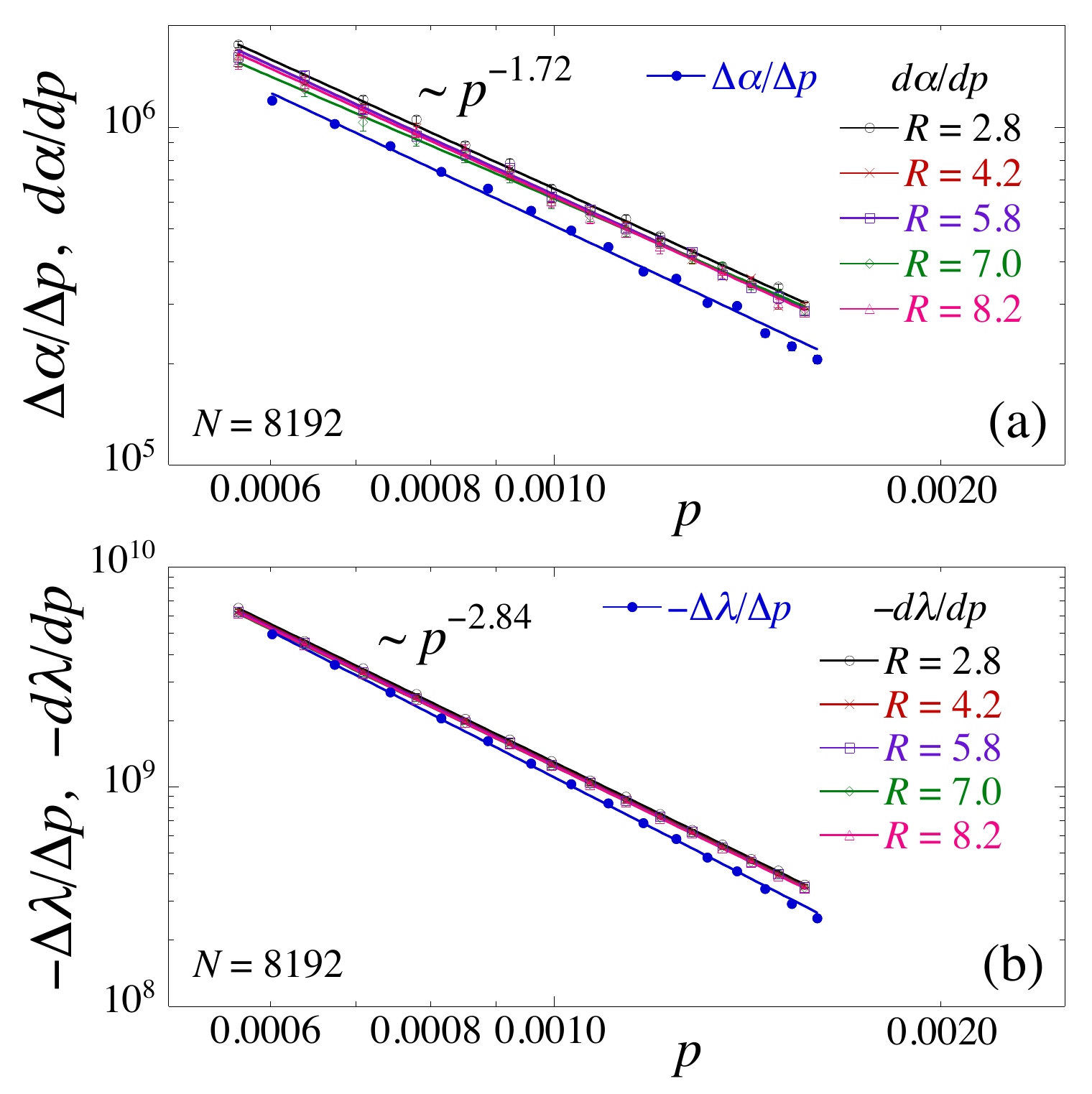}
\caption{(color online) (a) Comparison of $\Delta\alpha/\Delta p$ from the fit of the histogram ratio ${\cal R}$ to Eq.~(\ref{pRaR}) with $d\alpha/dp$ predicted by the covariance matrix $\mathbb{C}$ in Eq.~(\ref{eCinv}); $\Delta\alpha/\Delta p$ is computed assuming constant fit parameters for all cluster sizes $R$, while $d\alpha/dp$ is computed for the specific cluster sizes $R$ indicated.  (b) Similar comparison of $\Delta\lambda/\Delta p$ from the histogram ratio to $d\lambda/dp$ from the covariance matrix.  Results are plotted vs total system pressure $p$.
}
\label{Da-Dl-covars}
\end{figure}

Numerically computing the covariance matrix as in Ref.~\cite{WuTeitel}, in Fig.~\ref{Da-Dl-covars} we plot the $d\alpha/dp$ and $d\lambda/dp$ predicted by Eq.~(\ref{eCinv}) vs the system pressure $p$, for several different cluster radii $R$.  For comparison, on the same plot we also show $\Delta\alpha/\Delta p$ and $\Delta\lambda/\Delta p$ as obtained from our planar fit  to the histogram ratio ${\cal R}$, assuming constant fit parameters for all cluster sizes $R$ (as shown previously in Fig.~\ref{2Dplanar-params}).  We see good qualitative agreement, but quantitatively, the results from the histogram ratio are somewhat smaller than from the covariance matrix; $\Delta\alpha/\Delta p$ ranges from roughly 80\% to 75\% of $d\alpha/dp$, as pressure $p$ increases, while $
\Delta\lambda/\Delta p$ ranges from roughly 99\% to 80\% of $d\lambda/dp$ as $p$ increases.  Given the very good degree to which our data for the histogram ratio ${\cal R}$ is described by the flat plane of Eq.~(\ref{pRaR}), it is not clear why the agreement is not better.  We may speculate that additional macroscopic variables besides $\Gamma_R$ and $A_R$ might be needed for a more complete description of the ensemble \cite{WuTeitel,Song,Blumenfeld2}.   

\subsection{Gaussian approximation}
\label{gauss-2D}

As we did in Sec.~\ref{gauss-1D} for the distribution ${\cal P}(\Gamma_R|\tilde p)$, we can consider a Gaussian approximation to our joint distribution ${\cal P}(\Gamma_R, A_R|\tilde p)$.  Defining the two dimensional vector of observables $\mathbf{X}_R\equiv (\Gamma_R, A_R)$, we have,
\begin{equation}
{\cal P}(\Gamma_R,A_R|\tilde p)=\dfrac{1}{2\pi \sqrt{\mathrm{det}[\mathbb{C}]}}\mathrm{e}^{-\frac{1}{2}\boldsymbol\delta\mathbf{X}_R\cdot\mathbb{C}^{-1}\cdot\boldsymbol\delta\mathbf{X}_R},
\label{eGaussApprox}
\end{equation}
where $\mathbb{C}$ is the covariance matrix of Eq.~(\ref{eCovar}), and $\boldsymbol{\delta}\mathbf{X}_R \equiv \mathbf{X}_R-\langle\mathbf{X}_R\rangle$ is the fluctuation of the observables from their average.

\begin{figure}[h]
\includegraphics[width=3.5in]{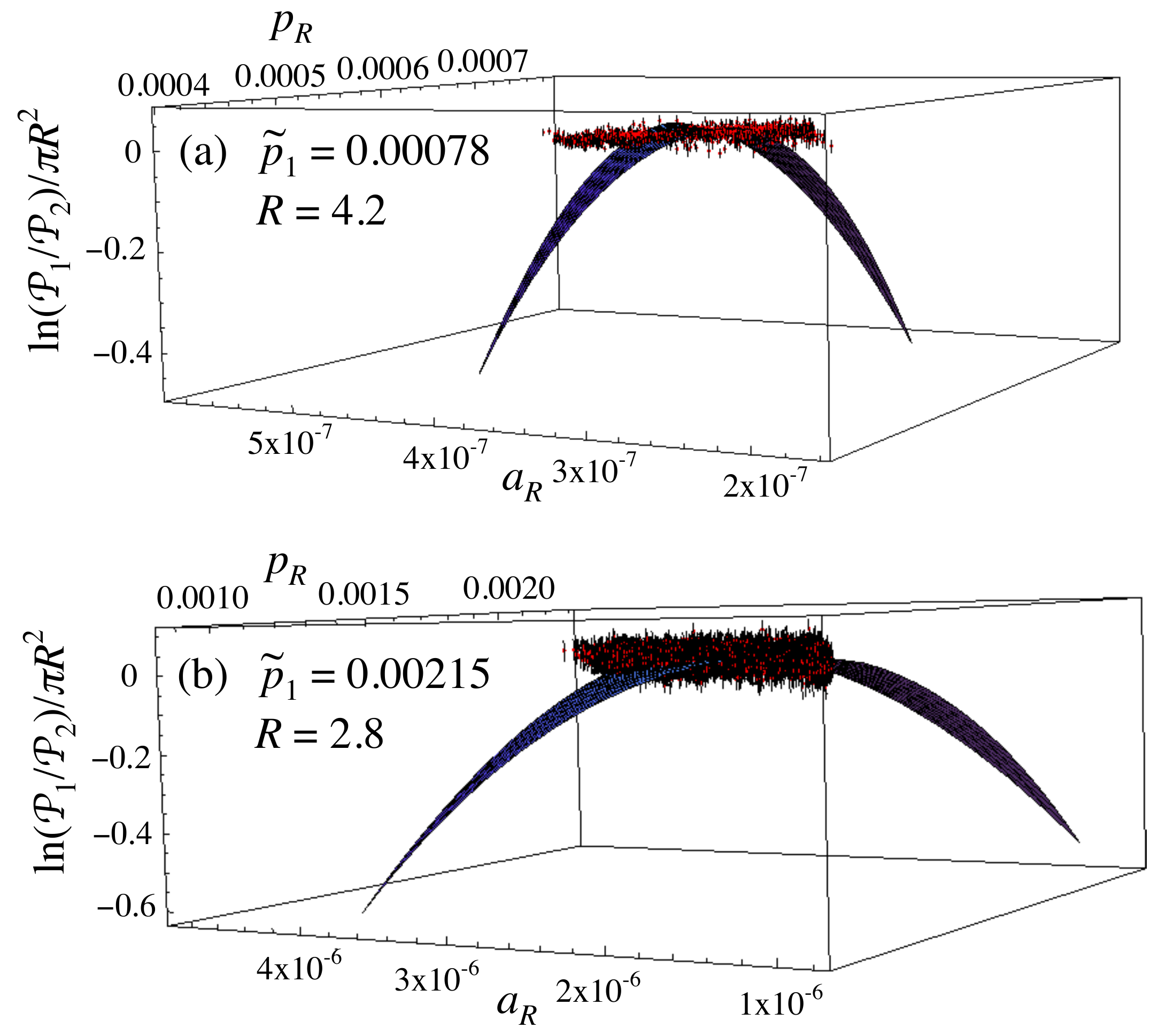}
\caption{(color online) Log histogram ratio ${\cal R}$ 
at neighboring values of the total system stress per particle $\tilde p_1$ and $\tilde p_2$, vs  $p_R=\Gamma_R/(\pi R^2)$ and $a_R=A_R/(\pi R^2)$.  Results are shown for (a) $\tilde p_1=0.00078$, cluster radius $R=4.2$ and (b) $\tilde p_1=0.00215$, cluster radius $R=2.8$.  Red points with error bars are our data for the numerically computed ${\cal R}$, while the curved surfaces shown are the predictions of the Gaussian approximation of Eq.~(\ref{eGauss-2D}).
}
\label{2D-gauss-surface}
\end{figure}

The histogram ratio ${\cal R}$ in this Gaussian approximation is then given by,
\begin{align}
{\cal R}&\equiv \dfrac{1}{\pi R^2}\ln\left[{\cal P}_1/{\cal P}_2\right]\nonumber\\[8pt]
&=\dfrac{1}{2\pi R^2} \Big[ \ln\big(\mathrm{det}[\mathbb{C}_2]/\mathrm{det}[\mathbb{C}_1]\big)
\label{eGauss-2D}\\[8pt]
&+
\boldsymbol\delta\mathbf{X}_{R2}\cdot\mathbb{C}^{-1}_2\cdot\boldsymbol\delta\mathbf{X}_{R2}-\boldsymbol\delta\mathbf{X}_{R1}\cdot\mathbb{C}^{-1}_1\cdot\boldsymbol\delta\mathbf{X}_{R1}\Big].
\nonumber
\end{align}
The quadratic forms in the above expression result in a parabolic surface rather than the flat plane expected for the Boltzmann distribution of Eq.~(\ref{PGA}).  To compare this surface against our numerical results, in Fig.~\ref{2D-gauss-surface} we show our data for ${\cal R}$, together with the surface predicted by Eq.~(\ref{eGauss-2D}), for (a) a cluster of radius $R=4.2$ at our smallest $\tilde p=0.00078$, and (b) a cluster of radius $R=2.8$ at our largest $\tilde p=0.00215$.   In both cases we see that the surface of the Gaussian approximation shows a clear curvature away from the ${\cal R}$ computed numerically from our overlapping histograms.  Unlike our results in Sec.~\ref{gauss-1D}, where the curvature of the Gaussian approximation gave a better description of the histogram ratio ${\cal R}$ than did the straight line of the stress ensemble, here the Gaussian approximation is yielding a curvature that is absent from the data.  The Boltzmann distribution of Eq.~(\ref{PGA}) is therefore clearly a better description of our data than the Gaussian approximation of Eq.~(\ref{eGaussApprox}).

\subsection{Relation to previous work}

The ideal that the Maxwell-Cremona force-tile area should play an important role in determining the stress distribution in two dimensional jammed packings was first put forward by Tighe and co-workers \cite{Tighe1,Tighe3,Tighe4}.  They, however, considered an idealized model known as the force-network ensemble (FNE) \cite{Snoeijer1, Snoeijer2,Tighe2} rather than the more realistic spatially disordered packings considered here.  The FNE is defined by noting that a mechanically stable packing above the jamming transition has an average particle contact number $\langle z\rangle$ that is larger than the isostatic value $z_\mathrm{iso}$ \cite{Liu+Nagel,OHern}.   For a fixed set of particle positions, when $\langle z\rangle > z_\mathrm{iso}$, the constraint of force balance on each particle under-determines the set of contact forces, and so there are many possible contact force configurations that can lead to a mechanically stable state, consistent with a given global stress tensor.  In the FNE one assumes that all such mechanically stable contact force configurations are equally likely, and posits that it is such contact force fluctuations, decoupled from fluctuations in the particle positions, that is the primary factor determining the distribution of stresses in the jammed packing.  The FNE thus considers only such contact force fluctuations for a given {\em fixed} set of particle positions.
Unlike the jammed packings considered in the present work, the FNE possesses no fluctuations in particle density nor system volume.  

In most of their computations for frictionless particles, Tighe and co-workers \cite{Tighe1,Tighe3,Tighe4} employed an FNE where the particles are constrained to sit at the sites of a regular triangular lattice, with forces acting between particles that share nearest neighbor bonds of the lattice.  In such a network each particle has a contact number $z=6$, well above the isostatic value $z_\mathrm{iso}=4$ that characterizes the jamming transition for frictionless circular disks in two dimensions \cite{Liu+Nagel,OHern} (the configurations in the present work have $\langle z\rangle$ ranging from 4.15 to 4.25 as $\tilde p$ increases).  In  their original work \cite{Tighe1}
Tighe et al. focused on  the distribution of the pressure on an individual single particle.  Expecting such a single particle property to obey a maximum entropy distribution is in effect making an ideal-gas-like assumption, where correlations between neighboring particles are ignored \cite{Tighe4}.  While they argue that this is reasonable for their triangular FNE, it is likely to be too simplistic for our disordered jammed packings, where the length scales measured in Fig.~\ref{quadFits-1oR}d suggest that correlations may extend over at least a few particle diameters for the range of stress considered here.

In Ref.~\cite{Tighe3}, however, Tighe and Vlugt consider the distribution of total stress within a {\em canonical }ensemble on finite triangular clusters of $N$ particles with non-periodic boundaries, computing the stress parameter $\alpha$ and the force-tile parameter $\lambda$  (this is denoted as ``$\gamma$" in their work) as a function of cluster size $N$.  They use a similar range of $N$ as the $\langle N_R\rangle$ we consider here.  Several clear differences exist between their results on finite clusters for the triangular FNE and our results for spatially disordered packings.   They find that  $\alpha$ and $\lambda$ are both positive.  In our work, where we can only compute the discrete derivatives with respect to global pressure, we find (see Fig.~\ref{2Dplanar-params}) $\Delta\alpha/\Delta p\sim p^{-1.8}$ and $\Delta\lambda/\Delta p \sim -p^{-3}$.  Integrating, and assuming that $\alpha,\lambda\to 0$ as $p\to\infty$, we conclude that $\lambda(p)>0$, but $\alpha(p)<0$. Furthermore, they found numerically that both $\alpha$ and $\lambda$ vary significantly with the cluster size, and that $\lambda$ vanishes as the cluster size increases.  We, however, find that both $\alpha$ and $\lambda$ approach non-zero constants as the cluster size $R$ increases. 

\begin{figure}[h]
\includegraphics[width=3.5in]{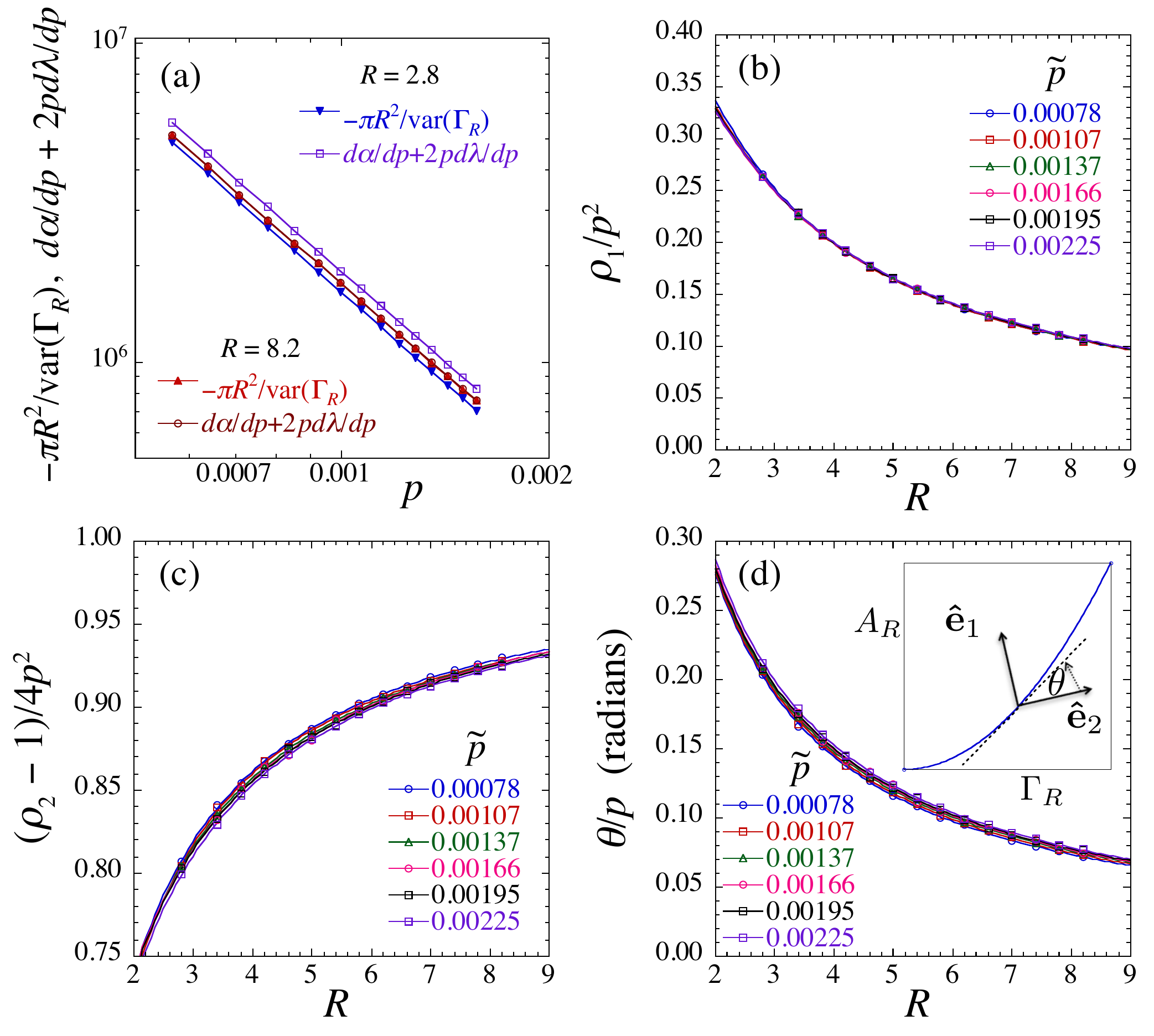}
\caption{(color online) (a) Comparison of $d\alpha/dp+2pd\lambda/dp$ with $-\pi R^2/\mathrm{var}(\Gamma_R)$ vs pressure $p$, so as to check the constraint of Eq.~(\ref{ealconstraint}).  Results are shown for our smallest cluster size, $R=2.8$ and our largest cluster, $R=8.2$. (b) and (c) Eigenvalues $\rho_1$ and $\rho_2$ of the scaled covariance matrix $\mathbb{\tilde C}$ of Eq.~(\ref{etildeC}).  Results for different $p$ collapse to a common curve when plotted as $\rho_1/p^2$ and $(\rho_2-1)/4p^2$ vs cluster radius $R$.
(d) Angle $\theta$ between the eigenvector $\mathbf{\hat e}_2$ corresponding to the non-vanishing eigenvalue $\rho_2$, and the tangent to the curve $\langle A_R\rangle = \langle \Gamma_R\rangle^2/(\pi R^2)$, as illustrated in the inset.  Data for different $p$ collapse to a common curve when plotted as $\theta/p$ vs $R$.  Solid lines in (b) and (d) are fits to the form $(c_1/R)(1-c_2/R)$, while those in (c) are fits to the form $c_0(1+c1/R+c_2/R^2)$. 
}
\label{eigen}
\end{figure}

Tighe and Vlugt \cite{Tighe3} argue that $\lambda\to 0$ as the cluster grows large because then fluctuations  in the cluster force-tile area $A_{\cal C}$ decay to zero, and hence $A_{\cal C}$ and $\Gamma_{\cal C}$ should no longer be regarded as independent observables that need to be independently constrained with separate Lagrange multipliers.  However, as we explain below, this argument does not appear to hold for our disordered soft disk packings.  Consider first the extreme limit where the cluster force-tile area is completely slaved to the cluster stress, i.e. $A_R=\Gamma_R^2/(\pi R^2)$ holds for each cluster configuration.  To lowest order in the fluctuations we then have $\delta A_R\equiv A_R-\langle A_R\rangle = 2\langle\Gamma_R\rangle \delta\Gamma_R/(\pi R^2)=2p\delta\Gamma_R$.  The covariance matrix of Eq.~(\ref{eCovar}) then becomes,
\begin{equation}
\mathbb{C}=\mathrm{var}(\Gamma_R)\mathbb{\tilde C},\quad\mathbb{\tilde C}\equiv\left[
\begin{array}{ll}
1&2p\\ 2p& 4p^2
\end{array}
\right].
\label{etildeC}
\end{equation}
$\mathbb{\tilde C}$ has eigenvalues $\rho_1=0$ and $\rho_2=1+4p^2$.  The eigenvector for $\rho_2$ in the two dimensional space of $(\Gamma_R,A_R)$ lies tangential to the curve $A_R=\Gamma_R^2/(\pi R^2)$, while the eigenvector for $\rho_1$ lies orthogonal to the curve.  Eq.~(\ref{eC1}) then yields the constraint,
\begin{equation}
\dfrac{d\alpha}{dp}+2p\dfrac{d\lambda}{dp}=-\dfrac{\pi R^2}{\mathrm{var}(\Gamma_R)}.
\label{ealconstraint}
\end{equation}
This result may also be obtained by taking the derivative with respect to pressure of Eq.~(5) in Tighe and Vlugt \cite{Tighe3}.
This constraint is well satisfied for our clusters, as we show in  Fig.~\ref{eigen}a by plotting both the left hand and right hand sides of Eq.~(\ref{ealconstraint}) vs pressure $p$.  For our smallest cluster size with $R=2.8$, the two quantities are fairly close, but for our biggest cluster size $R=8.2$ they are essentially equal.  

However the constraint of Eq.~(\ref{ealconstraint}) is not sufficient to uniquely determine $\alpha$ and $\lambda$.  Because $\rho_1=0$, $\alpha$ and $\lambda$ possess a degree of freedom such that we are free to shift to a new $\alpha^\prime=\alpha+g(p)$ and $\lambda^\prime=\lambda+h(p)$ for any functions $g$ and $h$ that satisfy $dg/dp=-2pdh/dp$.  One may use this freedom to choose $d\alpha/dp=-\pi R^2/\mathrm{var}(\Gamma_R)$ and $\lambda=0$, which is just the stress ensemble result of Eq.~(\ref{edalphadp}), or one can choose $d\lambda/dp=-\pi R^2/[2p\,\mathrm{var}(\Gamma_R)]$ and $\alpha=0$.  Indeed, Tighe and Vlugt \cite{Tighe3} explicitly show that, for a {\em periodic} FNE (where $A_N$ is slaved to $\Gamma_N$ as in Eq.~(\ref{ANGN})) in the {\em canonical} ensemble, either of these choices gives the same single particle pressure distribution if the system size $N$ is sufficiently large.  

We may note that the constraint of Eq.~(\ref{ealconstraint}) is the same as was found  in Sec.~\ref{s1DRatio}, if we take $\alpha$ and $\lambda$ as the parameters describing the distribution ${\cal P}(\Gamma_R|\tilde p)$ via Eq.~(\ref{RspR}).  In that case, we defined  $\bar\alpha$ in Eq.~(\ref{ebarDalpha}) such that in effect, $d\bar\alpha/dp\equiv d\alpha/dp+2pd\lambda/dp$, and we found in Fig.~\ref{barDalpha-Dp-1Dhist} excellent agreement between this and $-\pi R^2/\mathrm{var}(\Gamma_R)$, just as found now in Fig.~\ref{eigen}a from the distribution ${\cal P}(\Gamma_R,A_R|\tilde p)$.  One can show that the constraint of Eq.~(\ref{ealconstraint}) just ensures that the location and width of the peak in the stress distribution ${\cal P}(\Gamma_R|\tilde p)$ behaves correctly in a Gaussian approximation (which becomes more exact as $R$ increases), when the Boltzmann factor is a quadratic form as in Eq.~(\ref{eBoltz}).

For a finite cluster with non-periodic boundaries, however, fluctuations in $A_{\cal C}$ away from the average value at fixed $\Gamma_R$ may be small, but they are finite. Consequently $\rho_1>0$ is small but finite,  the covariance matrix $\mathbb{C}$ is invertible, the above freedom to vary $\alpha$ and $\lambda$ is broken, and a unique $\alpha(p)$ and $\lambda(p)$ result.  Where these unique $\alpha(p)$ and $\lambda(p)$ lie in the space of possibilities allowed by Eq.~(\ref{ealconstraint}) is determined in detail by such finite size effects.

To investigate this for the case of our soft disk packings, we explicitly compute the two eigenvalues $\rho_1$ and $\rho_2$ of the scaled covariance matrix $\mathbb{\tilde C}\equiv \mathbb{C}/\mathrm{var}(\Gamma_R)$ as a function of cluster radius $R$ and total system pressure $p$.  In Fig.~\ref{eigen}b we plot $\rho_1/p^2$ vs $R$.  The data for different $p$ collapse to a common curve that is very well fit by the form $(c_1/R)(1-c_2/R)$, and thus $\rho_1$ appears to vanish as $R$ gets large. We find $c_1=0.93\pm 0.01$ and $c_2=0.58\pm0.04$, where the errors here and in the following paragraph represent the variation in fit parameters found as $p$ varies.

Next we consider $\rho_2$.  Anticipating that $\rho_2$ should approach $1+4p^2$ at large $R$, we plot in Fig.~\ref{eigen}c
$(\rho_2-1)/4p^2$ vs $R$.  Again we find a fairly good collapse of the data for different $p$ to a common curve that is well fit by $c_0(1-c_1/R+c_2/R^2)$, with $c_0=0.999\pm 0.001$, $c_1=0.60\pm0.05$, and $c_2=0.20\pm 0.07$.  Thus $\rho_2$ indeed approaches $1+4p^2$ as $R$ increases.  Finally, we consider the orientation of the eigenvector $\mathbf{\hat e}_2$ for $\rho_2$ (the eigenvector $\mathbf{\hat e}_1$ for $\rho_1$ is necessarily orthogonal to this).  Defining $\theta$ as the angle between $\mathbf{\hat e}_2$ and the tangent to the curve $\langle A_R\rangle=\langle\Gamma_R\rangle^2/(\pi R^2)$, in Fig.~\ref{eigen}d we plot $\theta/p$ vs $R$.  Again we find a good collapse of the data for different $p$ to a common curve that is well fit by the form $(c_1/R)(1-c_2/R)$, showing that $\mathbf{\hat e}_2$ aligns parallel to the tangent to the curve as $R$ gets large; we find $c_1=0.62\pm 0.02$ and $c_2=0.20\pm0.02$.  Thus fluctuations in the direction orthogonal to the curve $\langle A_R\rangle=\langle\Gamma_R\rangle^2/(\pi R^2)$ vanish a factor of $1/R$ faster with increasing $R$ than do the fluctuations in the tangential direction.

%relative to the fluctuations in the tangential direction, and a scatter plot of $A_R$ vs $\Gamma_R$ such as in Fig.~\ref{2D-hist} will sweep out a line rather than a region of finite width, just as in the extreme case where $A_R$ is slaved to $\Gamma_R$.  

We can now use the results of Fig.~\ref{eigen} to write $d\alpha/dp$ and $d\lambda/dp$ in terms of the eigenvalues and eigenvectors of the scaled covariance matrix $\mathbb{\tilde C}$ of Eq.~(\ref{etildeC}). Projecting the vector $(1,2p)$ onto the eigenvectors $\mathbf{\hat e}_1$ and $\mathbf{\hat e}_2$ of $\mathbb{\tilde C}$ and applying Eq.~(\ref{eCinv}) we have,
\begin{align}
\dfrac{d\alpha}{dp}=-\dfrac{\pi R^2}{\mathrm{var}(\Gamma_R)}&\left[
\rho_1^{-1}\left(\sin^2\theta-2p\cos\theta\sin\theta\right)\right.\nonumber\\
&\left.+\rho_2^{-1}\left(\cos^2\theta+2p\cos\theta\sin\theta\right)\right]
\end{align}
\begin{align}
\dfrac{d\lambda}{dp}=-\dfrac{\pi R^2}{\mathrm{var}(\Gamma_R)}&\left[
\rho_1^{-1}\left(2p\sin^2\theta+\cos\theta\sin\theta\right)\right.\nonumber\\
&\left.+\rho_2^{-1}\left(2p\cos^2\theta-\cos\theta\sin\theta\right)\right].
\end{align}
To leading order, our results in Fig.~\ref{eigen} give $\rho_1\sim p^2/R$, $\rho_2\sim 1$, and $\theta\sim p/R$.  Inserting these into the above, we find that as $R$ increases, both $d\alpha/dp$ and $d\lambda/dp$ approach non-zero constants, with $1/R$ corrections that vanish as $R$ gets large.  For $d\alpha/dp$, the contribution from the projection onto $\mathbf{\hat e}_1$ is negative while the contribution from the projection onto $\mathbf{\hat e}_2$ is positive, and both are $\sim O(1)$ in magnitude.  Although the projection onto $\mathbf{\hat e}_1$ becomes vanishingly small as $R$ gets large (i.e. $\theta\to 0$), the prefactor $\rho_1^{-1}$ is diverging  so that the contribution from this term remains finite.
Thus $d\alpha/dp$ is determined by a balance between the two terms.
For $d\lambda/dp$ the contribution from the projection onto $\mathbf{\hat e}_1$ becomes $O(1/p)$ as $R$ gets large, while the contribution from the projection onto $\mathbf{\hat e}_2$ becomes $O(p)$.  Hence it is the projection onto $\mathbf{\hat e}_1$ that dominates $d\lambda/dp$ for small $p$ approaching the jamming transition.  Thus, although the fluctuations in direction $\mathbf{\hat e}_1$ are decaying more rapidly as a function of cluster size $R$ than are the fluctuations in the direction $\mathbf{\hat e}_2$, nevertheless the fluctuations along $\mathbf{\hat e}_1$ continue to give significant, non-vanishing, contributions to both $d\alpha/dp$ and $d\lambda/dp$ even as the cluster size gets large.  This conclusion is contrary to the qualitative argument of Tighe and Vlugt.

%****************
The analysis of Tighe and Vlugt for the triangular FNE proceeds differently from our own approach here.  Rather than analyze the stress distribution on a finite cluster embedded within a larger microcanonical (i.e. fixed $\Gamma_N$) system as we do, they consider a finite cluster on its own within a canonical ensemble, and determine $\alpha$ and $\lambda$ so as to get the desired $\langle\Gamma\rangle$ and $\langle A\rangle$ for the cluster.  It is possible that the differences they observe, as compared to our own work, might be a consequence of the differing ensembles used; equivalence of ensembles is only expected in the thermodynamic limit.
Or it may be that fluctuations in the FNE are sufficiently different from soft disk packings so as to yield a different balance between the contributions from $\mathbf{\hat e}_1$ vs $\mathbf{\hat e}_2$, and so select qualitatively different values for $\alpha$ and $\lambda$ from among the family of choices allowed by Eq.~(\ref{ealconstraint}).  

We note, in this regard, that the behavior of $\mathrm{var}(\Gamma)$ appears to be different in the two models.  In Ref.~\cite{Tighe4}, Tighe and Vlugt show that, for a cluster of $N$ particles in the canonical FNE, $\mathrm{var}(\Gamma)=2\langle\Gamma\rangle^2/(\Delta z N)$.  Here $\Delta z=\langle z\rangle - z_\mathrm{iso}$, which for the harmonic soft-core interaction used here is believed to scale with system pressure as $\Delta z\sim p^{1/2}$ \cite{WuTeitel-hyper,Wyart}.  Taking $\langle\Gamma\rangle =pV$, we get for the FNE, $\mathrm{var}(\Gamma)/V\sim p^{3/2}$.  In contrast, for clusters of radius $R$ embedded in our soft disk packings, we have previously found  from numerical simulations \cite{WuTeitel} that $\mathrm{var}(\Gamma_R)/(\pi R^2)\sim p^{1.9}$, for the range of pressure and cluster sizes considered here; it is of course possible that the power-law 1.9 is only an effective value that could change if we probed closer to the jamming transition.
To clarify the difference between the FNE and soft disk packings, it would be interesting to compute the covariance matrix of stress and force-tile area for the FNE and do a similar analysis as in Fig.~\ref{eigen}, however such a computation lies outside the scope of the present work.
%****************

\begin{figure}[h]
\includegraphics[width=3.5in]{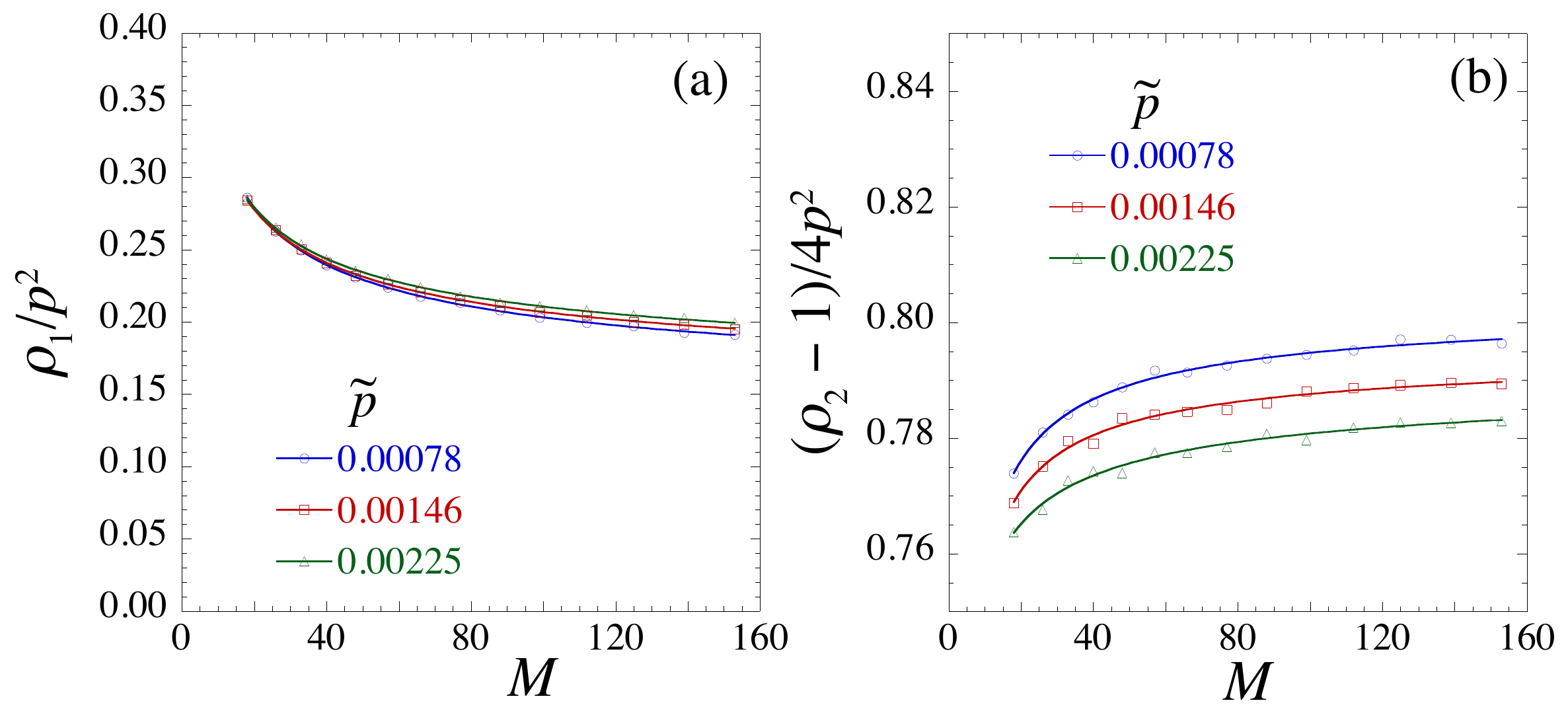}
\caption{(color online) (a) and (b) Eigenvalues $\rho_1$ and $\rho_2$ of the scaled covariance matrix $\mathbb{\tilde C}$ of Eq.~(\ref{etildeC}) for clusters defined by a fixed number of particles $M$ plotted as $\rho_1/p^2$ and $(\rho_2-1)/4p^2$ vs pressure $p$.  Unlike the case for clusters defined by a fixed radius $R$, and shown in Fig.~\ref{eigen}, here we find that both eigenvalues approach finite, non-zero, constants as $M$ increases.
}
\label{eigen-fixedM}
\end{figure}

Finally, it is interesting to note that if we define our clusters by a fixed number of particles $M$, rather than a fixed radius $R$ \cite{WuTeitel}, then we find that both eigenvalues $\rho_1$ and $\rho_2$  go to  finite non-zero constants as $M$ increases, hence fluctuations remain comparable in all directions in the $(\Gamma_M,A_M)$ plane. Our results are shown in Fig.~\ref{eigen-fixedM}.   However, we find that $d\alpha/dp$ and $d\lambda/dp$, as computed from the covariance matrix for such fixed $M$ clusters, behave qualitatively the same as found for the fixed $R$ clusters; although we find a somewhat stronger dependence on the cluster size $M$ than we do for clusters of fixed radius $R$, both $d\alpha/dp$ and $d\lambda/dp$ approach limiting non-zero values as $M$ increases and display similar power-law behaviors with pressure $p$ as found in Fig.~\ref{Da-Dl-covars} for the clusters of fixed $R$.

%For the FNE, where the contact network structure is by definition held fixed and the contact number $z$ remains constant, $\alpha(p)$ and $\lambda(p)$ have a trivial dependence on the global system pressure, $\alpha(p)=\alpha^\prime/p$ and $\lambda(0)=\lambda^\prime/p^2$.

\subsection{Relation to the stress ensemble}
\label{sRelation}

Our analysis of the histogram ratio ${\cal R}$ of the joint distribution ${\cal P}(\Gamma_R,A_R|\tilde p)$ thus clearly suggests that ${\cal P}(\Gamma_R,A_R|\tilde p)$ has the form of Eq.~(\ref{PGA}), with a Boltzmann factor $\mathrm{exp}[-\alpha\Gamma_R -\lambda A_R]$.  In this section we explore how this form may give rise to the marginalized distribution ${\cal P}(\Gamma_R|\tilde p)=\int dA_R{\cal P}(\Gamma_R,A_R|\tilde p)$, which was found in Sec.~\ref{sStress} to have the quadratic Boltzmann factor of Eq.~(\ref{eBoltz}), $\mathrm{exp}[-\alpha\Gamma_R-\lambda\Gamma_R^2/(\pi R^2)]$.  We consider how the parameters $\alpha$ and $\lambda$ of Eq.~(\ref{eBoltz}) for ${\cal P}(\Gamma_R|\tilde p)$ may be related to the parameters $\alpha$ and $\lambda$ of Eq.~(\ref{PGA}) for ${\cal P}(\Gamma_R, A_R|\tilde p)$.  For clarity, in this section we will denote the former parameters as bold-faced $\boldsymbol{\alpha}$ and $\boldsymbol{\lambda}$.

%This form then gives a natural explanation for the quadratic term in the Boltzmann factor of Eq.~(\ref{eBoltz}), found in Sec.~\ref{sStress} for the stress distribution ${\cal P}(\Gamma_R|\tilde p)$.  To see this, we follow the approach of Tighe and Vlugt \cite{Tighe3}, but point out several corrections to their analysis that apply to our soft disk system.

We follow the approach of Tighe and Vlugt \cite{Tighe3,Tighe4}.  We can write for the number of states,
\begin{equation}
\Omega_R(\Gamma_R,A_R)=\Omega_R(\Gamma_R)\Psi_R(A_R|\Gamma_R)
\end{equation}
where $\Psi_R(A_R|\Gamma_R)$ is the fraction of possible states with force-tile area $A_R$, when the cluster stress is constrained to the value $\Gamma_R$.  Note, $\int dA_R\Psi_R(A_R|\Gamma_R)=1$ and so $\Omega_R(\Gamma_R)=\int dA_R\Omega_R(\Gamma_R,A_R)$.  By assumption, $\Omega_R(\Gamma_R, A_R)$, and hence
$\Omega_R(\Gamma_R)$ and $\Psi_R(A_R|\Gamma_R)$ are independent of the total system stress per particle $\tilde p=\Gamma_N/N$.
Note, the conditional density of states $\Psi_R(A_R|\Gamma_R)$ is {\em not} in general the same as the conditional probability for the cluster to have $A_R$ given the cluster stress is $\Gamma_R$; the conditional probability ${\cal P}(A_R|\Gamma_R;\tilde p)\equiv{\cal P}(\Gamma_R,A_R|\tilde p)/{\cal P}(\Gamma_R|\tilde p)$ is given by, 
\begin{equation}
{\cal P}(A_R|\Gamma_R;\tilde p)=\dfrac{\Psi_R(A_R|\Gamma_R)\mathrm{e}^{-\lambda(\tilde p)A_R}}{\int dA_R\Psi_R(A_R|\Gamma_R)\mathrm{e}^{-\lambda(\tilde p)A_R}},
\label{ePAgivenG}
\end{equation}
and does depend on the total system stress $\tilde p$.  Only when $\lambda\to 0$, i.e. $\tilde p\to \infty$, do we have ${\cal P}(A_R|\Gamma_R;\tilde p) = \Psi_R(A_R|\Gamma_R)$.

We can now express the marginal distribution ${\cal P}(\Gamma_R|\tilde p)$ by integrating the joint distribution over the force-tile area $A_R$,
\begin{align}
{\cal P}(\Gamma_R|\tilde p)&\equiv\int dA_R{\cal P}(\Gamma_R,A_R|\tilde p)\nonumber\\
&=\dfrac{\Omega_R(\Gamma_R)\mathrm{e}^{-\alpha(\tilde p)\Gamma_R}}{Z_R(\tilde p)} 
\int dA_R\Psi_R(A_R|\Gamma_R)\mathrm{e}^{-\lambda(\tilde p)A_R}.
\label{ePGmarg}
\end{align}
Tighe and Vlugt then argue \cite{Tighe3,Tighe4} that $\Psi_R(A_R|\Gamma_R)$ should be sharply peaked about its average.  Defining the average,
\begin{equation}
\langle A_R(\Gamma_R)\rangle\equiv \int dA_R\Psi_R(A_R|\Gamma_R)A_R, 
\end{equation} 
we would then expect,
\begin{equation}
{\cal P}(\Gamma_R|\tilde p)\approx\dfrac{\Omega_R(\Gamma_R)}{Z_R}\mathrm{e}^{-\alpha(\tilde p)\Gamma_R-\lambda(\tilde p)\langle A_R(\Gamma_R)\rangle}.
\label{ePGapprox}
\end{equation}

To proceed, we now need an expression for $\langle A_R(\Gamma_R)\rangle$.  We do not have direct access to $\Psi_R(A_R|\Gamma_R)$, but we can numerically measure the conditional probability ${\cal P}(A_R|\Gamma_R;\tilde p)$ and hence compute the conditional average,
\begin{equation}
\langle A_R|\Gamma_R;\tilde p\rangle \equiv \int dA_R{\cal P}(A_R|\Gamma_R;\tilde p)A_R.
\end{equation}
By Eq.~(\ref{ePAgivenG}), the desired $\langle  A_R(p_R)\rangle$ is just the large $\tilde p$ (i.e. $\lambda\to 0$) limit of $\langle A_R|\Gamma_R;\tilde p\rangle$.

\begin{figure}[h]
\includegraphics[width=3.5in]{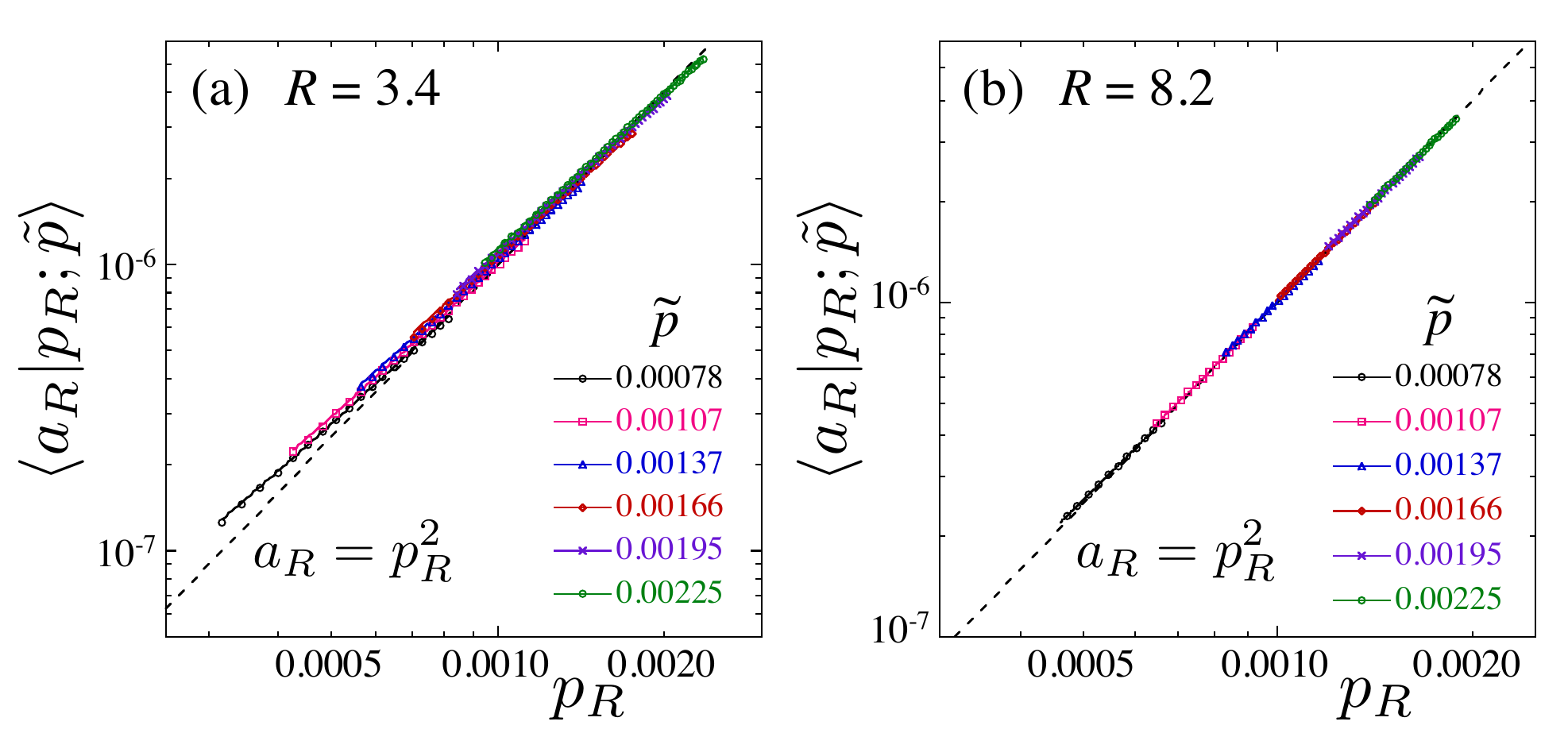}
\caption{(color online)  Intensive conditional average force-tile area $\langle a_R|p_R;\tilde p\rangle$ for clusters of radius $R$, vs the cluster pressure $p_R=\Gamma_R/(\pi R^2)$, for different values of the total system stress per particle $\tilde p=\Gamma_N/N$; $a_R=A_R/(\pi R^2)$.  Results are shown for clusters of size (a) $R=3.4$ and (b) $R=8.2$. Solid lines are fits to $c_1 p_R+c_2p_R^2$.  Dashed lines are $a_R=p_R^2$.
}
\label{AofG}
\end{figure}

In Fig.~\ref{AofG} we plot the intensive version of this conditional average, $\langle a_R|p_R;\tilde p\rangle = \langle A_R|\Gamma_R;\tilde p\rangle/(\pi R^2)$ vs $p_R=\Gamma_R/(\pi R^2)$, for several different values of the global stress per particle $\tilde p=\Gamma_N/N$.  In panel a we show results for clusters of radius $R=3.4$, while in panel b we show results for $R=8.2$.  In both panels the dashed line is the curve $a_R=p_R^2$, as would be expected if fluctuations away from the average in both $a_R$ and $p_R$ were negligible (see Eq.~(\ref{eap2})).  We see  that the  data is approaching this dashed line as either $\tilde p$ or $R$ increases. 

To determine the limiting behavior, we fit our data to a quadratic form,
\begin{equation}
\langle a_R|p_R;\tilde p\rangle = c_1p_R+c_2p_R^2,
\label{eaofp}
\end{equation}
which gives the solid lines in Fig.~\ref{AofG}.  If we denote the large $\tilde p$ (i.e. $\lambda\to 0$) limits of $c_1$ and $c_2$ by $c_{1\infty}$ and $c_{2\infty}$, we then have,
\begin{equation}
\langle A_R(\Gamma_R)\rangle=c_{1\infty}\Gamma_R+c_{2\infty}\Gamma_R^2/(\pi R^2).
\end{equation}
Substituting into Eq.~(\ref{ePGapprox}) then yields the quadratic Boltzmann factor for the distribution ${\cal P}(\Gamma_R|\tilde p)\propto \mathrm{exp}[-\boldsymbol{\alpha}\Gamma_R-\boldsymbol{\lambda}\Gamma_R^2/(\pi R^2)]$, with,
\begin{equation}
\boldsymbol{\alpha}(\tilde p)=\alpha(\tilde p)+c_{1\infty}\lambda(\tilde p),\quad
\boldsymbol{\lambda}(\tilde p)=c_{2\infty}\lambda(\tilde p),
\end{equation}
or equivalently, comparing parameter differences at $\tilde p_1$ and $\tilde p_2=\tilde p_1+\Delta \tilde p$,
\begin{equation}
\boldsymbol{\Delta\alpha}=\Delta\alpha+c_{1\infty}\Delta\lambda,\quad
\boldsymbol{\Delta\lambda}=c_{2\infty}\Delta\lambda.
\label{ealcomparedif}
\end{equation}

\begin{figure}[h]
\includegraphics[width=3.5in]{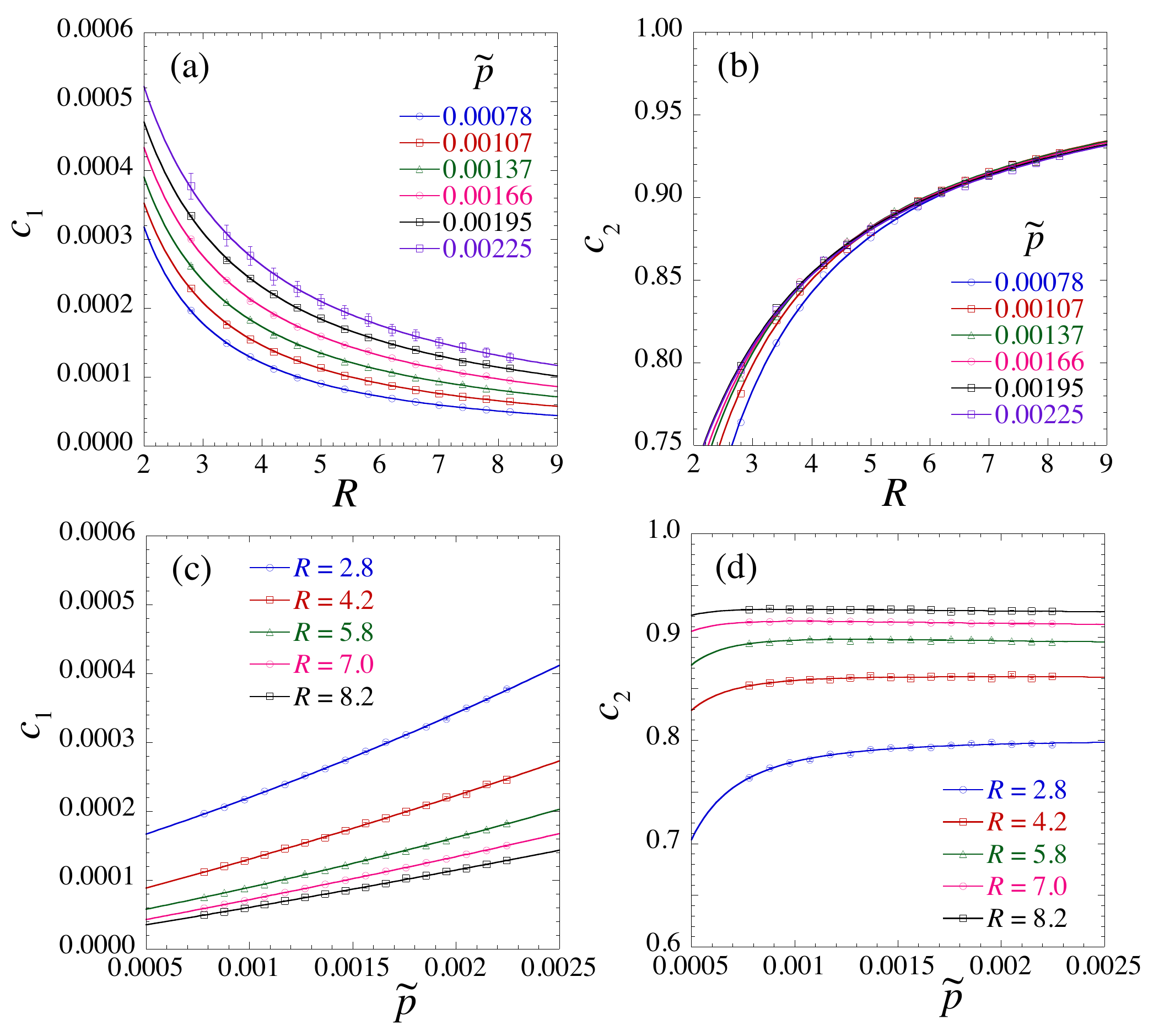}
\caption{(color online) (a) and (b): Parameters $c_1$ and $c_2$ of Eq.~(\ref{eaofp}), giving the dependence of the conditional force-tile area $a_R$ on cluster pressure $p_R$, vs cluster radius $R$, for different values of the total system stress per particle $\tilde p$.  Solid lines in (a) are fits to $c_1=(u_1/R)(1+u_2/R)$; in (b) solid lines are fits to $c_2=1 + v_1 /R+v_2/R^2$.  (c) and (d): Parameters $c_1$ and $c_2$ vs $\tilde p$, for different values of $R$.  Solid lines in $(c)$ are fits to a quadratic form.
}
\label{c1-c2}
\end{figure}

In Fig.~\ref{c1-c2}a,b we plot the resulting values of $c_1$ and $c_2$ as a function of cluster radius $R$, for several different values of the total stress per particle $\tilde p$.  The solid lines in panel a are fits to the form $c_1=(u_1/R)(1+u_2/R)$, while the solid lines in panel b are fits to the form $c_2=1 + v_1 /R+v_2/R^2$; these fits are excellent.  We thus conclude that, as the cluster size $R\to\infty$, then $c_1\to 0$ and $c_2\to 1$ for any $\tilde p$, and so $c_{1\infty}=0$ and $c_{2\infty}=1$.  In this limit, we have $\boldsymbol{\Delta\alpha}=\Delta\alpha$ and $\boldsymbol{\Delta\lambda}=\Delta\lambda$.  

For a cluster of finite radius $R$, however, the situation is less clear.  In Fig.~\ref{c1-c2}c,d we plot $c_1$ and $c_2$ vs $\tilde p$ for several different cluster sizes $R$.  We see from panel d that $c_2$ approaches a constant value as $\tilde p$ increases, as can also be seen in panel b.  However panel c shows that, for all $R$,  $c_1$ continues to increase as $\tilde p$ increases, for the entire range of $\tilde p$ we consider; solid lines in panel c are fits to a quadratic form.  Thus, at finite $R$, the limiting $\tilde p\to\infty$ value of $c_1$, i.e. $c_{1\infty}$, is unclear.  However, from Fig.~\ref{2Dplanar-params} we have $\Delta\alpha>0$ while $\Delta\lambda<0$.  Since $c_{1\infty}>0$, hence from Eq.~(\ref{ealcomparedif}) we expect $\boldsymbol{\Delta\alpha}<\Delta\alpha$.

\begin{figure}[h]
\includegraphics[width=3.5in]{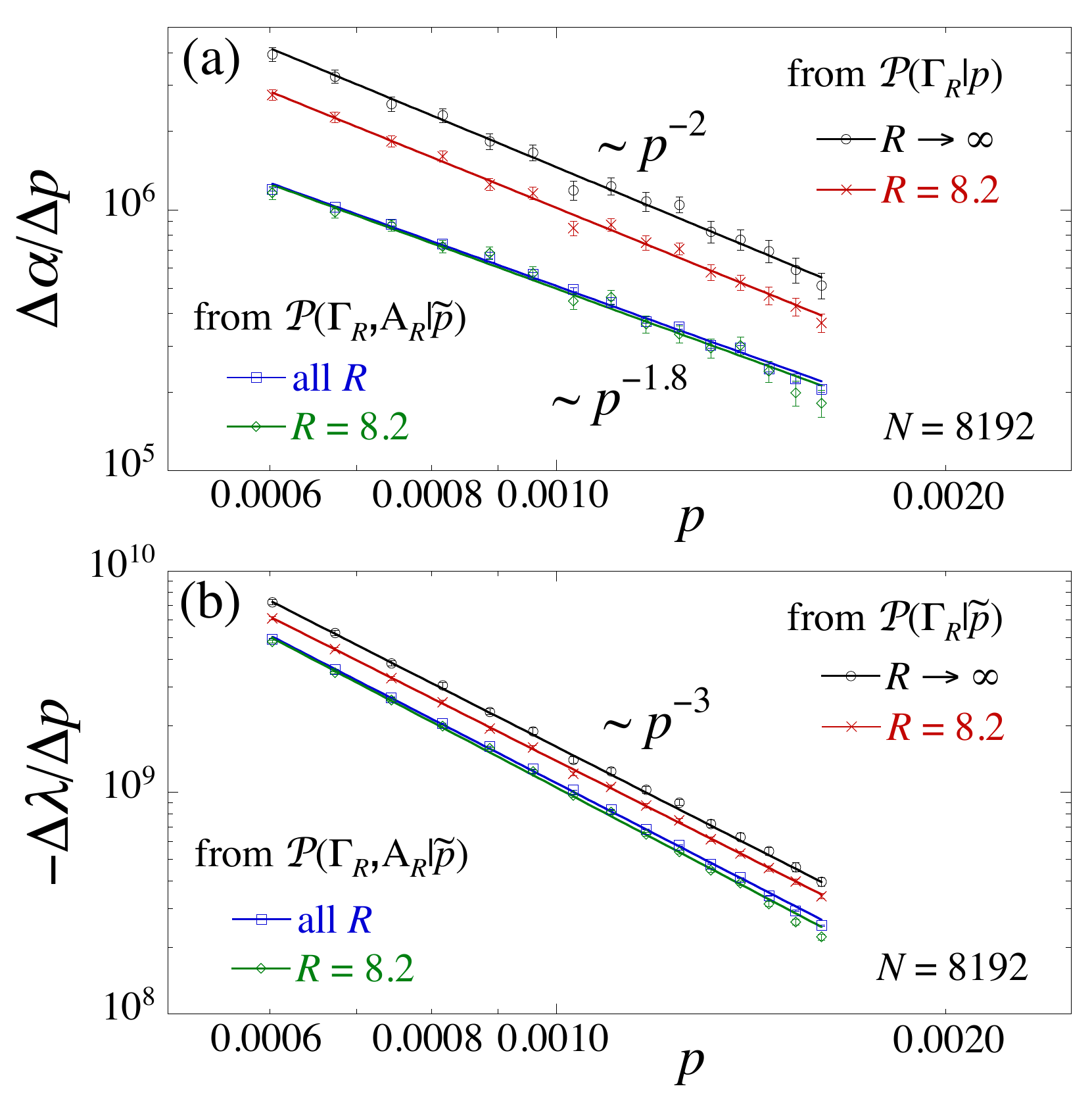}
\caption{(color online) Comparison of results for (i) $\Delta\alpha/\Delta p$ and $\Delta\lambda/\Delta p$ obtained from the log histogram ratio of ${\cal P}(\Gamma_R|\tilde p)$, with (ii) $\Delta\alpha/\Delta p$ and $\Delta\lambda/\Delta p$ obtained from the log histogram ratio of ${\cal P}(\Gamma_R,A_R|\tilde p)$.  For (i), we show results for cluster sizes $R\to\infty$ and $R=8.2$, as previously shown in Figs.~\ref{quadFits-1oR}a,b. For (ii), we show results for the case where we fit to all sizes $R$ simultaneously, and for the specific case $R=8.2$, as previously shown in Figs.~\ref{2Dplanar-params}a,b.
}
\label{1D-2D-compare} 
\end{figure}

In Fig.~\ref{1D-2D-compare} we explicitly compare our results for (i) $\boldsymbol{\Delta\alpha}$ and $\boldsymbol{\Delta\lambda}$ obtained from the log histogram ratio of ${\cal P}(\Gamma_R|\tilde p)$, with (ii) $\Delta\alpha$ and $\Delta\lambda$ obtained from the log histogram ratio of ${\cal P}(\Gamma_R,A_R|\tilde p)$.  For (i), we replot our results for $\boldsymbol{\Delta\alpha}/\Delta p$ and $-\boldsymbol{\Delta\lambda}/\Delta p$ vs pressure $p$, for the cases $R\to\infty$ and $R=8.2$ as previously shown in Figs.~\ref{quadFits-1oR}a,b. For (ii), we replot our results for $\Delta\alpha/\Delta p$ and $-\Delta\lambda/\Delta p$ vs $p$, for the case where we fit to all sizes $R$ simultaneously, and for the specific case $R=8.2$ as previously shown in Figs.~\ref{2Dplanar-params}a,b; as noted earlier, for (ii) there is essentially no dependence observed on the cluster size $R$.  

For $R\to \infty$, the arguments above give $c_{1\infty}=0$, $c_{2\infty}=1$, and so by Eq.~(\ref{ealcomparedif}) we expect (i) and (ii) to be equal.  For finite $R=8.2$, we have $c_{2\infty}\lesssim 1$, $c_{1\infty}>0$, and so we expect $\boldsymbol{\Delta\alpha}/\Delta p < \Delta\alpha/\Delta p$ as well as $-\boldsymbol{\Delta\lambda}/\Delta p < -\Delta\lambda/\Delta p$.  However in Fig.~\ref{1D-2D-compare} we see that the reverse is true.  In panel b we see that $-\Delta\lambda/\Delta p$ and $-\boldsymbol{\Delta\lambda}/\Delta p$ are close in value and both scale roughly as $p^{-3}$; but $-\Delta\lambda/\Delta p$ is {\em smaller} than $-\boldsymbol{\Delta\lambda}/\Delta p$, with the difference being about $20\%$ of $-\boldsymbol{\Delta\lambda}/\Delta p$ for the case $R=8.2$.  In panel a we see that the difference between $\Delta\alpha/\Delta p$ and $\boldsymbol{\Delta\alpha}/\Delta p$ is more substantial; the power-law dependence on $p$ is close, but slightly different, and $\Delta\alpha/\Delta p$ is about half the value of $\boldsymbol{\Delta\alpha}/\Delta p$ for the case $R=8.2$.

We are not certain of the reason for the lack of agreement between (i) and (ii) observed in Fig.~\ref{1D-2D-compare}.  One possible concern is the validity of the approximation going from Eq.~(\ref{ePGmarg}) to Eq.~(\ref{ePGapprox}).  To try to test this, we construct the conditional density of states $\Psi_R(A_R|\Gamma_R)$ from the conditional probability ${\cal P}(A_R|\Gamma_R;\tilde p)$ using,
\begin{equation}
\Psi_R(A_R|\Gamma_R)=C\mathrm{e}^{\lambda(\tilde p)A_R}{\cal P}(A_R|\Gamma_R;\tilde p),
\label{ePsifromP}
\end{equation}
where the constant $C$ is chosen to normalize $\int dA_R\Psi_R(A_R|\Gamma_R)=1$.

\begin{figure}[h]
\includegraphics[width=3.5in]{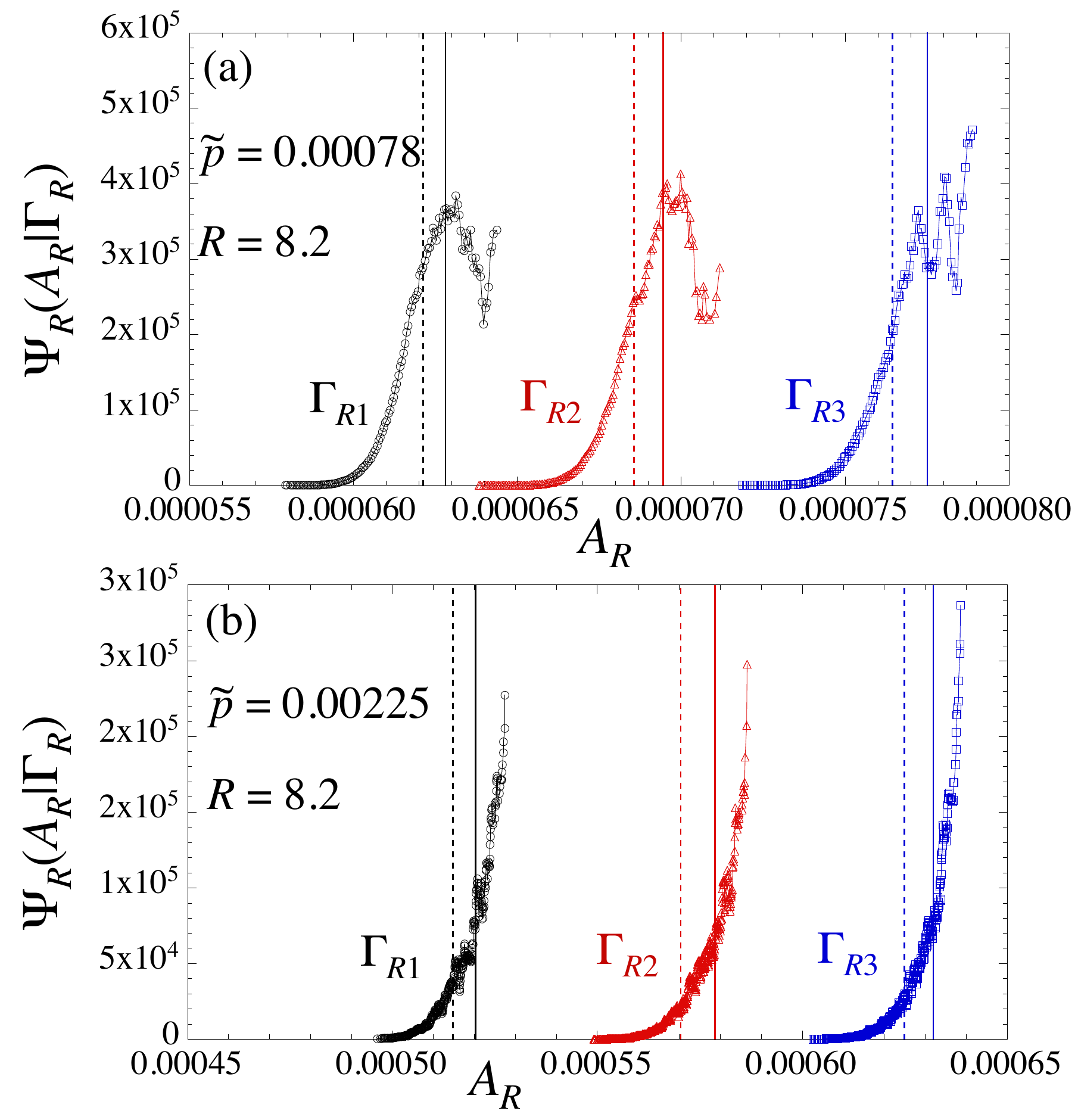}
\caption{(color online) Conditional density of states $\Psi_R(A_R|\Gamma_R)$ vs force-tile area $A_R$, for clusters of radius $R=8.2$.  Results are shown for total system stress per particle (a) $\tilde p=0.00078$ and (b) $\tilde p=0.00225$.  In each panel results are shown for three different values of the cluster stress, $\Gamma_{R1}=\langle \Gamma_R\rangle -[\mathrm{var}(\Gamma_R)]^{1/2}$, $\Gamma_{R2}=\langle \Gamma_R\rangle$, and $\Gamma_{R3}=\langle \Gamma_R\rangle +[\mathrm{var}(\Gamma_R)]^{1/2}$, where the average and variance of $\Gamma_R$ is computed at the corresponding value of $\tilde p$.  Solid vertical lines locate the average $\langle A_R(\Gamma_R)\rangle\equiv\int dA_R\Psi_R(A_R|\Gamma_R)A_R$, while dashed vertical lines locate the conditional average $\langle A_R|\Gamma_R;\tilde p\rangle = \int dA_R{\cal P}(A_R|\Gamma_R;\tilde p)A_R$.
}
\label{Psi} 
\end{figure}

To evaluate Eq.~(\ref{ePsifromP}) we need the value of $\lambda(\tilde p)$, whereas our histogram ratio method only directly gives $\Delta\lambda=\lambda(\tilde p_2)-\lambda(\tilde p_1)$.  To obtain $\lambda(\tilde p)$ we fit our results for $\Delta\lambda/\Delta \tilde p$ to a power-law, and then integrate the power-law assuming $\lambda\to 0$ as $\tilde p\to\infty$.  This approach has a measure of uncertainty since we cannot be sure our fitted power-law is a valid expression for all $\tilde p$ above the largest $\tilde p$ we have simulated.  In Fig.~\ref{Psi} we plot the resulting $\Psi_R(A_R|\Gamma_R)$ vs $A_R$ for the specific case of our largest cluster size $R=8.2$.  In panel a we show results for our smallest $\tilde p=0.00078$, and in panel b we show results for our largest $\tilde p=0.00225$.  For each $\tilde p$ we show results for three different values of $\Gamma_R$, roughly equal to $\langle\Gamma_R\rangle$, $\langle\Gamma_R\rangle\pm [\mathrm{var}(\Gamma_R)]^{1/2}$.  The solid vertical lines indicate the values of $\langle A_R(\Gamma_R)\rangle$, obtained by numerically integrating $\Psi_R(A_R|\Gamma_R)A_R$.   The dashed vertical lines indicate the values of $\langle A_R|\Gamma_R;\tilde p\rangle$, obtained by numerically integrating ${\cal P}(A_R|\Gamma_R;\tilde p)A_R$.  We see that these averages do not in general lie at a sharp peak of $\Psi_R(A_R|\Gamma_R)$.  It thus may be that the approximation above, going from Eq.~(\ref{ePGmarg}) to (\ref{ePGapprox}), gives the qualitative explanation for the quadratic $\Gamma_R^2$ term in the Boltzmann factor of Eq.~(\ref{eBoltz}) for ${\cal P}(\Gamma_R|\tilde p)$, but is not sufficiently accurate to allow a quantitative determination of $\boldsymbol{\Delta\alpha}$ and $\boldsymbol{\Delta\lambda}$ from $\Delta\alpha$ and $\Delta\lambda$.  
To our knowledge, a similar direct comparison of the joint distribution ${\cal P}(\Gamma_R,A_R|\tilde p)$ to the marginal distribution ${\cal P}(\Gamma_R|\tilde p)$ has not been made for the FNE.
%; Tighe et al.'s work \cite{Tighe1,Tighe3} involves only analysis of the distribution ${\cal P}(\Gamma|\tilde p)$.

\section{Summary}
\label{sSummary}

We have used numerical simulations to study  the distribution of stresses on compact finite sub-clusters of particles embedded within an athermal, two dimensional, mechanically stable packing of soft-core frictionless disks, at fixed isotropic total stress above the jamming transition.  Our clusters are defined as the set of particles whose centers lie within a randomly placed circle of radius $R$.
We have investigated whether this stress distribution is consistent with a maximum entropy hypothesis, such as commonly applies to thermodynamic systems in equilibrium.

We have tested in detail the {\em stress ensemble} formalism of Henkes et al. \cite{Henkes1,Henkes2} in which, for isotropic systems, the trace of the extensive stress tensor, $\Gamma_R$, is the key parameter.  Since $\Gamma_R$ is a conserved quantity, additive over disjoint subsystems, the stress ensemble predicts that the cluster stress distribution ${\cal P}(\Gamma_R)$ involves a Boltzmann factor, $\mathrm{exp}[-\alpha\Gamma_R]$, with $\alpha$ an inverse temperature-like quantity fixed by the parameters of the global system in which the cluster is embedded.  We have found that our measured stress distribution is not consistent with this prediction, but that rather ${\cal P}(\Gamma_R)$ involves a Boltzmann factor which includes a quadratic term in the stress, $\mathrm{exp}[-\alpha\Gamma_R-\lambda\Gamma_R^2/(\pi R^2)]$.  We have shown that this quadratic Boltzmann factor is a better explanation of our data than a simple Gaussian approximation, and we have presented arguments as to why previous work \cite{Henkes1, Henkes2} failed to detect this quadratic term.

We have then tested ideas due to Tighe and co-workers \cite{Tighe1,Tighe3,Tighe4} that a correct statistical description of the stress distribution must take into account a second extensive conserved quantity, the Maxwell-Cremona force-tile area $A_R$.  Measuring the joint distribution ${\cal P}(\Gamma_R,A_R)$, we find that it is indeed well described by a Boltzmann factor, $\mathrm{exp}[-\alpha\Gamma_R-\lambda A_R]$, as predicted by the maximum entropy hypothesis.  For our total system of $N$ particles with periodic boundary conditions, the average global force-tile area $A_N$ is a deterministic function of the global stress $\Gamma_N$. This implies that the parameters $\alpha$ and $\lambda$ cannot be chosen independently of each other, but are related to each other parametrically via the global pressure $p=\Gamma_N/V$, or equivalently the total stress per particle $\tilde p=\Gamma_N/N$. Using a histogram ratio method, we have determined the discrete derivatives $\Delta\alpha/\Delta p$ and $\Delta\lambda/\Delta p$ for clusters of different size $R$ at different values of the stress per particle $\tilde p$, and find these quantities have negligible dependence on $R$ for the cluster sizes we consider here.

Tighe and co-workers tested  \cite{Tighe1,Tighe3,Tighe4} their ideas on a highly idealized model of a jammed packing, the triangular force-network ensemble. 
In this model particles sit on the sites of a triangular lattice, and fluctuations in the contact forces are decoupled from fluctuations in the particle positions.  Originally, their work \cite{Tighe1} focused on the pressure distribution on individual single particles in large periodic systems.  Later, Tighe and Vlugt \cite{Tighe3} considered the distribution of stress in smaller non-periodic clusters of particles, using a canonical ensemble.  Several differences appear to exist between our results for soft disk packings and these earlier results for the FNE.  We find $\alpha <0$, whereas Tighe and co-workers find $\alpha >0$.  Further, Tighe and Vlugt \cite{Tighe3} find that the parameters $\alpha$ and $\lambda$ vary significantly with the number of particles $N$ in their cluster, and that $\lambda\to 0$ as $N$ grows large.  We find that $\alpha$ and $\lambda$ both approach finite values as the clusters grow large.

It is unclear if these differences have to do with the different ways in which the cluster ensemble is created,  or if the behavior of soft disk packings is just poorly approximated by the FNE, and the two have different structural properties.  We note that consistency tests we have carried out for the soft disk packings, such as (i) the comparison of parameters obtained by the ratio method vs obtained from fluctuations via the covariance matrix as discussed in Sec.~\ref{sfluc2D}, and (ii) the comparison of parameters obtained from the distribution ${\cal P}(\Gamma_R|\tilde p)$ vs those obtained from the joint distribution ${\cal P}(\Gamma_R,A_R|\tilde p)$ as discussed in Sec.~\ref{sRelation}, have yet to be performed for the FNE.  If such tests were carried out on the FNE, it might help to clarify the relation between the two models.

To conclude, we find that the distribution of stress in finite clusters of frictionless granular particles embedded in a two dimensional isotropic, mechanically stable, packing above jamming is well described by the maximum entropy hypothesis, provided one identifies all relevant conserved variables, in this case $\Gamma_R$ and $A_R$.  

\section*{Acknowledgements}

This work has been supported by NSF Grant No. DMR-1205800.  Computations were carried out at the Center for Integrated Research Computing at the University of Rochester.  We wish to thank B. Chakraborty, S. Henkes, B. Tighe and P. Olsson for helpful discussions.


\begin{thebibliography}{99}

\bibitem{Liu+Nagel}A.~J.~Liu and S.~R.~Nagel, Annu. Rev. Condens. Matter Phys. {\bf 1}, 347, (2010).

\bibitem{OHern}C.~S.~O'Hern, L.~E.~Silbert, A.~J.~Liu, and S.~R.~Nagel, Phys. Rev. E {\bf 68}, 011306 (2003).

\bibitem{Edwards}S.~F.~Edwards and R.~B.~S.~Oakeshott, Physica A 157, 1080 (1989); S.~F.~Edwards, D.~V.~Grinev,
Phys. Rev. E {\bf 58}, 4758 (1998); R.~Blumenfeld, S.~F.~Edwards, Phys. Rev. Lett. {\bf 90}, 114303.
(2003)

\bibitem{Henkes1}S.~Henkes, C.~S.~O'Hern, and B.~Chakraborty, Phys. Rev. Lett. {\bf 99}, 038002 (2007).

\bibitem{Henkes2}S.~Henkes and B.~Chakraborty, Phys. Rev. E {\bf 79}, 061301 (2009).

\bibitem{Blumenfeld}R.~Blumenfeld and S.~F.~Edwards, J. Phys. Chem. B {\bf 113}, 3981 (2009).

\bibitem{Tighe1}B.~P.~Tighe, A.~R.~T.~van~Eerd, and T.~J.~H.~Vlugt, Phys. Rev. Lett. {\bf 100}, 238001 (2008).

\bibitem{Tighe3}B.~P.~Tighe and T.~J.~H.~Vlugt, J. Stat. Mech.: Theory Exp. P01015 (2010).

\bibitem{Tighe4}B.~P.~Tighe and T.~J.~H.~Vlugt, J. Stat. Mech.: Theory Exp. P04002 (2011).

\bibitem{Lechenault}F.~Lechenault, F.~da~Cruz, O.~Dauchot, and E.~Bertin, J. Stat. Mech.: Theory Exp. P07009 (2006).

\bibitem{McNamara}S.~McNamara, P.~Richard, S.~K.~de Richter, G.~Le~Ca{\"e}r, and R.~Delannay, Phys. Rev. E {\bf 80}, 031301 (2009).

\bibitem{Puckett}J.~G.~Puckett and K.~E.~Daniels, Phys. Rev. Lett. {\bf 110}, 058001 (2013).

\bibitem{Schroter}S.-C.~Zhao and M.~Schr{\"o}ter, Soft Matter {\bf 10}, 4208 (2014).

\bibitem{LeesEdwards}D.~J.~Evans and G.~P.~Morriss, {\it Statistical Mechanics of Non-equilibrium Liquids} (Academic, London, 1990).

\bibitem{Dagois}S.~Dagois-Bohy, B.~P.~Tighe, J.~Simon, S.~Henkes, and M.~van~Hecke, Phys. Rev. Lett. {\bf 109}, 095703 (2012).

\bibitem{WuTeitel}Y.~Wu and S.~Teitel, Phys. Rev. E {\bf 91}, 022207 (2015).

\bibitem{VagbergFSS}D.~V{\aa}gberg, D.~Valdez-Balderas, M.~A.~Moore, P.~Olsson, and S.~Teitel, Phys. Rev. E {\bf 83}, 030303(R) (2011).

\bibitem{WuTeitel-hyper}Y.~Wu and S.~Teitel, preprint arXiv:1506.01948 (2015).

\bibitem{NR}W.~H.~Press, S.~A.~Teukolsky, W.~T.~Vetterling and B.~P.~Flannery, {\it Numerical Recipes}  3rd ed. (Cambridge University Press, New York, NY, 2007).

\bibitem{note0}Our reason for choosing clusters with a fixed radius $R$, rather than a fixed number of particles $M$, is detailed in Ref.~\cite{WuTeitel}.

\bibitem{Plischke}M.~Plischke and B.~Bergersen, {\em Equilibrium Statistical Physics} 2nd ed. (World Scientific, Singapore, 1994).

\bibitem{Dean}D.~S.~Dean and A.~Lef{\`e}vre, Phys. Rev. Lett., {\bf 90}, 198301 (2003).

\bibitem{Maxwell}J.~C.~Maxwell, Phil. Mag. {\bf 27}, 250 (1864).

\bibitem{Ball}R.~C.~Ball and R.~Blumenfeld, Phys. Rev. Lett. {\bf 88} 115505 (2002).

\bibitem{Song}K.~Wang, C.~Song, P.~Wang, and H.~A.~Makse, Europhys. Lett. {\bf 91}, 68001 (2010) and Phys. Rev. E {\bf 86}, 011305 (2012).

\bibitem{Blumenfeld2}R.~Blumenfeld, J.~F.~Jordan, and S.~F.~Edwards, Phys. Rev. Lett. {\bf 109}, 238001 (2012).

\bibitem{Snoeijer1}J.~H.~Snoeijer, T.~J.~H.~Vlugt, M.~van~Hecke and W.~van~Saarloos, Phys. Rev. Lett. {\bf 92}, 054302 (2004).

\bibitem{Snoeijer2}J.~H.~Snoeijer, T.~J.~H.~Vlugt, W.~G.~Ellenbroek, M.~van~Hecke  and J.~M.~J.~van~Leeuwen, Phys. Rev. E {\bf 70}, 061306 (2004).

\bibitem{Tighe2}B.~P.~Tighe, J.~H.~Snoeijer, T.~J.~H.~Vlugt, and M.~van~Hecke, Soft Matter  {\bf 6}, 2908 (2010).

\bibitem{Wyart}M.~Wyart, S.~R.~Nagel, and T.~A.~Witten, Europhys. Lett. {\bf 72}, 486 (2005).


\end{thebibliography}
\end{document}